\documentclass[11pt]{article}
\usepackage{graphicx,amsmath,amssymb,epsf,fdag}
\usepackage{latexsym,bm, slashed}
\usepackage{xcolor}
\definecolor{dark}{rgb}{0.10,0.2,0.3}
\definecolor{magenta}{rgb}{0.7,0.1,0.3}
\definecolor{purpure}{rgb}{0.5,0.15,0.3}
\usepackage[font=small,format=plain,labelfont=bf,up,textfont=it,up]{caption}
\usepackage{hyperref, cite}
\hypersetup{colorlinks,
citecolor=blue,
filecolor=blue,
linkcolor=magenta,
urlcolor=purpure,hyperfootnotes=true,pdftex}

\begin{document}

\title{ \bf The quark induced Mueller-Tang jet impact factor at next-to-leading order
%\footnote{Preprint numbers: xxx}
}

\author{ M.~Hentschinski$^1$, J.~D.~Madrigal Mart\'inez$^2$, B.~Murdaca$^3$,  A.~Sabio Vera$^{4,5}$ 
\bigskip \\
{ $^1$~Physics Department, Brookhaven National Laboratory, Upton, NY, 11973, USA.}
\\
{ $^2$~Institut de Physique Th{\'e}orique, CEA Saclay, F-91191 Gif-sur-Yvette, France.}
\\
{ $^3$~Dipartimento di Fisica,   Universit\`a della Calabria, and} \\
{
Istituto Nazionale di Fisica Nucleare, Gruppo collegato di Cosenza,} \\{ I-87036 Arcavacata di Rende, Cosenza, Italy.}\\
{ $^4$~Departamento de F{\' 	\i}sica Te{\' o}rica \& Instituto de F{\' \i}sica Te{\' o}rica UAM/CSIC,}\\
{Universidad Aut{\' o}noma de Madrid,  Cantoblanco E-28049 Madrid, Spain.}\\
{ $^5$~CERN, Geneva, Switzerland.}
}
%\date{}
\maketitle

\begin{abstract}
  We present the NLO corrections for the quark induced forward production of a jet with
  an associated rapidity gap.  We make use of Lipatov's QCD high energy effective action
  to calculate the real emission contributions to the so-called Mueller-Tang impact factor. We combine 
  them with the previously calculated virtual corrections and verify ultraviolet and collinear
  finiteness of the final result.
\end{abstract}

\section{Introduction}
\label{sec:intro}

A very interesting test of QCD in the high energy limit is provided
by dijet events with associated rapidity gaps.  As originally pointed out by Mueller
and Tang \cite{Mueller:1992pe}, this type of events, when the tagged jets are far apart in 
rapidity, allow for the study of 
the Balitsky-Fadin-Kuraev-Lipatov (BFKL) hard pomeron~\cite{BFKL1} at finite momentum
transfer $t \neq 0$.  Absence of hadronic activity over a large region in
rapidity $\Delta y_{\text{gap}}$ suggests that an important contribution to the dijet
cross-section is due to configurations with color singlet exchange in
the $t$-channel. Such exchange is well described by the non-forward BFKL Green's function with finite momentum transfer.
Unlike the case of zero momentum transfer, which describes the rise of total cross-sections
and  has been investigated for a number of observables (see {\it e.g.}~\cite{LEP, HERA, LHCdijet})  
the BFKL dynamics with finite momentum transfer remains relatively unexplored.
\begin{figure}[thb]
  \centering
\parbox{.29\textwidth}{\includegraphics[width = .29\textwidth]{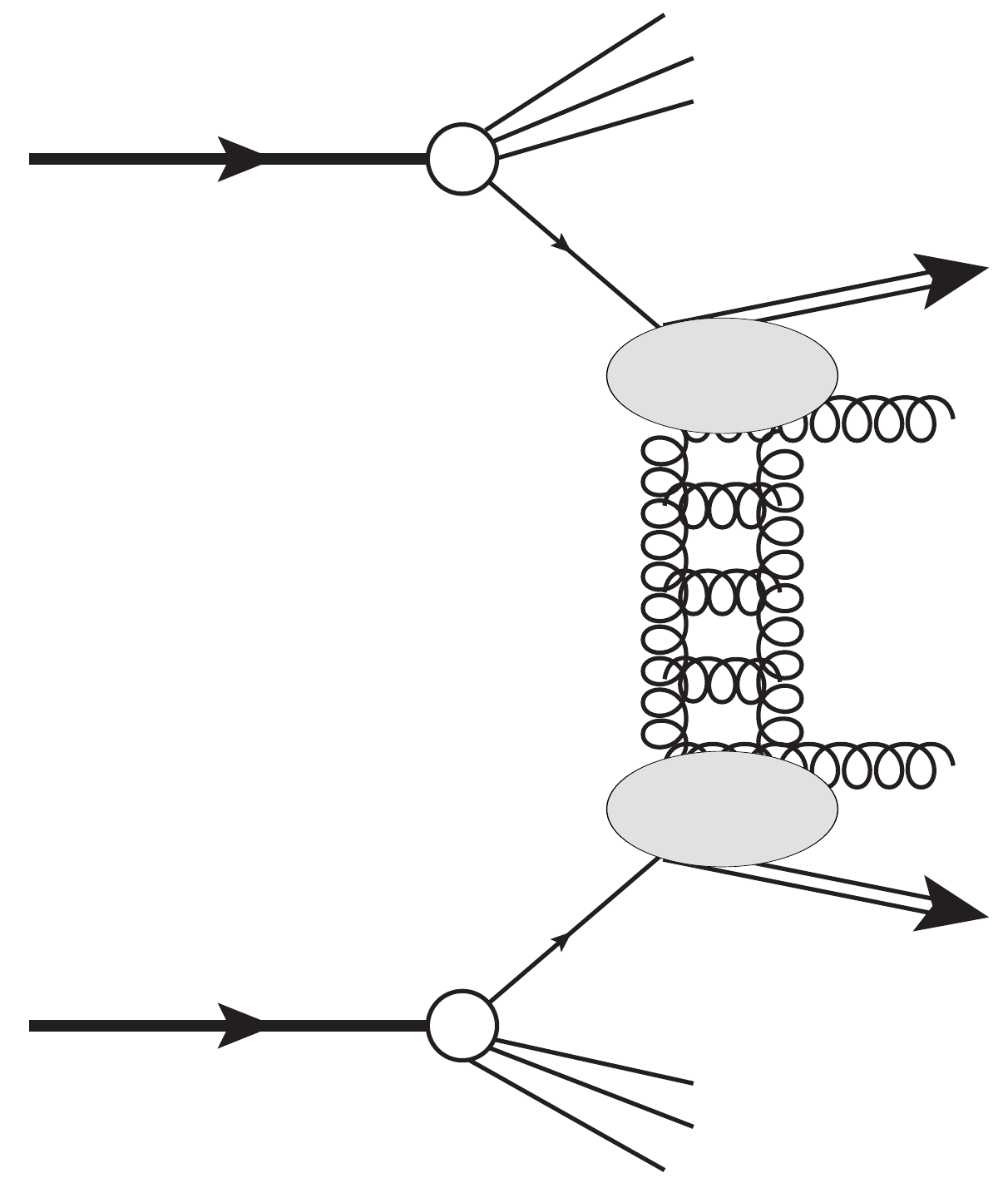}}
\parbox{.29\textwidth}{\includegraphics[width = .29\textwidth]{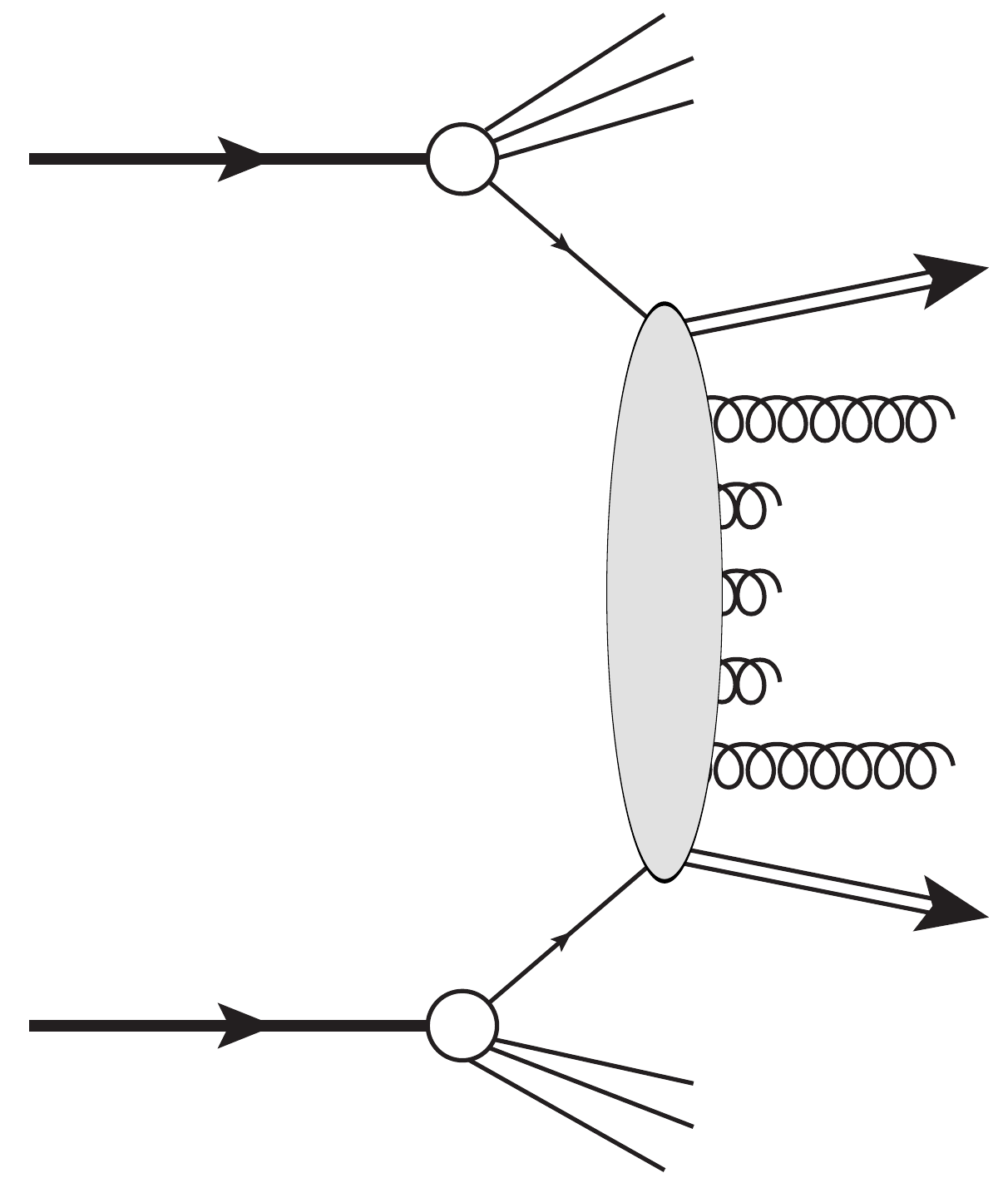}}
\parbox{.29\textwidth}{\includegraphics[width = .29\textwidth]{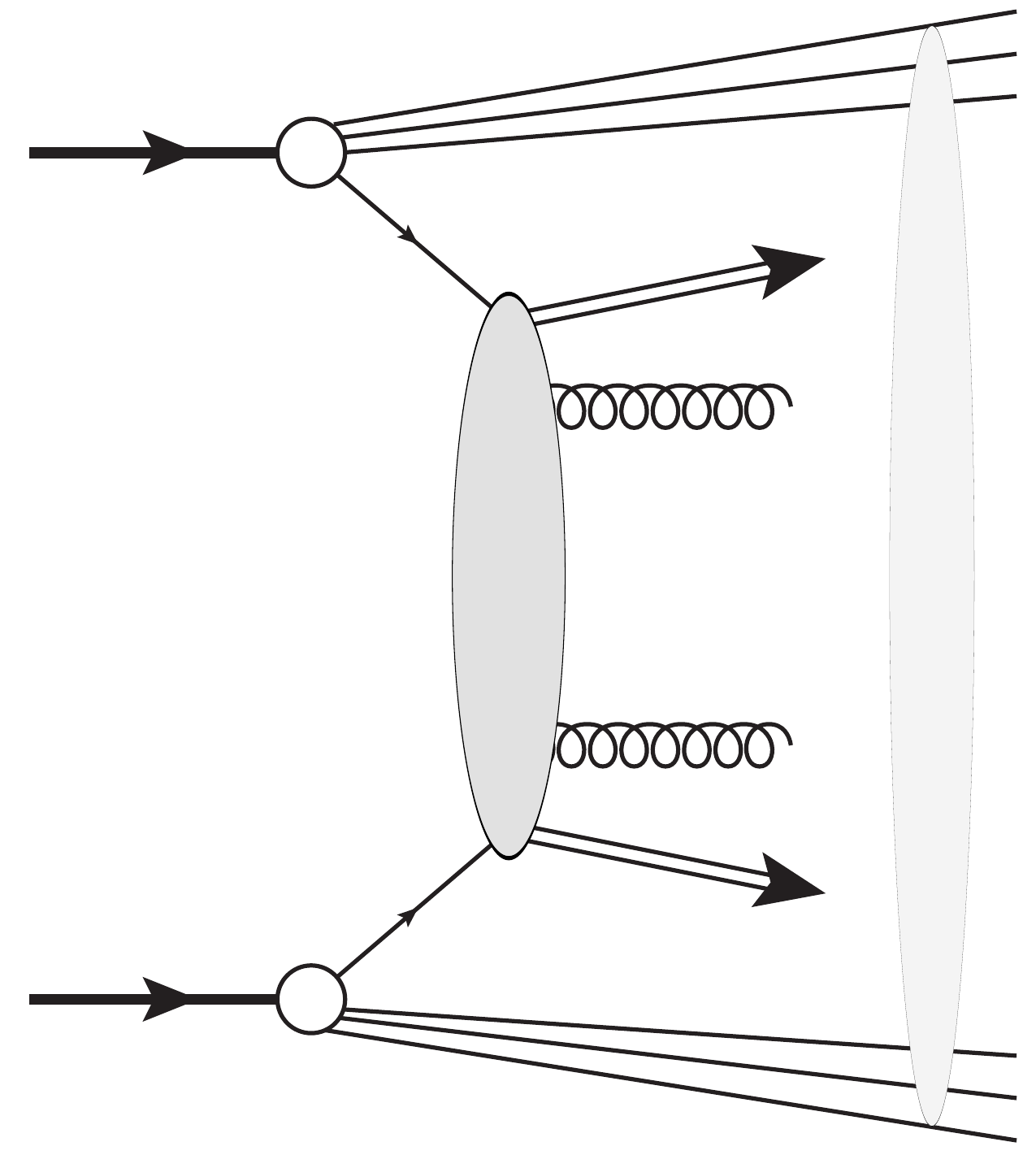}}
\\
\parbox{.29\textwidth}{\center (a)}
\parbox{.29\textwidth}{\center (b)}
\parbox{.29\textwidth}{\center (c)}

  \caption{Contributions to the Mueller-Tang cross-section{: a) color singlet exchange; b) emissions with $p_T$ smaller   than the experimental resolution in the rapidity gap; }c) soft rescattering of the hadron remnants which destroy the rapidity gap.}
  \label{fig:different_contributions}
\end{figure}
While dijets with associated rapidity gaps allow to access such dynamics, precise
phenomenology remains a challenging task. 

The configurations of
interest with color singlet exchange,
Fig.~\ref{fig:different_contributions}.a, which do not generate any
emission into the gap, compete with color exchange contributions where
emissions are allowed up to a scale set by the experimental resolution
$E_{\text{gap}}$ of the rapidity gap definition, 
Fig.~\ref{fig:different_contributions}.b. Moreover jet-gap-jet events
are affected by soft rescatterings of the proton remnants which
destroy the rapidity gap and lead to a violation of collinear
factorization, see Fig.~\ref{fig:different_contributions}.c; for
further details we refer to \cite{gapStudies} and references therein.

 In this work we study the  color
 singlet $t$-channel exchange contribution within the framework of
 high energy factorization. A complete description is currently only
 available at leading logarithmic (LL) accuracy, where terms enhanced
 by the gap size $(\alpha_s \Delta y_{\text{gap}})^n$ are resummed to
 all orders in the strong coupling $\alpha_s$.  Phenomenological
 studies, including a comparison to data by the D0 and CDF
 collaboration at Tevatron/Fermilab have been performed in
 \cite{Enberg:2001ev, Motyka:2001zh} and later on by
 \cite{Chevallier:2009cu}, where a subset of the NLO corrections was included.  
 Given the  importance of the NLO corrections to both impact factors and Green's
 function found in the forward limit, a similar study in the non-forward case is mandatory.  
 High precision in the calculations is even more pressing since the BFKL driven color singlet exchange 
 needs to be isolated from other competing contributions. In particular, the study of the possible effects due to non-perturbative gap survival
 probability factors, makes an accurate description of the
 perturbative sub-process crucial for the correct understanding of the diffractive observables.  While the NLO non-forward BFKL
 kernel is well known \cite{Fadin:1998fv}, both in momentum and
 configuration space \cite{dipole}, the NLO corrections for the impact
 factors are only available at the level of virtual corrections {\it
   i.e.} for elastic parton-parton scattering amplitudes
 \cite{Fadin:1999df, Fadin:1999de}.

 Here we calculate the NLO impact factors for quark induced jets with color singlet exchange
 using Lipatov's effective action \cite{LevSeff}. The determination of
 the gluon induced jets will be addressed in a follow-up paper \cite{toappear}. The calculation of higher order 
 corrections from the effective action approach \cite{LevSeff} has been successfully explored by our 
 group in recent
 years. In particular, both NLO impact factors for quark
 \cite{quarkjet} and gluon induced forward jets \cite{gluonjet} and
 the gluon Regge trajectory up to two loops \cite{traject} have been 
 obtained making use of Lipatov's effective action and a set of
 supplementary calculational rules. In the following we will use this
 framework for the determination of the missing real NLO corrections
 which will be then combined at partonic level with the already known
 virtual corrections. Introducing a jet definition and integrating over the
 real emission phase space, we finally verify that all remaining
 singularities are removed by renormalization of the QCD Lagrangian
 and collinear factorization. While, in general, infrared
 finiteness is to be expected, it presents an important result in the present context, 
 given the notoriously complicated perturbative QCD environment
 for jet-gap-jet events.

The outline is as follows: Sec.~\ref{sec:def} provides a
definition of the NLO Mueller-Tang jet impact factor and  
Sec.~\ref{sec:eff} contains a short review of the
high energy effective action. In Sec.~\ref{sec:partonic} we give some
details on the derivation of the leading order Mueller-Tang impact
factor and the real next-to-leading order corrections from the
effective action.  Sec.~\ref{sec:jet} addresses the definition of the
quark induced Mueller-Tang jet vertex at NLO within
collinear factorization. In Sec.~\ref{sec:concl} we summarize our
results, already presented in \cite{letter},  and provide an outlook for future work. Two appendices gather
additional material concerning the high energy limit of the NLO impact
factor (Appendix~\ref{sec:central_prod}) and explicit results for the
inclusive (perturbative) Pomeron - quark impact factor
(Appendix~\ref{sec:incl}).

\section{The NLO Mueller-Tang impact factor - definition}
\label{sec:def}

We are interested in the  hadron-hadron scattering process
\begin{align}
  \label{eq:process}
  h(p_A) + h(p_B) \to  J_1(p_{J,1}) +   J_2(p_{J,2}) + \text{gap},
\end{align}
with two jets produced in the final state separated by a large rapidity gap, 
 which is characterized by no hadronic activity in the detectors.  In addition we limit ourselves
to color singlet exchange in the the $t$-channel. As outlined in the
introduction, the latter constraint is at first a theoretical one and
it remains a task for future phenomenological analysis to
determine those observabels for which this configuration is
dominant.  For an interesting proposal in this direction see
\cite{Marquet:2012ra}.

With quark exchange in the $t$-channel suppressed by a factor $ \sim
\exp (-\Delta y_{\text{gap}}) $, color singlet exchange appears for
large $\Delta y_{\text{gap}}$ for the first time at
$\mathcal{O}(\alpha_s^4)$. It constitutes therefore a NNLO correction,
relative to the conventional dijet cross-section.  While this is
beyond the reach of current exact calculations, the presence of a
large rapidity gap suggests that a description of this process in
terms of high energy factorized amplitudes can provide a good
approximation to the full result.  To this end we define light-cone
vectors as rescaled light-like momenta of the incoming hadrons $n^\pm
= 2 p_{A,B}/\sqrt{s}$ with $s= 2 p_A \cdot p_B$. Assuming massless
jets, the Sudakov decomposition of the external particle momenta reads
\begin{align}
  \label{eq:SudaAMP}
  p_A & = p_A^+ \frac{n^-}{2}, & 
p_{J,1} & = \sqrt{{\bm k}_{J,1}^2}  \left(  e^{y_{J,1}}\frac{n^-}{2} +   e^{-y_{J,1}} \frac{n^+}{2} \right) + {\bm k}_{J,1}; \notag \\
  p_B & = p_B^- \frac{n^+}{2}, & p_{J,2} & =  \sqrt{{\bm k}_{J,2}^2}  \left(  e^{y_{J,2}}\frac{n^-}{2} +   e^{-y_{J,2}} \frac{n^+}{2} \right) + {\bm k}_{J,2}; 
\end{align}
with $({\bm k}_{J, i}, y_{J,i})$, $i = 1,2$ being the transverse momenta and
rapidity of the jet.  To obtain the hadronic dijet cross-section,
we  first calculate the corresponding partonic cross-sections. For
quark induced jets we need the leading order (LO) high energy limit
of the process
\begin{align}
  \label{eq:partonicLO}
  q(p_a) + q(p_b) \to q(p_1) + q(p_2),
\end{align}
with color singlet exchange in the $t$-channel. In the high energy
limit $\hat{s} \to \infty$ with $\hat{s} = 2 p_a \cdot p_b$ and in $d
= 4 + 2 \epsilon$ dimensions, the partonic cross-section, rederived in
Sec.~\ref{sec:LO}, reads
\begin{align}
  \label{eq:Born}
  d \hat{\sigma}_{ab} &=  h_{q,a}^{(0)} h_{q,b}^{(0)} \left[ \int \frac{d^{2 + 2\epsilon} {\bm l}_1}{\pi^{1 + \epsilon}} \frac{1}{{\bm l}_1^2 ({\bm k} - {\bm l}_1)^2} \right]\left[ \int \frac{d^{2 + 2\epsilon} {\bm l}_2}{\pi^{1 + \epsilon}} \frac{1}{{\bm l}_2^2 ({\bm k} - {\bm l}_2)^2} \right] d^{2 + 2\epsilon} {\bm k} \, .
\end{align}
Here $h_q^{(0)}$ denotes the LO impact factor. Resummation
of $\Delta y_{\text gap}$ enhanced terms to all orders in the strong
coupling $\alpha_s$ is then achieved through replacing the transverse
gluon propagators with the non-forward BFKL Green's function $G( {\bm
  l}, {\bm l}', {\bm q}, s/s_0)$, where the latter is obtained as a
solution to the non-forward BFKL equation. 

The resummed cross-section
takes the form
\begin{align}
  \label{eq:Born}
  d \hat{\sigma}_{ab}^{\text{res.}} &=  
h_{q,a}^{{(0)}} h_{q,b}^{{(0)}} \left[ \int \frac{d^{2 + 2
    \epsilon}
    {\bm l}_1} {{\pi}^{1 + \epsilon}} \int   \frac{d^{2 + 2 \epsilon}
    {\bm l}_1'}{{\pi}^{1 + \epsilon}}
G\left( {\bm l}_1, { \bm l}_1', {\bm k},  \frac{\hat{s}}{s_0} \right)
 \right]
\notag \\
& 
\qquad \qquad \qquad \qquad
\left[ \int \frac{d^{2 + 2 \epsilon} {\bm l}_2}{\pi^{1 + \epsilon}}
\int \frac{d^{2 + 2\epsilon} {\bm l}'_2}{\pi^{1 + \epsilon}}
G\left( {\bm l}_2, {\bm l}_2', {\bm k}, \frac{\hat{s}}{s_0}\right)
 \right] d^{2 + 2\epsilon} {\bm k}.
\end{align}
In this expression $s_0$ denotes the reggeization scale, which
parametrizes the scale uncertainty due to the all order
resummation. Constraining the $s_0$ dependence is an additional
benefit of a complete NLO treatment while the natural choice for $s_0$
is $\ln (\hat{s}/s_0) = \Delta y_{\text{gap}}$.  Apparently both
transverse integrals in Eq.~\eqref{eq:Born} are divergent and a
suitable infrared regulator is needed. This divergence is in principle
also present in the (LO) Green's function.  However, in the asymptotic
limit $\ln (\hat{s}/s_0) \to \infty$, the dependence on the infrared
regulator vanishes and the result turns out to be finite
\cite{Motyka:2001zh}. The combination with an approach resumming 
logarithms in the jet transverse momentum and the gap resolution
$E_{\text{gap}}$, including a matching of singularities at finite
perturbative orders of the BFKL Green's function has been discussed in
\cite{Forshaw:2005sx}.  In the following we assume that these
singularities are addressed in a suitable way, either through a
suitable matching and/or by working in the strict asymptotic limit
$\ln (\hat{s}/s_0) \to \infty$. In particular, the integrals over
transverse momentum are assumed to yield a finite result.

To calculate the NLO impact factors it is needed to determine
both the 1-loop corrections to the process~(\ref{eq:partonicLO})
and the leading order process
\begin{align}
  \label{eq:partonicNLOreal}
   q(p_a) + q(p_b) \to q(p_1) + q(p_2) + g(q),
\end{align}
with color singlet exchange in one of the $t$-channels $t_1 = (p_a -
p_1)^2$ and $t_2 = (p_b - p_2)^2$. The 1-loop corrections to
Eq.~(\ref{eq:partonicLO}) have been obtained in
\cite{Fadin:1999df}. As the non-forward BFKL Green's function
generates no real emissions, the entire $s_0$ dependence is for this
particular process contained in the virtual corrections to the impact
factors. As verified in \cite{Fadin:1999df}, the $s_0$ dependence
cancels if the all-order Green's function is truncated at NLO.

Both at LO and NLO, it is necessary to restrict
the phase space of the final state system, to avoid particle emissions into the
forbidden gap region.  To be more precise, we will require that the
invariant mass of the diffractive system in the forward region of each
proton to be smaller than a certain upper cut-off $M_{X,
  \text{max}}^2$, set by experiment.   At LO, contributions to the diffractive system
are due to initial state radiation, which is encoded in the parton
distribution functions, while at NLO  this includes 
also contributions from the produced partonic system. For the
diffractive system in the forward region of the proton with momentum
$p_A$ we find (the $t$-channel momentum is $k = p_b - p_2$) at
partonic level:
\begin{align}
  \label{eq:2}
  \hat{M}_X^2 & = (p_a + k)^2 =  p_a^+ k^- - {\bm k}^2\; .
\end{align}
At LO $\hat{M}_X^2 = 0$, while $\hat{M}_X^2$ assumes a
finite value at NLO.  It is related to the diffractive mass at
hadronic level through
\begin{align}
  \label{eq:hadronic}
  \hat{M}_X^2 & =   x_1    M_X^2 - (1-x_1){\bm k}^2 \; .
\end{align}
The constraint $M_X^2 <  M_{X, \text{max}}^2$ leads then to the  following lower bound on the proton  momentum fraction of the incoming parton
\begin{align}
  \label{eq:x0}
  x_1 > x_0 = \frac{{\bm k}^2}{M_{X, \text{max}}^2 + {\bm k}^2} \;,
\end{align}
while at NLO, in addition, the phase space of the partonic quark-gluon
final state system is constrained.

To obtain the dijet cross-section from the partonic process, we
further require a function which selects the configurations
contributing to the particular choice of jet definition made in an
experiment from the full partonic final state phase space.  Formally, 
this is achieved through the convolution with a distribution $S_J$,
which contains the details about the chosen jet
algorithm. Schematically, the partonic differential cross-section reads
\begin{align}
  \label{eq:djetsigma}
\frac{d \hat{\sigma}_J}{  d J_1 d J_2   d^2{\bm k}}  &= 
\,{d \hat{\sigma}} \, \otimes \, S_{J_1}  S_{J_2}  \,
\, \; ,
\end{align}
with  $d J_i = d^2 {\bm p}_{J_i} d y_{J_i}$  the jet phase space and $ {\bm k}$ the $t$-channel transverse momentum transfer. 
At leading order, ${\bm k}$ coincides with the transverse momentum of the jet and 
the jet functions are trivial. They map each of the final state quarks with one of the jets through
\begin{align}
  \label{eq:S0}
  S^{(2)}_{J_i} ( {\bm p}_i, x_i) &  
 =  x_i \, \delta \left(x_i - \frac{|{\bm k}_{J,i}|{e^{y_{J,i}}}}{\sqrt{s}} \right)\delta^{2 + 2\epsilon} ({\bm p}_i - {\bm k}_{J_i}),
 & i   &= 1,2 \;.
\end{align}
In particular, due to the large rapidity gap spanned between the two
jets, the quark with momentum $p_i$ corresponds directly to the jet
with momentum $p_{J,i}$, $i=1,2$. The full NLO treatment
will be addressed in Sec.~\ref{sec:jet}.  We stress that in addition to the phase space of the two jets, the cross-section Eq.~\eqref{eq:djetsigma} is also
differential in the (transverse) momentum transfer through the gap
region. As long as LO impact factors are used, the
cross-section describes essentially elastic quark-quark scattering and
this momentum transfer is identical to the transverse jet momentum. As
soon as the diffractive system contains other objects than the jet,
scattering is no longer elastic and both transverse momenta
deviate. In principle, this quantity is measurable as the total
transverse momentum of the diffractive systems. In practical
applications, the cross-section of Eq.~\eqref{eq:djetsigma} is probably
too differential and one would prefer to integrate over some of the
variables. A possibility is to make the deviation from purely elastic
scattering explicit through the following parametrization of the jet
transverse momenta
\begin{align}
  \label{eq:parametrizePT}
  {\bm k}_{J, 1} & = {\bm p} + \Delta {\bm p}, & {\bm k}_{J, 2} & = -{\bm p} + \Delta {\bm p}, 
\end{align}
and to integrate both over $\Delta {\bm p }$ and ${\bm k}$ as well as the overall rapidity $y = y_1 + y_2$, resulting into  a cross-section differential in the mean transverse momentum $2{\bm p} = {\bm k}_{J,1} - {\bm k}_{J,2}$ and the rapidity difference of the two jets. Here we focus on the determination of impact factors and  leave such details concerning the construction of suitable  observables out of the  (maximal) differential cross-section as a task for a future phenomenological studies.

To the end of  describing the full hadronic process of Eq.~(\ref{eq:process})
we convolute the partonic process in Eq.~(\ref{eq:djetsigma}) with parton
distribution functions.  In principle, the use of collinear
factorization can be questioned for this type of processes since soft
re-scattering of the hadron remnants can destroy the rapidity gap and
lead to its violation.  At partonic level,
violation of factorization is manifest through the divergent
transverse integrals in Eq.~(\ref{eq:Born}) at finite $\ln
\hat{s}/s_0$.  In the following we use the working assumption that all
initial state collinear singularities of the impact factors can be
consistently absorbed through the conventional redefinition of the
parton distribution functions.  We will demonstrate in
Sec.~\ref{sec:jet} that this is actually the case and that the impact
factor itself is a well-defined quantity.  The final cross-section
then takes the following form
\begin{align}
\label{eq:1}
  \frac{ d \sigma}{ dJ_1 \, dJ_2\,\,d^2{\bm k} } & =  \sum_{l,k = q,\bar{q}} \int_0^1 d x_1 \, \int_0^1 d x_2 \, \,
  f_{l/p}^{\text{gap}}(x_1, \mu_f) \, \,  f_{k/p}^{\text{gap}} (x_2, \mu_f)  \, \,  H_{kl}(x_1, x_2, \mu_f),
 \end{align}
 where we suppressed for the collinear coefficient $H_{kl}$ the
 dependence on the final state variables. At LO, $H_{qq}$
 coincides with the partonic cross-section in Eq.~(\ref{eq:djetsigma}), while the 
 NLO treatment  requires at first the identification of initial state
 collinear singularities.  Furthermore, we added to each of the (anti-)
 quark distribution functions the superscript `gap'. This is meant to
 indicate that these distributions do not necessarily coincide with
 standard pdfs. In phenomenological applications they may be
 calculated from the standard pdfs using phenomenological gap
 survival probability factors and/or restricting to certain
 combinations of observables which are insensitive to possible soft
 rescatterings, see {\it e.g.} \cite{Marquet:2012ra}.

 \section{The High-Energy Effective Action}
\label{sec:eff}

For the calculation of the missing real NLO corrections, we make use
of Lipatov's high
energy effective action \cite{LevSeff}. Within this framework, QCD amplitudes are in the high energy limit decomposed into gauge invariant sub-amplitudes which are localized in rapidity space and describe the coupling  of quarks ($\psi$), gluon ($v_\mu$) and ghost ($\phi$) fields  to a new degree of freedom, the reggeized gluon field $A_\pm (x)$. The latter  is introduced as a convenient tool to reconstruct the complete QCD amplitudes in the high energy limit out of the sub-amplitudes restricted to small rapidity intervals.

Lipatov's effective action is obtained by adding an induced term $
S_{\text{ind.}}$ to the QCD action $S_{\text{QCD}}$,
\begin{align}
  \label{eq:effac}
S_{\text{eff}}& = S_{\text{QCD}} +
S_{\text{ind.}}\; ,
\end{align}
where the induced term $ S_{\text{ind.}}$ describes the coupling of
the gluonic field $v_\mu = -it^a v_\mu^a(x)$ to the reggeized gluon
field $A_\pm(x) = - i t^a A_\pm^a (x)$.  High energy factorized
amplitudes reveal strong ordering in plus and minus components of
momenta which is reflected in the following kinematic constraint
obeyed by the reggeized gluon field:
\begin{align}
  \label{eq:kinematic}
  \partial_+ A_- (x)& = 0 = \partial_- A_+(x).
\end{align}

Even though the reggeized gluon field is charged under the QCD gauge
group SU$(N_c)$, it is invariant under local gauge transformation
$\delta A_\pm = 0$.  Its kinetic term and the gauge invariant coupling
to the QCD gluon field are contained in the induced term
\begin{align}
\label{eq:1efflagrangian}
  S_{\text{ind.}} = \int \text{d}^4 x \,
\text{tr}\left[\left(W_-[v(x)] - A_-(x) \right)\partial^2_\perp A_+(x)\right]
+\text{tr}\left[\left(W_+[v(x)] - A_+(x) \right)\partial^2_\perp A_-(x)\right],
\end{align}
with 
\begin{align}
  \label{eq:funct_expand}
  W_\pm[v(x)] =&
v_\pm(x) \frac{1}{ D_\pm}\partial_\pm,% =  
%- \frac{1}{g} \partial_\pm  U[v_\pm]
%= v_\pm - g  v_\pm\frac{1}{\partial_\pm} v_\pm + g^2 v_\pm
%\frac{1}{\partial_\pm} v_\pm\frac{1}{\partial_\pm} v_\pm - \ldots
&
D_\pm & = \partial_\pm + g v_\pm (x).
\end{align}

\begin{figure}[htb]
    \label{fig:subfigures}
   \centering
   \parbox{.7cm}{\includegraphics[height = 1.8cm]{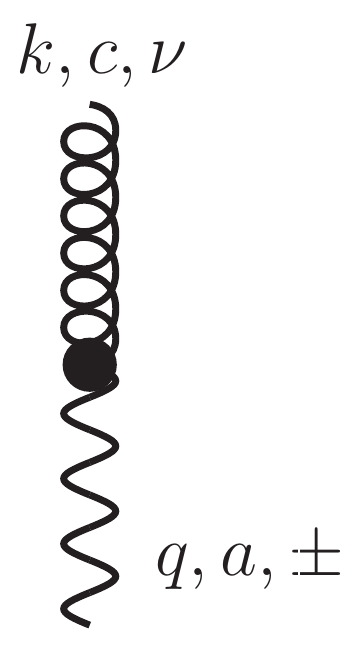}} $=  \displaystyle 
   \begin{array}[h]{ll}
    \\  \\ - i{\bm q}^2 \delta^{a c} (n^\pm)^\nu,  \\ \\  \qquad   k^\pm = 0.
   \end{array}  $ 
 \parbox{1.2cm}{ \includegraphics[height = 1.8cm]{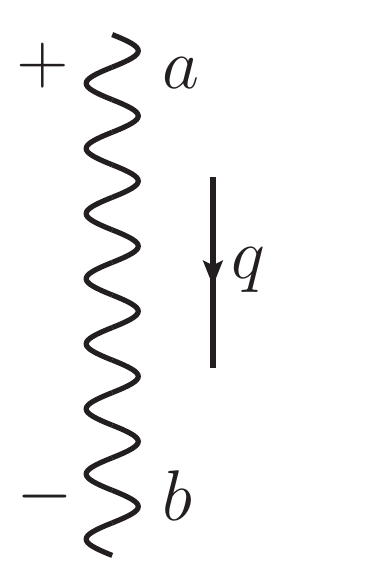}}  $=  \displaystyle    \begin{array}[h]{ll}
    \delta^{ab} \frac{ i/2}{{\bm q}^2} \end{array}$ 
 \parbox{1.7cm}{\includegraphics[height = 1.8cm]{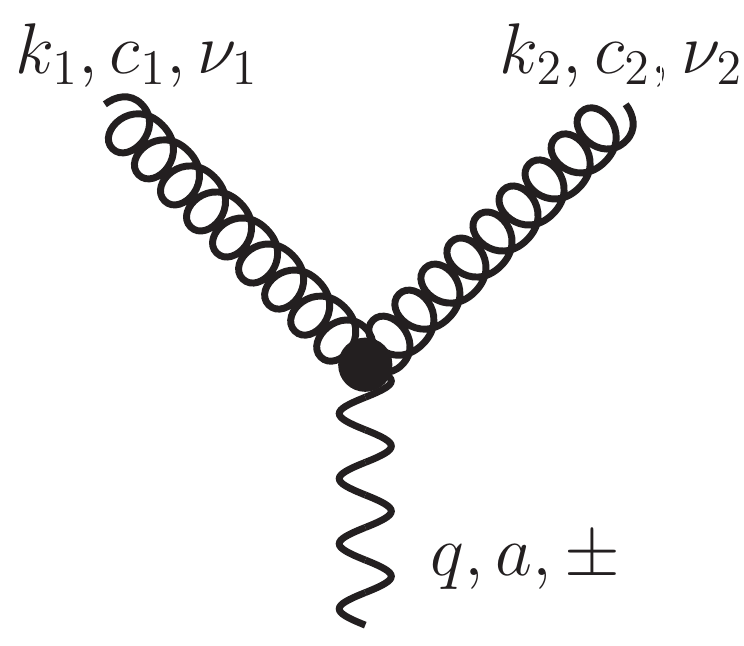}} $ \displaystyle  =  \begin{array}[h]{ll}  \\ \\ g f^{c_1 c_2 a} \frac{{\bm q}^2}{k_1^\pm}   (n^\pm)^{\nu_1} (n^\pm)^{\nu_2},  \\ \\ \quad  k_1^\pm  + k_2^\pm  = 0
 \end{array}$
 \\
\parbox{4cm}{\center (a)} \parbox{4cm}{\center (b)} \parbox{4cm}{\center (c)}

 \caption{\small Feynman rules for the lowest-order effective vertices of the effective action. Wavy lines denote reggeized fields and curly lines gluons. }
\label{fig:3}
\end{figure}

For a more in depth discussion of the effective action we refer to the
recent review \cite{review}. Due to the induced term in
Eq.~(\ref{eq:effac}), the Feynman rules of the effective action
comprise, apart from the usual QCD Feynman rules, the propagator of
the reggeized gluon and an infinite number of so-called induced
vertices.  Vertices and propagators needed for the current study are
collected in Fig.~\ref{fig:3}.
Determination of NLO corrections using this effective action approach has been addressed recently  to a certain extent, through the explicit calculation of the NLO corrections to both quark  \cite{quarkjet} and gluon \cite{gluonjet} induced forward jets (with associated radiation) as well as the determination of the gluon Regge trajectory up to 2-loops \cite{traject}. These previous applications have all in common that they are, at amplitude level, restricted to a color octet projection and, therefore, single reggeized gluon exchange. Due to the particular color structure of the reggeized gluon field, which is restricted to the anti-symmetric color octet, see Fig.~\ref{fig:3} and \cite{LevSeff, Hentschinski:2011xg},  color singlet exchange requires to go beyond a single reggeized gluon exchange and to consider the two reggeized gluon exchange contribution.

\section{The high energy factorized cross-section at partonic level}
\label{sec:partonic}

\subsection{The Mueller-Tang jet  cross-section at LO}
\label{sec:LO}

\begin{figure}[th]
  \centering
% \parbox{5cm}{\center \includegraphics[width= .25\textwidth]{jetjetMTXsec.pdf}}
 \parbox{4cm}{\center \includegraphics[height = 3cm]{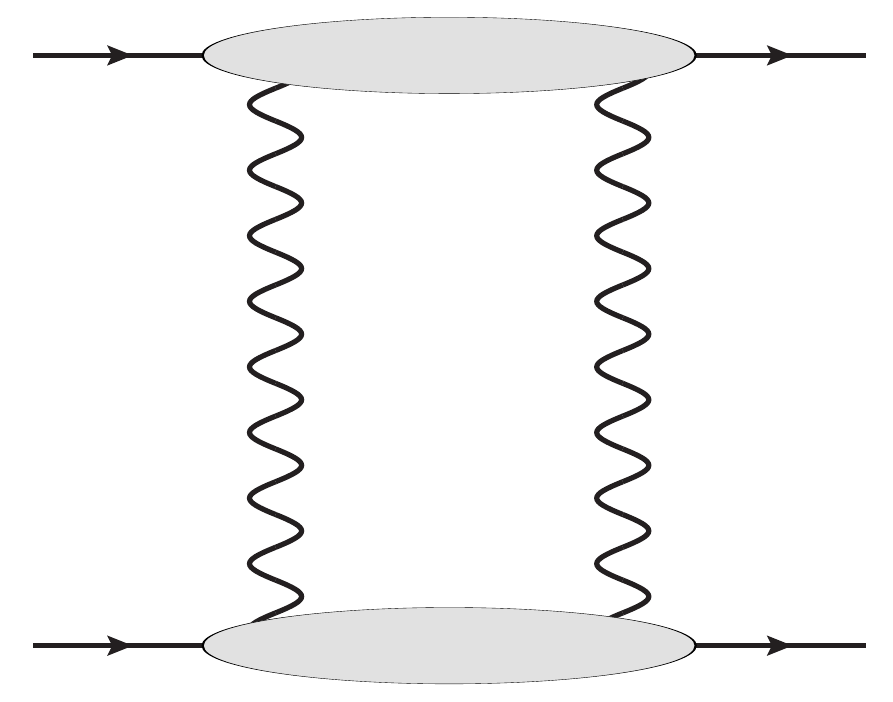}}
\parbox{.7\textwidth}{\center\includegraphics[width = 2.5cm]{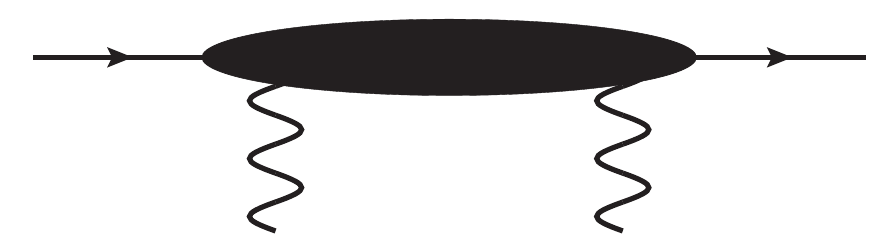} $=$  
\includegraphics[width = 2.5cm]{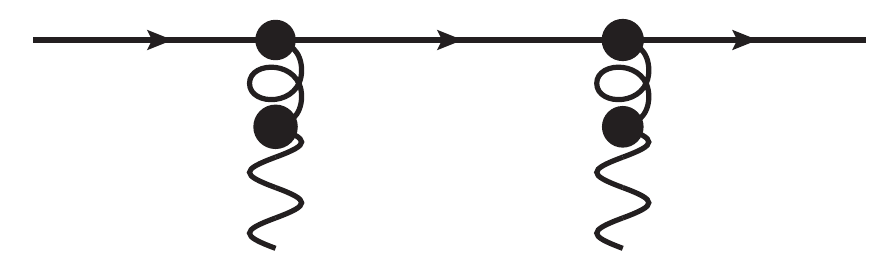} + \includegraphics[width = 2.5cm]{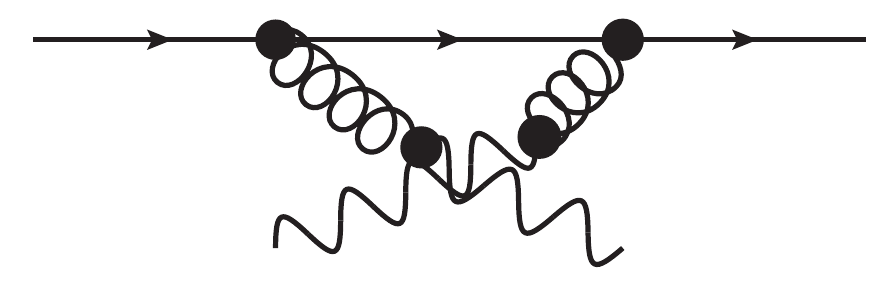}  }  

\parbox{4cm}{\center (a)}\parbox{.7\textwidth}{\center (b)}
  \caption{a) The LO amplitude for quark induced jets in the high energy approximation. The 2 reggeized gluon state  in the $t$-channel is projected on the color singlet. b) LO diagrams which describe within the effective action the coupling of the two-reggeized gluon state to the quark.}
  \label{fig:schematic}
\end{figure}

The Mueller-Tang jet impact factor at leading order can  be
determined from the elastic scattering amplitude $q(p_a) + q(p_b) \to
q(p_1) + q(p_2)$ with color singlet exchange.  In the high energy
limit, such a color singlet exchange is within the
effective action -- to LO in the strong coupling $\alpha_s$
-- provided by the $t$-channel exchange of two reggeized gluons in the
color singlet state, Fig.~\ref{fig:schematic}.b. With the Sudakov decomposition of the external momenta,
\begin{align}
  \label{eq:SudaAMP}
  p_a & = p_a^+ \frac{n^-}{2} & p_1 & = p_a^+ \frac{n^-}{2} + k^- \frac{n^+}{2} + {\bm k} \notag \\
  p_b & = p_b^- \frac{n^+}{2} & p_2 & = p_b^- \frac{n^+}{2} - k^+ \frac{n^-}{2} - {\bm k},
\end{align}
where the kinematic constraint, Eq.~\eqref{eq:kinematic}, has been
taken into account, the Mandelstam invariants read
\begin{align}
  \label{eq:sANDt}
  s & = p_a^+ p_b^- & t = - {\bm k}^2 \; ,
\end{align}
and the   quark-quark scattering amplitude with color singlet
exchange  is at leading order given by
\begin{align}
    \label{eq:mplus2}
    i\mathcal{M}  = \frac{1}{2 \cdot 2!} \int \frac{dl^+ dl^-}{(2\pi)^2} \int \frac{d^{2+2\epsilon} {\bm l}}{(2\pi)^{2+2\epsilon}} 
 i \tilde{\mathcal{M}}^{de}_{q2r_+^* \to q}  P^{de} \cdot  i \tilde{\mathcal{M}}^{d'e'}_{q2r_-^* \to q}  P^{d'e'} \frac{(i/2)^2}{{\bm l}^2({\bm l} - {\bm k})^2} \; ,
\end{align}
with 
\begin{align}
  \label{eq:project_exact}
  \mathcal{P}^{ab, a'b'} & =   P^{ab}  P^{a'b'}, & P^{ab} &= \frac{\delta^{ab}}{\sqrt{N_c^2 -1}},
\end{align}
being the projector onto the color singlet.   The Sudakov decomposition of the  momenta of the  sub-process\footnote{$r_\pm^*=$ reggeized gluon with the index `$\pm$' referring to its  polarization vector $n^\pm$, see also Fig.~\ref{fig:3}.}
$q(p_a) + r_+^*(l_1) + r_+^*(k-l_1) \to q(p)$ 
reads
 \begin{align}
   \label{eq:sudaBORN}
   p_a &= p_a^+ \frac{n^-}{2} &  p &=  p_a^+\frac{n^-}{2} + k^- \frac{n^+}{2} + {\bm k} \notag \\
 l_1 & = l_1^- \frac{n^+}{2} + {\bm l}_1 & k&=  k^-\frac{n^+}{2} + {\bm k},
 \end{align}
with
\begin{align}
  \label{eq:qgqg2}
 i \tilde{\mathcal{M}}^{ed}_{q2r_+^* \to q} \mathcal{P}^{ed} &=-g^2
 \bar{u}(p)    \fdag{n}^+  u(p_a)   \frac{ \delta_{i_1i_a}C_F }{\sqrt{N_c^2 -1}}
 \cdot   
\bigg[ \frac{  2 i }{ l^- - \frac{{\bm l}^2  - i\epsilon}{p_a^+}} 
+ 
\frac{ 2   i   }{ k^- - l^- - \frac{({\bm k}- {\bm l})^2 - i\epsilon}{p_a^+}} \bigg].
\end{align}

Due to Eq.~\eqref{eq:kinematic}, the entire dependence on the
longitudinal loop momenta $l^-$ and $l^+$ is contained in the $qr^*r^*
\to q$ amplitudes. Note that this observation holds also for the case
where higher order corrections to the $qr^*r^* \to q$ amplitude are
included and/or there are additional particles in the final  state.  Due to this
property it is possible to express Eq.~\eqref{eq:mplus2} as a
transverse loop integral alone,
\begin{align}
    \label{eq:mip}
    i\mathcal{M}  =  \int \frac{d^{2+2\epsilon} {\bm l}}{(2\pi)^{2+2\epsilon}}  \phi_{qq,a} \phi_{qq,b}  \frac{1}{{\bm l}^2({\bm l} - {\bm k})^2},
\end{align}
 with
\begin{align}
  \label{eq:impa_def}
  { i}\phi_{qq} & =    \int \frac{d l^-}{8 \pi}  i \tilde{\mathcal{M}}^{ab}_{qr^*r^* \to q} P^{ab} = -  \delta_{i_1 i_a}\frac{g^2}{4}\frac{C_f}{\sqrt{N_c^2 - 1}}  \bar{u}_{(\lambda)}(p) \fdag{n}^+ u_{(\lambda)}(p_a)  .
\end{align}
For later use we also give the result for the leading order gluon
impact factor
\begin{align}
  \label{eq:gluon_lo_imp}
  i\phi_{gg} =\delta_{c_1 c_a} {g^2}\frac{C_a}{\sqrt{N_c^2 - 1}} p_a^+ { \bm \epsilon}_{(\lambda)}^*(p) \cdot { \bm  \epsilon}_{(\lambda)}(p_a),
\end{align}
where gluon polarization vectors in the `right-handed light cone gauge' have been used. The latter obey the conditions
\begin{align}
  \label{eq:rhg1}
  \epsilon_{(\lambda)}(p, n^+) \cdot p & = 0 &  \epsilon_{(\lambda)}(p, n^+) \cdot n^+ & = 0,
\end{align}
and can be parametrized as
\begin{align}
  \label{eq:rhgpara}
   \epsilon^\mu_{(\lambda)}(p, n^+) & = \frac{{\bm \epsilon}_{(\lambda)} \cdot {\bm p}}{p^+} (n^+)^\mu +
 {\bm \epsilon}_{(\lambda)}^\mu.
\end{align}
Using these results, we obtain the LO partonic differential
cross-section for dijets with color singlet exchange as
\begin{align}
  \label{eq:Born}
  d \hat{\sigma}_{ab} &=  h_{q,a}^{(0)} h_{q,b}^{(0)} \left[ \int \frac{d^{2 + 2\epsilon} {\bm l}_1}{\pi^{1 + \epsilon}} \frac{1}{{\bm l}_1^2 ({\bm k} - {\bm l}_1)^2} \right]\left[ \int \frac{d^{2 + 2\epsilon} {\bm l}_2}{\pi^{1 + \epsilon}} \frac{1}{{\bm l}_2^2 ({\bm k} - {\bm l}_2)^2} \right] d^{2 + 2\epsilon} {\bm k},
\end{align}
with
\begin{align}
  \label{eq:XsecLOimp}
  h_q^{(0)} &= \frac{1}{2} \sum_{\text{spin}} \frac{1}{N_c} \sum_{\text{color}} 
\int \frac{d k^-}{2^{-\epsilon}  p_a^+ (2 \pi) (4 \pi)^{2 +2 \epsilon} } \left| \phi_q \right|^2 d\Phi^{(1)}
%\notag \\
%&
=
C_f^2 h^{(0)}
\end{align}
and
\begin{align}
  \label{eq:H0}
  h^{(0)} &=  
 \frac{\alpha_{s, \epsilon}^2  2^\epsilon   }{\mu^{4\epsilon}\Gamma^2(1-\epsilon)(N_c^2 -1)},
\end{align}
in agreement with \cite{Mueller:1992pe}. In the above formula, the
1-particle phase space
\begin{align}
  \label{eq:1pt_ps}
  d\Phi^{(1)} & = 2 \pi \delta( p_ak^- - {\bm k}^2)
\end{align}
and the  dimensionless strong coupling in $d = 4 + 2 \epsilon$
\begin{align}
  \label{eq:MSbar}
 \alpha_{s, \epsilon} = \frac{g^2\mu^{2\epsilon}\Gamma(1-\epsilon) }{(4\pi)^{1 + \epsilon}}
\end{align}
have been used. In the same way we find for the gluonic case
\begin{align}
  \label{eq:H0g}
  h^{(0)}_g & =% \frac{C_a^2}{N_c^2 -1}  \frac{N_c^2 -1}{2 (N^2_c -1)}  (2 + 2 \epsilon) 4 (p_a^+)^2  \frac{1}{2 p_a^+}   \cdot   \frac{1}{p_a^+}\int \frac{d k^-}{2 \pi} 2\pi \delta (k^- - \frac{{\bm p}^2}{p_a^+}) \frac{g^4 }{4(4 \pi)^{1 + \epsilon} (2 \pi)^{1 + \epsilon}} 
%\notag \\
%&=
%\frac{\alpha_s^2 C_a^2 2^\epsilon (1 + \epsilon)  }{\mu^{4\epsilon}\Gamma^2(1-\epsilon)(N_c^2 -1)} %  =
 h^{(0)} (1 + \epsilon) C_a^2 .
\end{align}

As pointed out in Sec.~\ref{sec:def}, both transverse integrals in
Eq.~\eqref{eq:Born} are divergent.  A more detailed study of this
singularity in the context of the high energy effective action is left
as a task for future research.  As we will see below, the presence of
this singularity does not affect the determination of the NLO jet
impact factors, which is the main goal of this paper.

% ein kurzer Check: eq.11 vom gluon jet. ohne den factor k2 (den wir ebenso in den  reggeized gluon exchange stecken koenneten), ist unser H0 genau h0^2/4

%color 1/Nc tr(1234) = 1/4 Nc^2 d12 d34 -> PPPP -> (Nc^2-1)/4 Nc^2 = Cf^2/(Nc^2 -1)
% here  1/Nc tr(PPPP) = Cf^2/(Nc^2 -1)
%      1/Nc tr(12)  = 1/(2Nc) d12 ->  PP 

% moegliche erklaerung: 1/2 vom 

\subsection{The real  NLO  corrections to the impact factors}
\label{sec:me}
\begin{figure}[thb]
  \centering
  \parbox{3cm}{\center \includegraphics[height=2.5cm]{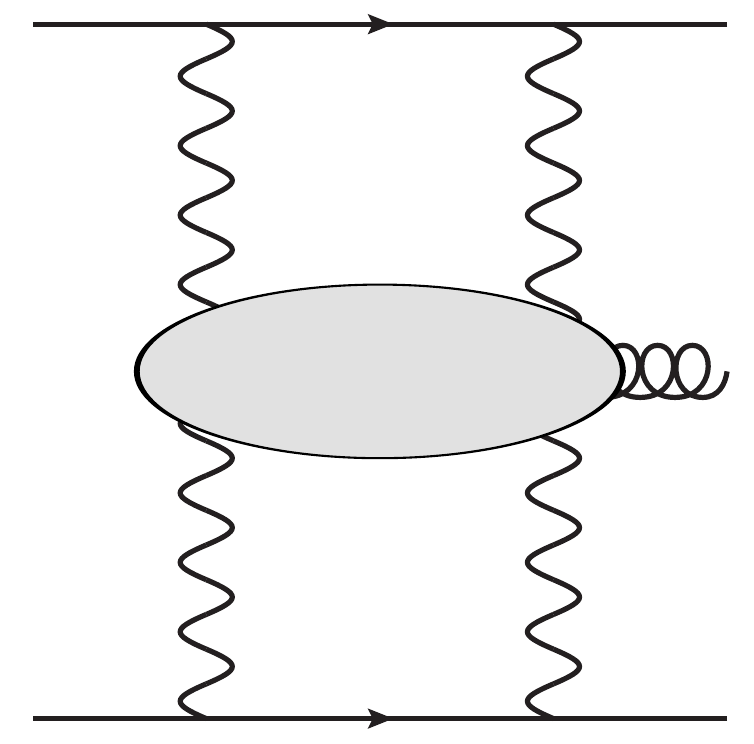}} 
  \parbox{3cm}{\center \includegraphics[height=2.5cm]{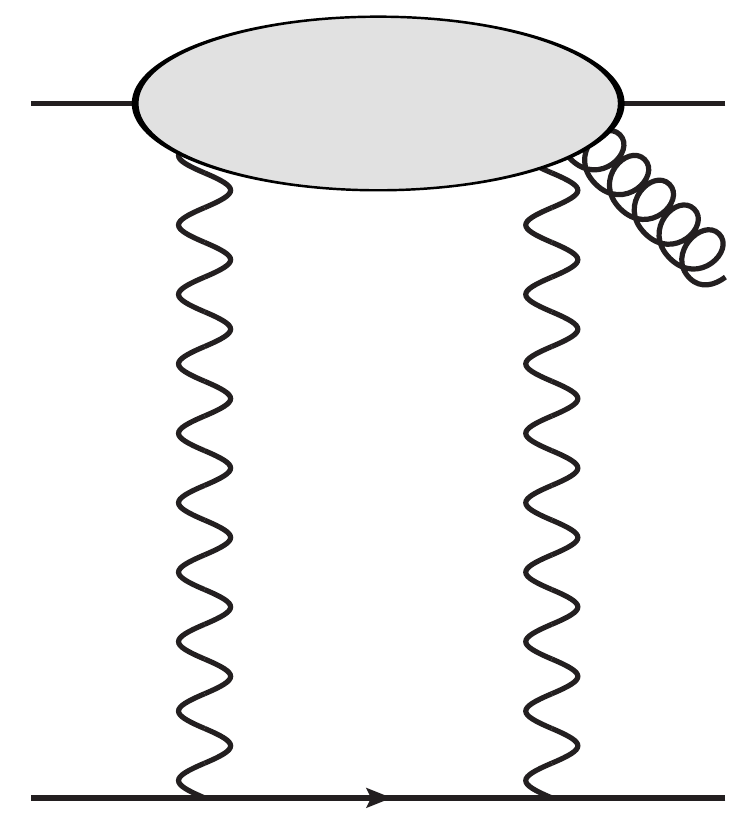}}
  \parbox{3cm}{\center \includegraphics[height=2.5cm]{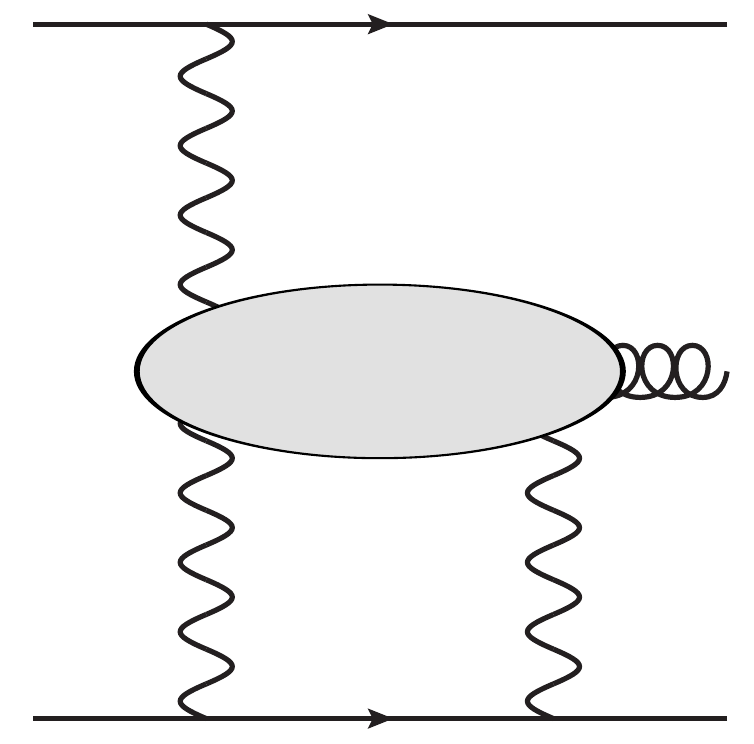}}
  \parbox{3cm}{\center \includegraphics[height=2.5cm]{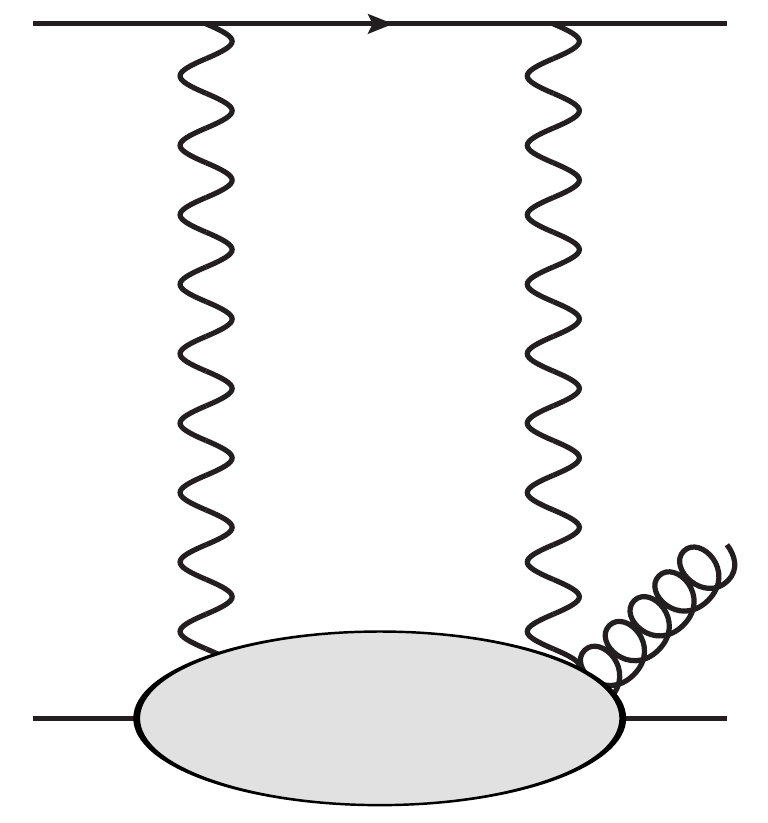}}      
  \parbox{3cm}{\center \includegraphics[height=2.5cm]{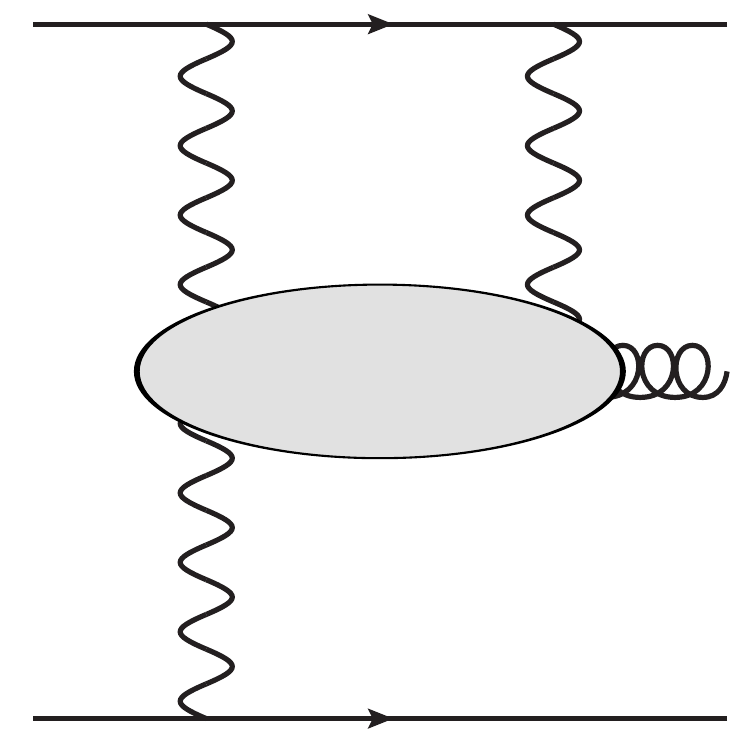}} \\

  \parbox{3cm}{\center (a) }
  \parbox{3cm}{\center (b) }
  \parbox{3cm}{\center (c) }
  \parbox{3cm}{\center (d) }
  \parbox{3cm}{\center (e) }

  \caption{Different types of  real NLO corrections.}
  \label{fig:realNLO}
\end{figure}

\begin{figure}[htb]
  \centering
 \parbox{.3\textwidth}{ \center \includegraphics[width = .2\textwidth]{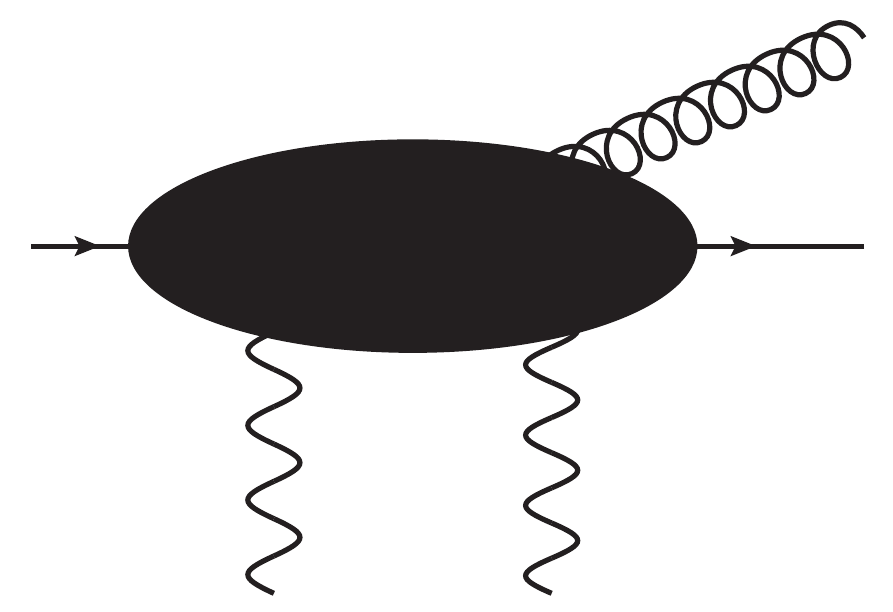} } 
 \parbox{.3\textwidth}{\center \includegraphics[width = .2\textwidth]{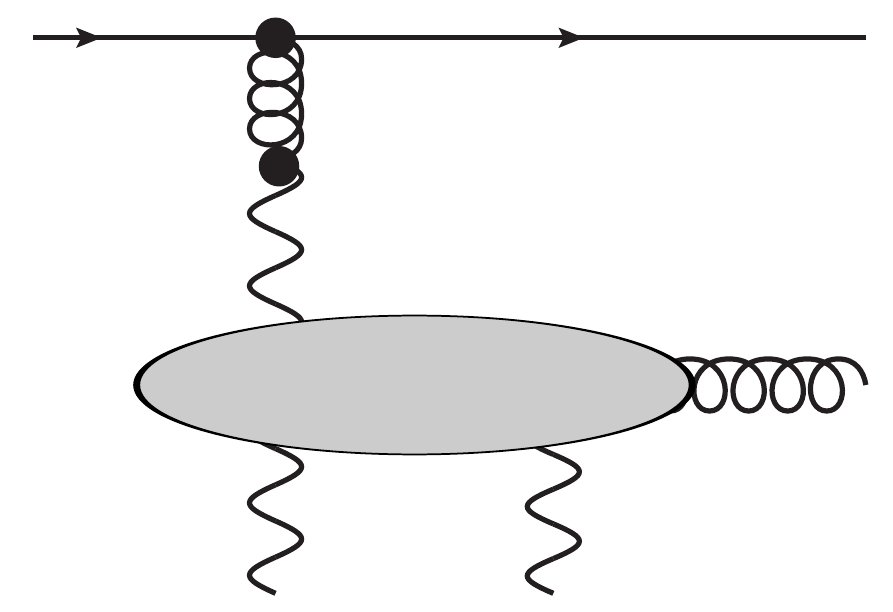}}
\parbox{.3\textwidth}{\center  \includegraphics[width = .145\textwidth]{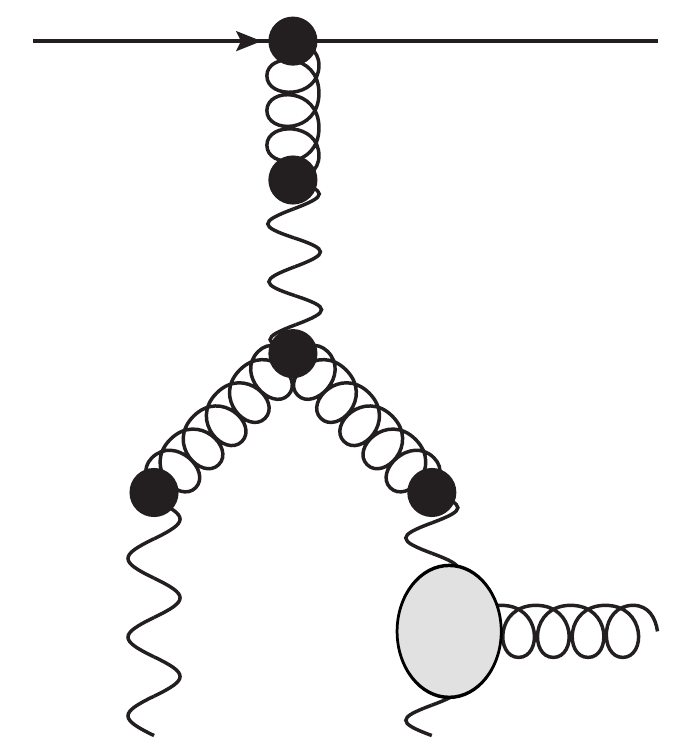}}

 \parbox{.3\textwidth}{ \center (a)}  \parbox{.3\textwidth}{ \center (b)}  \parbox{.3\textwidth}{ \center (c)}
  \caption{Different reggeized gluon diagrams contributing to the real corrections to the Mueller-Tang impact factor. (a) Quasi-elastic and (b) central production diagram. (c) Diagram with a reggeized gluon - 2 reggeized gluon splitting. The gray blob denotes an effective coupling known as the Lipatov vertex. For a derivation from the high energy effective action see \cite{quarkjet}. Those contributions can be shown to vanish identically, if the light-cone denominator is treated with a symmetric pole prescription as proposed in \cite{Hentschinski:2011xg}.}
  \label{fig:different_ReggeDiagrams}
\end{figure}

To determine the real NLO corrections it is necessary to study the process of Eq.~(\ref{eq:partonicNLOreal}) within high energy factorization. 
Fig.~\ref{fig:realNLO} provides a list of high energy factorized
amplitudes with color singlet exchange. They can be loosely classified
into two contributions: those with reggeized gluon exchange in both
$t$-channels (Fig.~\ref{fig:realNLO}.a, c, e), corresponding to gluon emission at central
rapidities and those where the additional gluon is emitted in the
quasi-elastic region of one of the quarks (Fig.~\ref{fig:realNLO}.b, d).
Among the former class, Fig.~\ref{fig:realNLO}.a is immediately absent
due to the decoupling of the anti-symmetric color octet from the two
reggeized gluon state; combined with projection of one of the
$t$-channels on the color singlet, the corresponding diagrams vanish
by color algebra.  As we are interested in events with large rapidity
gaps, also Fig.~\ref{fig:realNLO}.c and Fig.~\ref{fig:realNLO}.e will
not contribute to the jet impact factor. These contributions become
relevant if the size of the diffractive system formed by the gluon and
{\it e.g.} the upper quark (in the case of Fig.~\ref{fig:realNLO}.c)
is large and a resummation of logarithms $\ln M_X^2$ becomes
mandatory. Here we are not interested in such configurations and we
will not pursue further this idea.  These contributions provide
however a cross-check on the diagrams of interest,
Fig.~\ref{fig:realNLO}.b and Fig.~\ref{fig:realNLO}.d which describe
emissions in the quasi-elastic region. In the limit of large invariant
mass of the gluon and the upper/lower final state quark in
Fig.~\ref{fig:realNLO}.b/d, this
contribution is required to turn into the factorized expression
Fig.~\ref{fig:realNLO}.c/e. 
 The  central production vertex is  well known from the literature,
both using conventional methods \cite{BKP} and the effective action
\cite{Braun}, see also \cite{thesis}. For completeness its calculation will be briefly discussed in Appendix \ref{sec:central_prod}. In principle
there exist further contributions such as
Fig.~\ref{fig:different_ReggeDiagrams}.c which contain an explicit
splitting of a single reggeized gluon into two reggeized
gluons. Contributions containing such splittings can be shown to
vanish after integration over the longitudinal loop momentum $l^-$ and
 therefore will not be considered here.

In the following we determine the quasi-elastic subprocess emission of 
Fig.~\ref{fig:different_ReggeDiagrams}.a. 
To this end we note that  the diagrams in the
black blobs are understood to contain no internal reggeized gluon
lines. For the determination of reggeized gluon - particle vertices, the reggeized gluon must be  therefore treated as a background
field, see also the discussion in \cite{quarkjet, gluonjet, traject} for further details. In particular, Fig.~\ref{fig:different_ReggeDiagrams}.b and  diagrams such as   Fig.~\ref{fig:different_ReggeDiagrams}.c  are not a subset of the Feynman diagrams contributing to  Fig.~\ref{fig:different_ReggeDiagrams}.a.

\subsection{The quasi-elastic corrections}
\label{sec:realNLO}
\begin{figure}[htb]
%  \centering
  \parbox{2cm}{\includegraphics[width=2cm]{mt_nloX.pdf}} =
 \parbox{2cm}{\includegraphics[width=2cm]{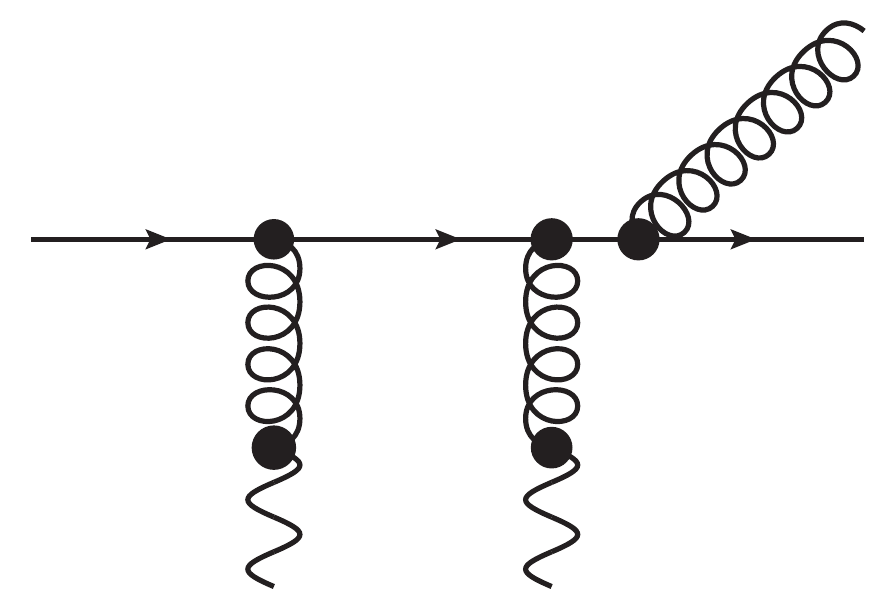}}+
  \parbox{2cm}{\includegraphics[width=2cm]{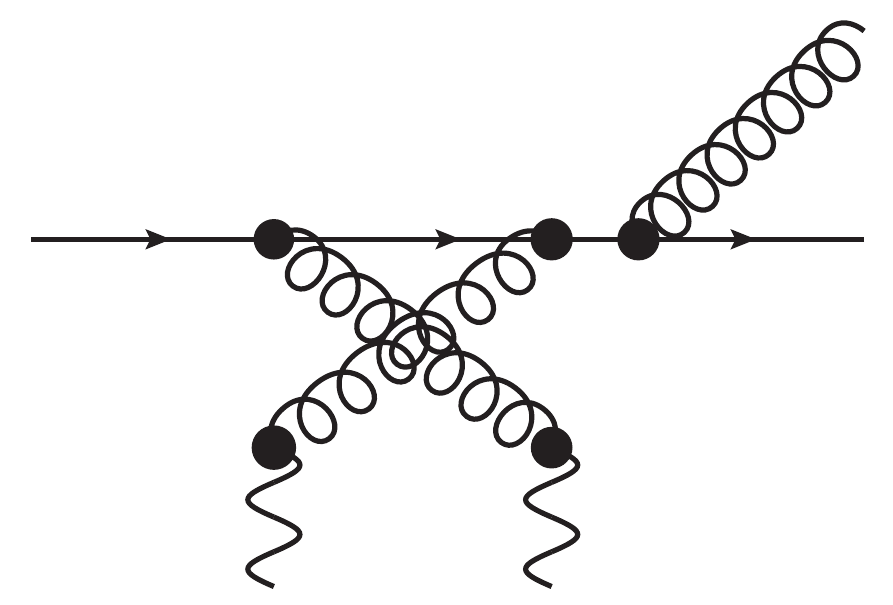}}+
 \parbox{2cm}{\includegraphics[width=2cm]{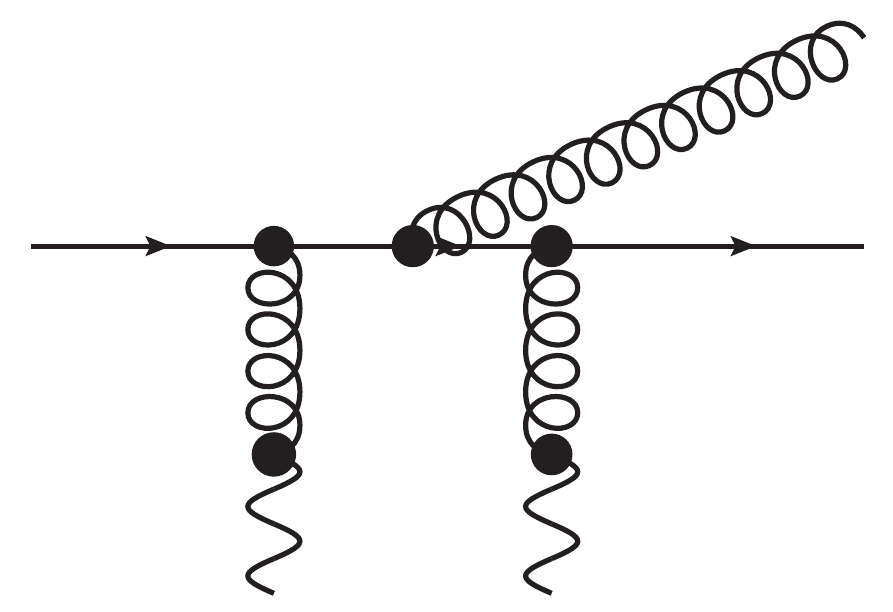}}+
  \parbox{2cm}{\includegraphics[width=2cm]{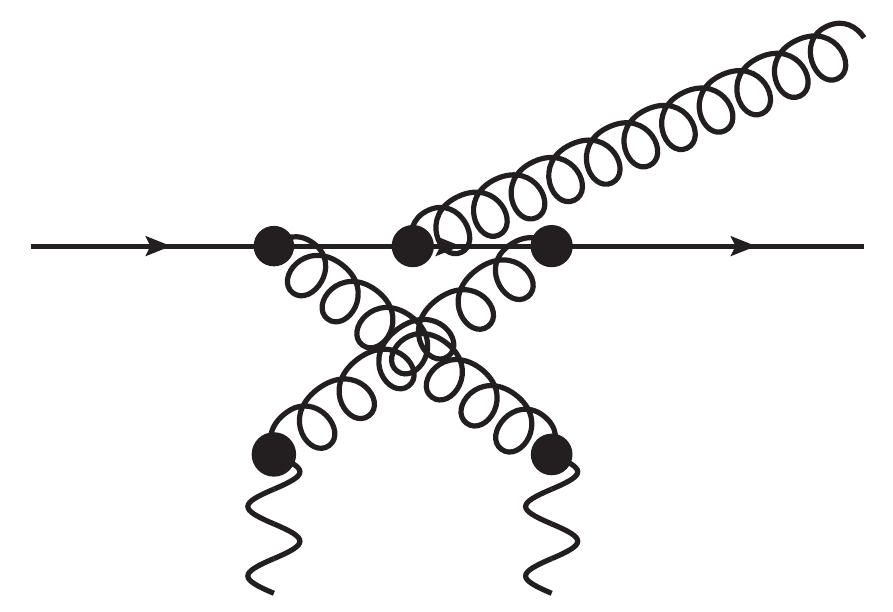}}
 +
\parbox{2cm}{\includegraphics[width=2cm]{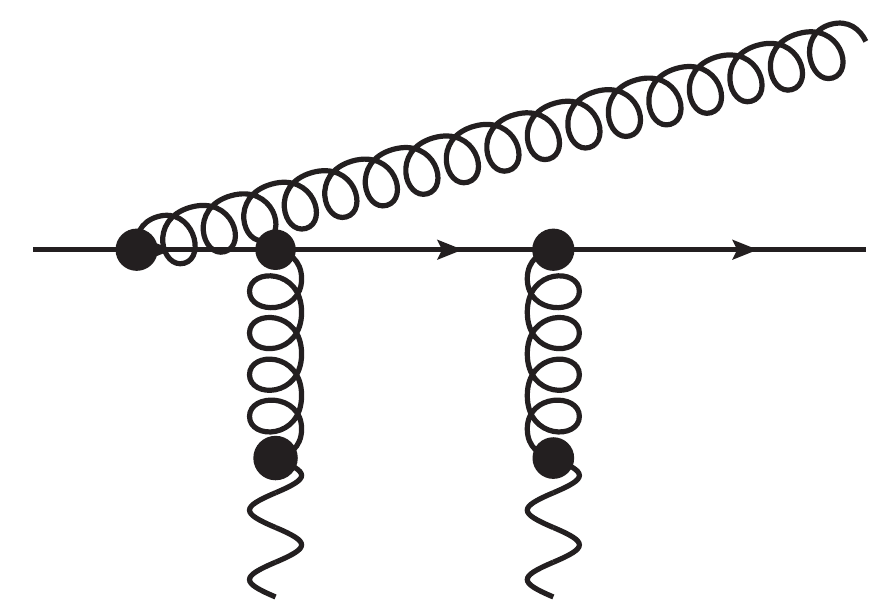}}  \\ +
  \parbox{2cm}{\includegraphics[width=2cm]{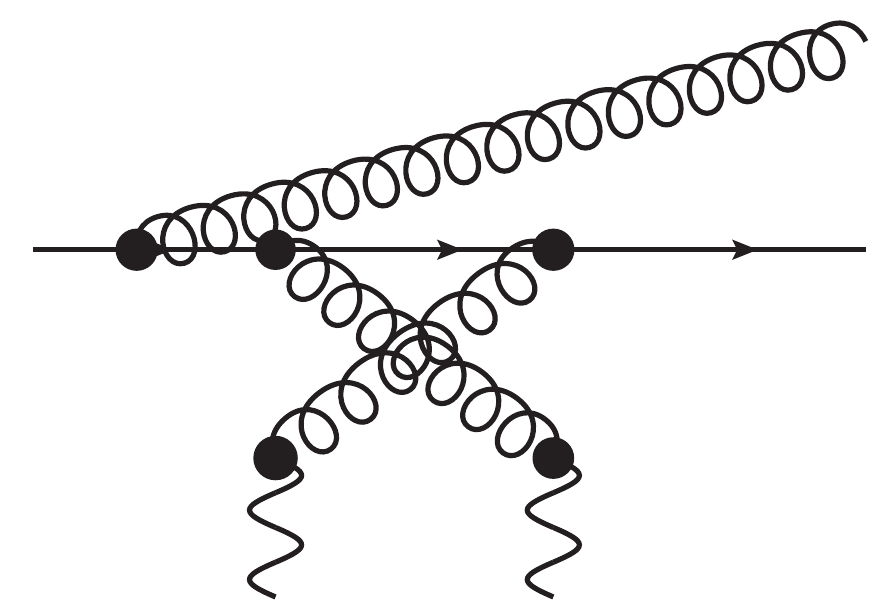}}+ 
\parbox{2cm}{\includegraphics[width=2cm]{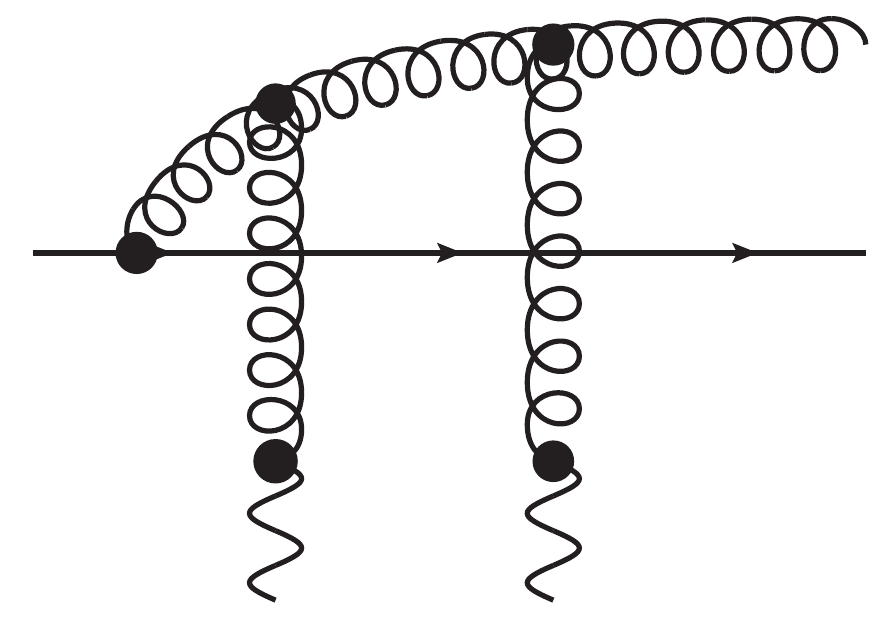}}+
  \parbox{2cm}{\includegraphics[width=2cm]{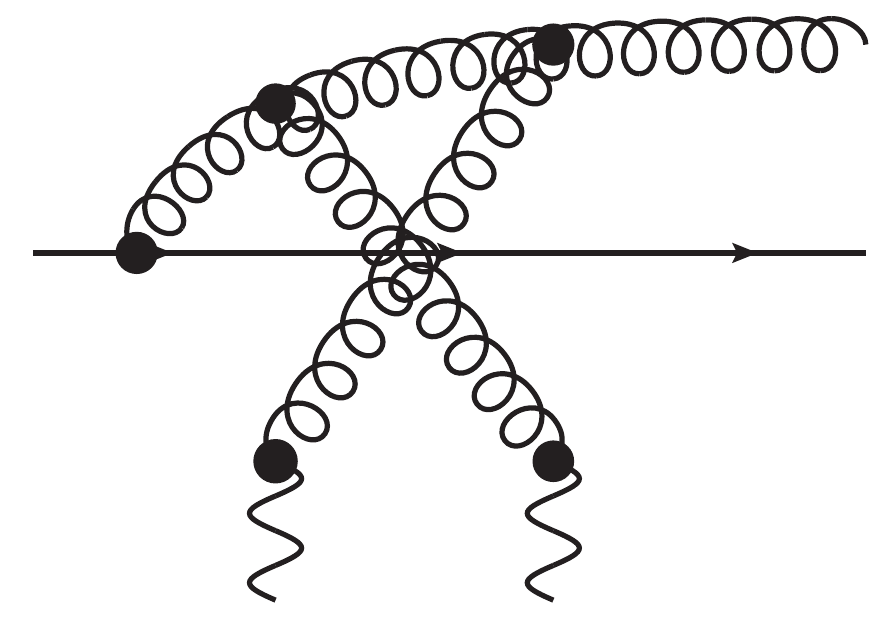}}+
 \parbox{2cm}{\includegraphics[width=2cm]{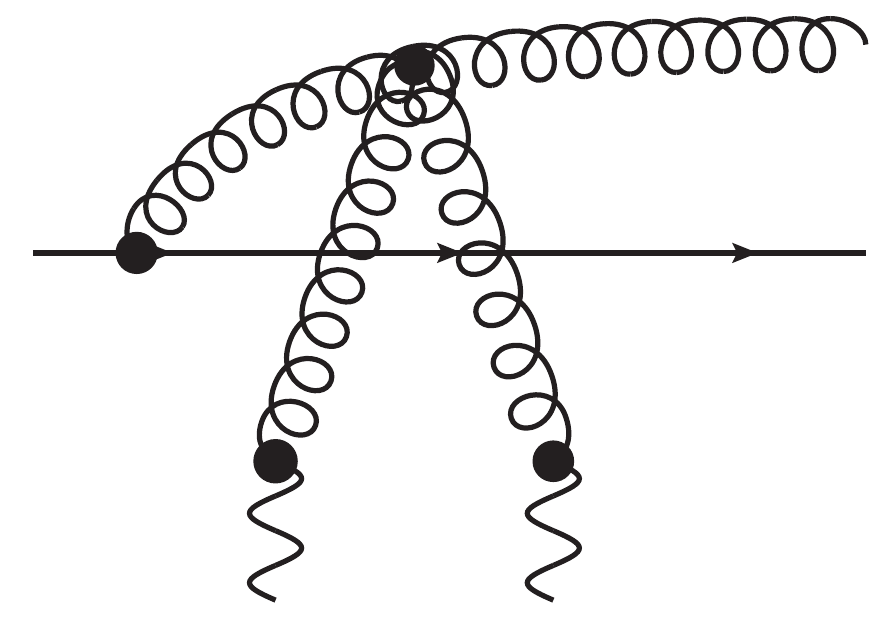}}+ 
\parbox{2cm}{\includegraphics[width=2cm]{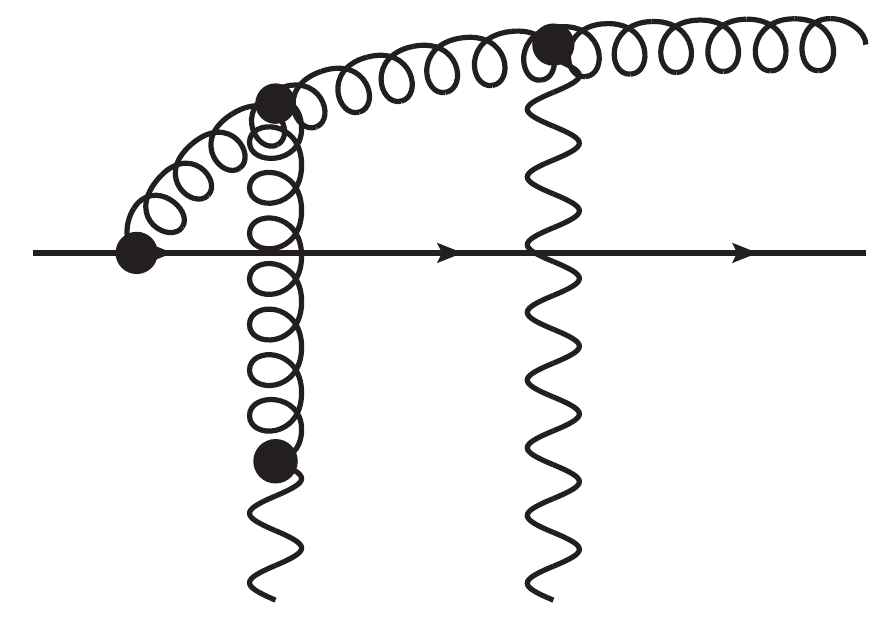}}+
  \parbox{2cm}{\includegraphics[width=2cm]{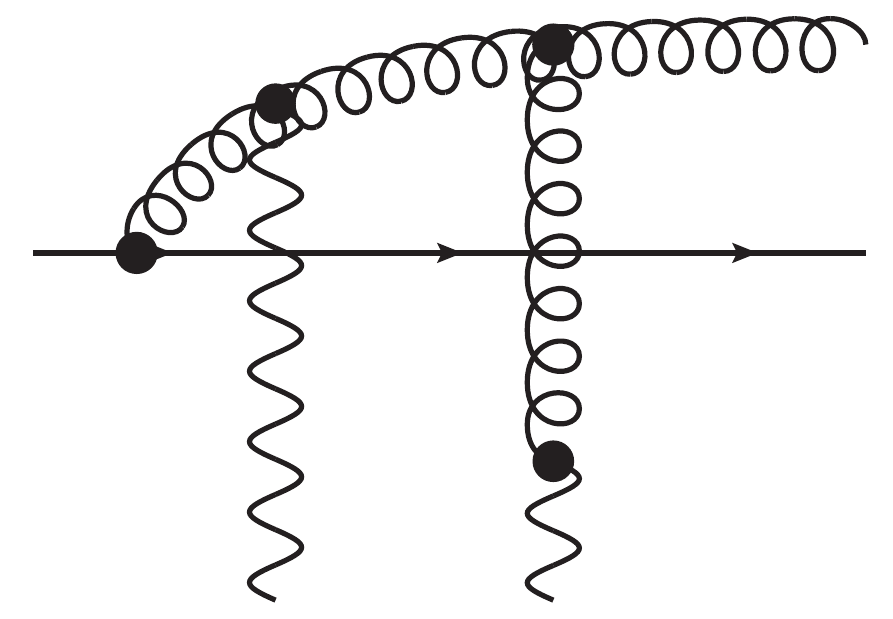}} \\ 
+
 \parbox{2cm}{\includegraphics[width=2cm]{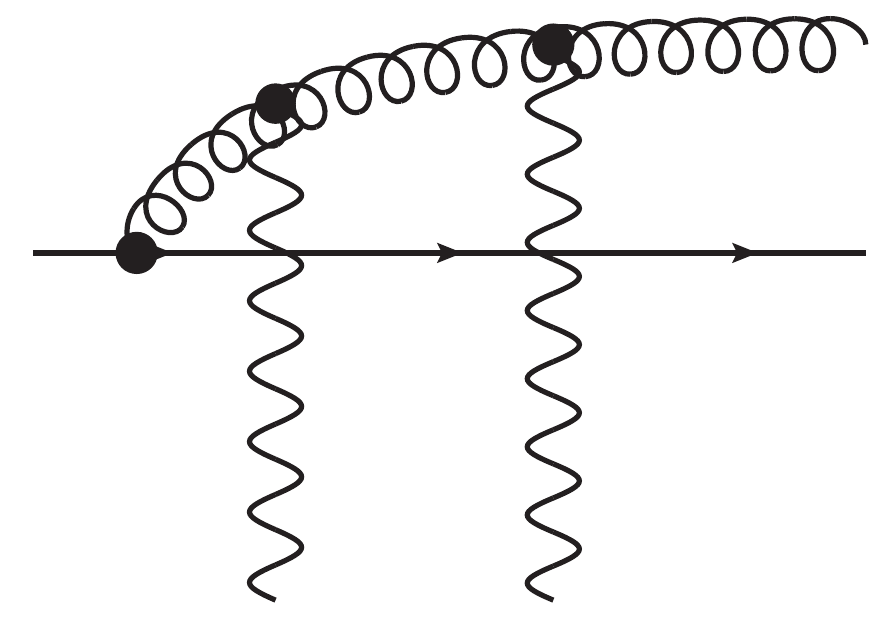}}+
 \parbox{2cm}{\includegraphics[width=2cm]{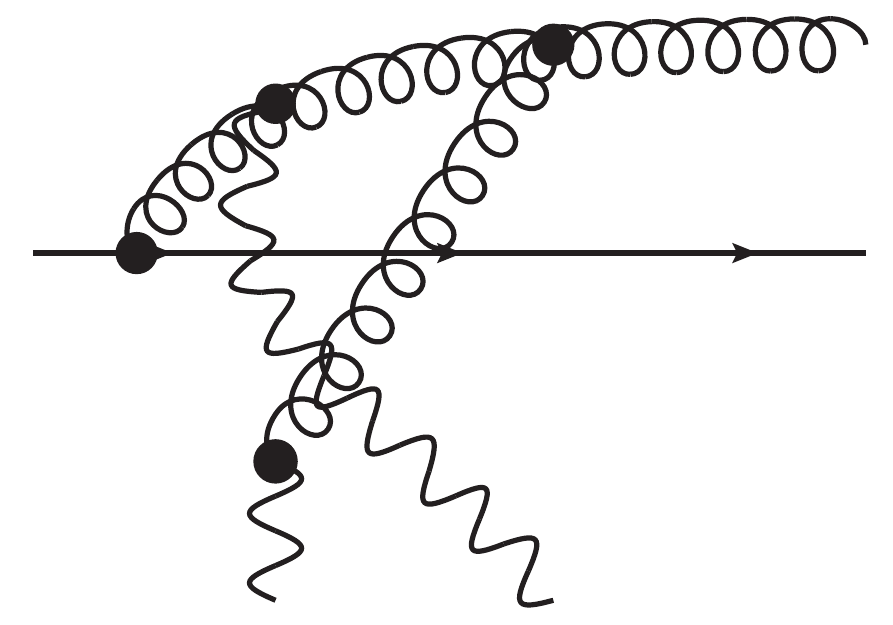}}+
 \parbox{2cm}{\includegraphics[width=2cm]{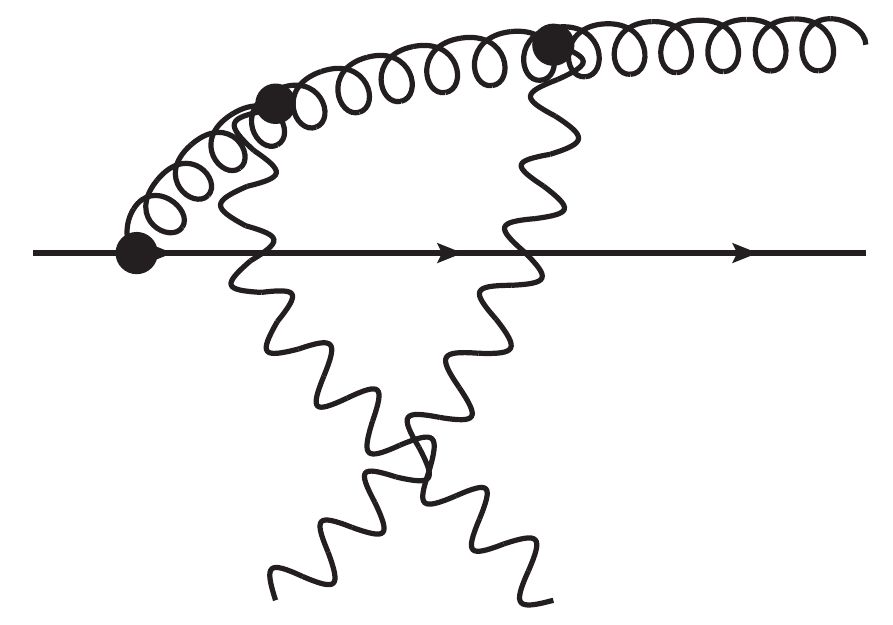}}+
 \parbox{2cm}{\includegraphics[width=2cm]{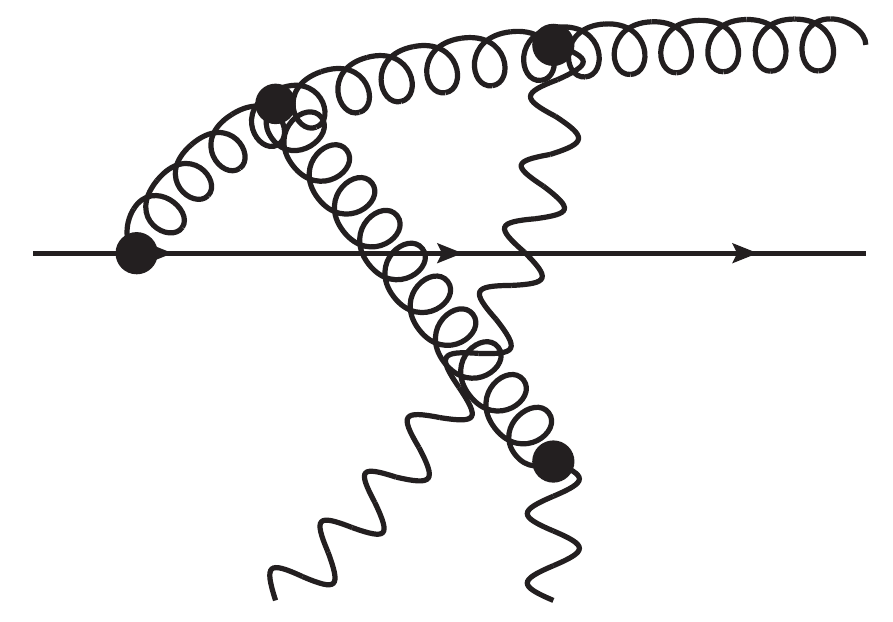}}+
  \parbox{2cm}{\includegraphics[width=2cm]{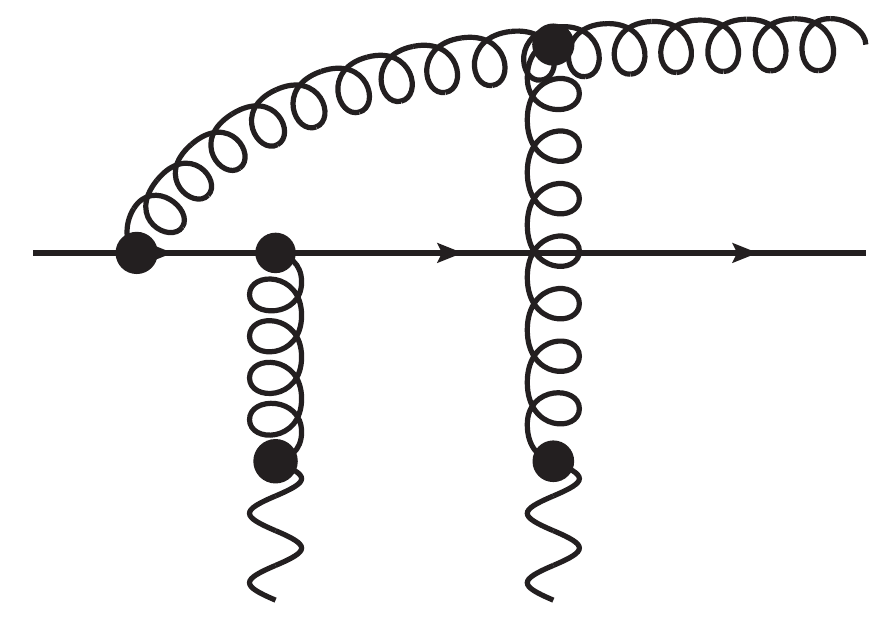}}+
 \parbox{2cm}{\includegraphics[width=2cm]{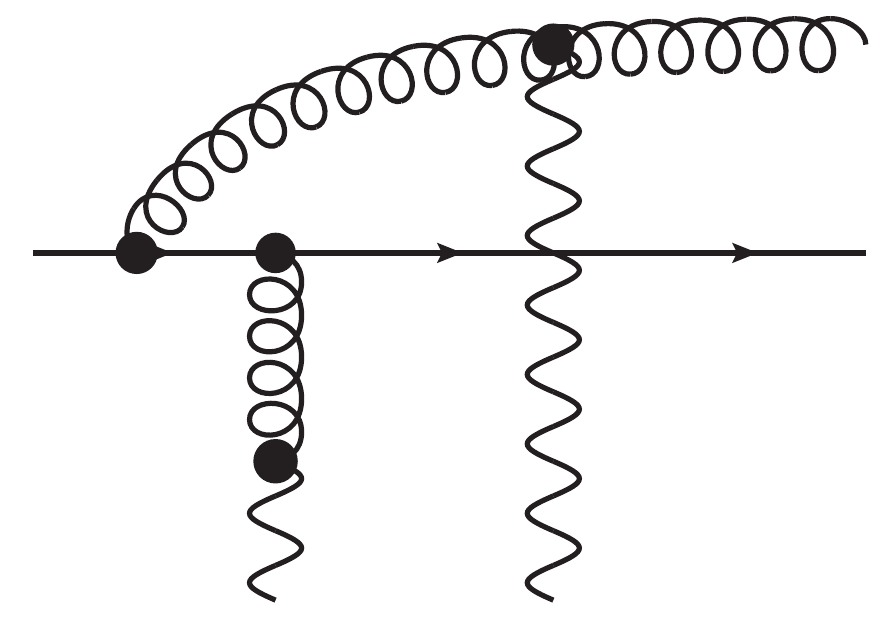}}
\\
+
  \parbox{2cm}{\includegraphics[width=2cm]{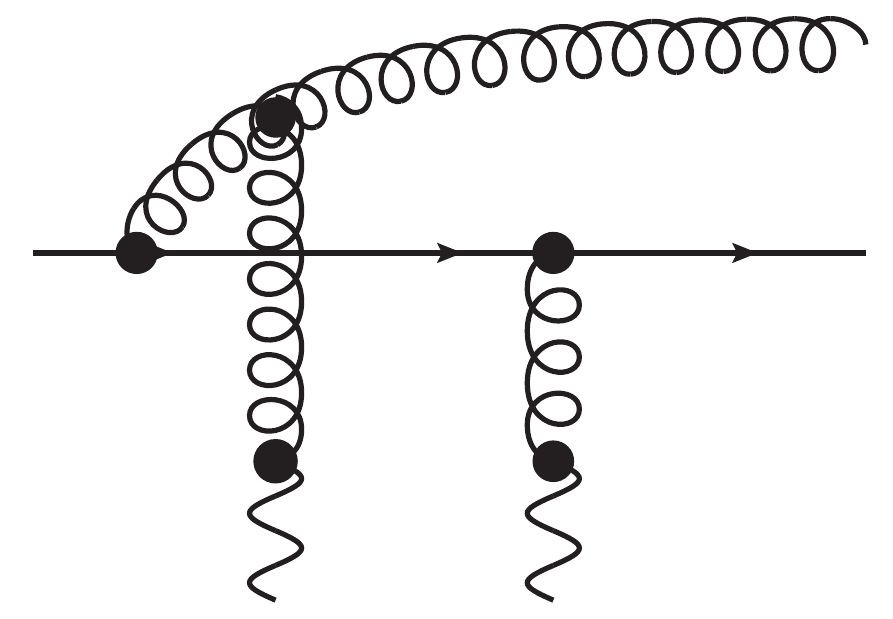}}+
\parbox{2cm}{\includegraphics[width=2cm]{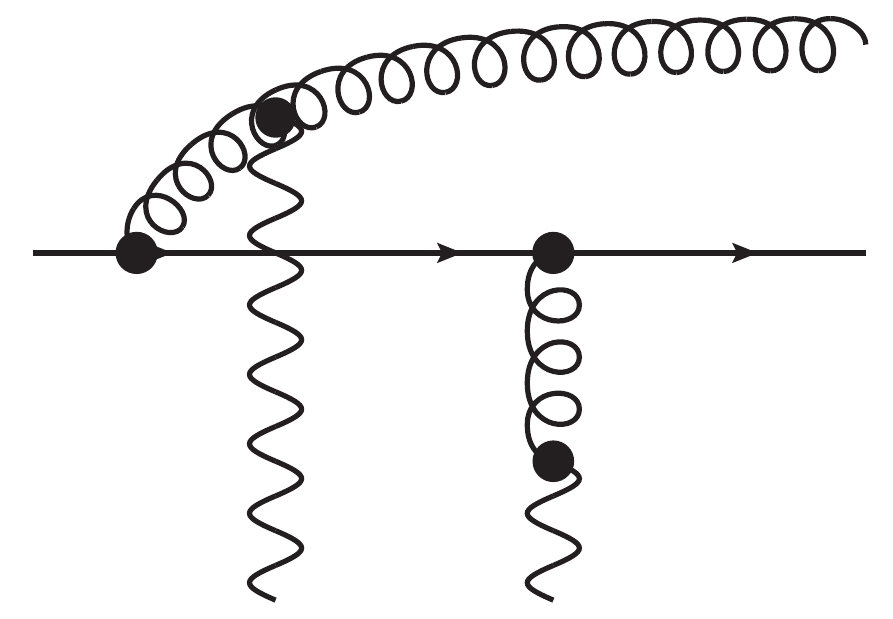}}+
\parbox{2cm}{\includegraphics[width=2cm]{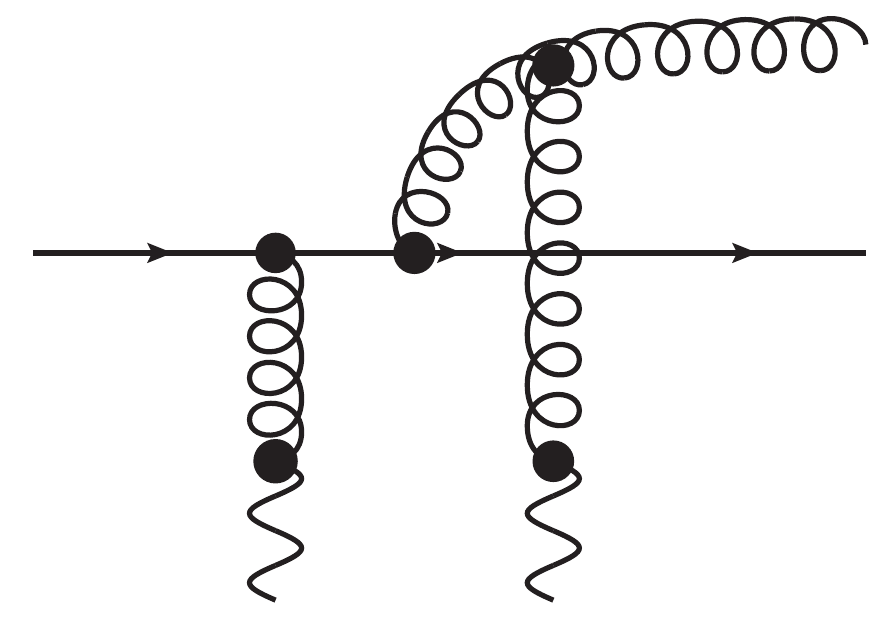}}+
 \parbox{2cm}{\includegraphics[width=2cm]{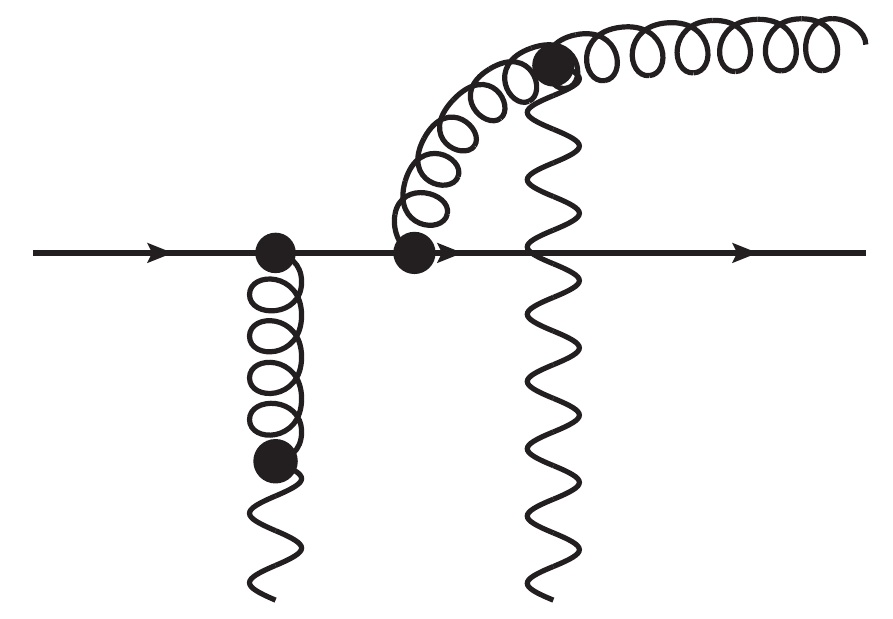}}+ 
  \parbox{2cm}{\includegraphics[width=2cm]{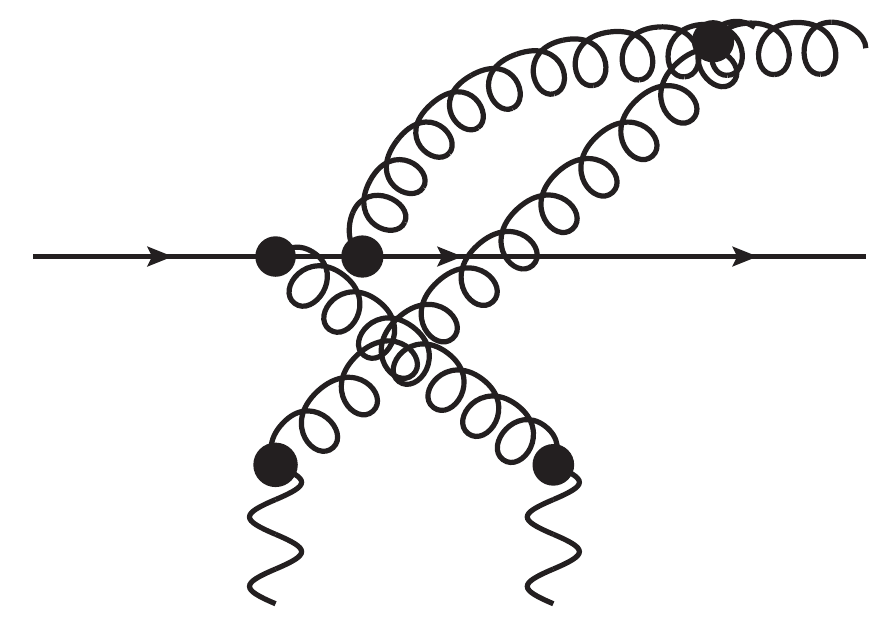}}+
\parbox{2cm}{\includegraphics[width=2cm]{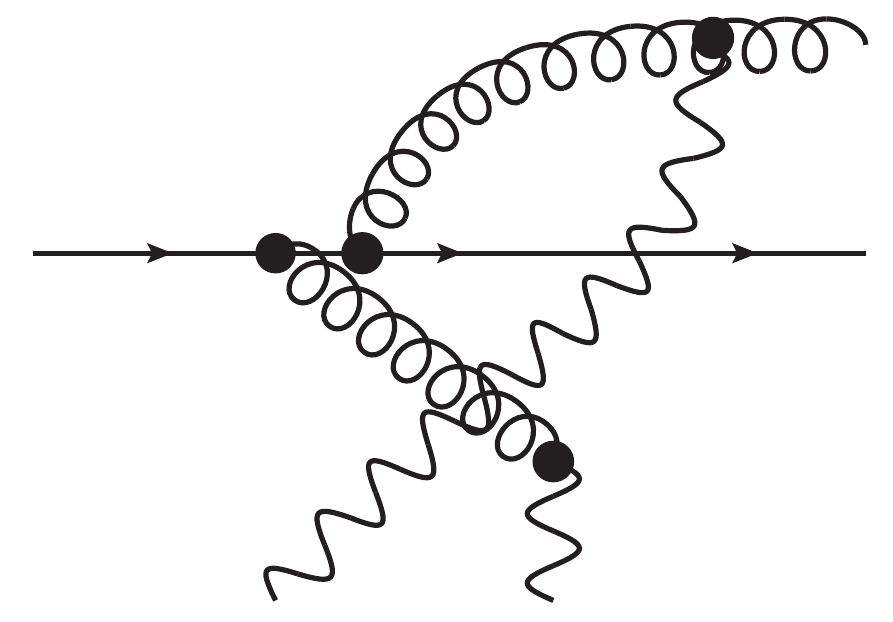}}.

  \caption{Real NLO diagrams: the quark + 2 reggeized gluon  $\to$ quark gluon amplitude.}
  \label{fig:realNLOdiag}
\end{figure}

%% introduce a Sudakov decomposition of external particles 
%% check the definition of hte ai and their acutal expressions
%% check consistency of hte final expression with the factorized result
To extract the real corrections to the jet impact factor it is therefore
sufficient to study the contribution corresponding to
Fig.~\ref{fig:realNLO}.b. As in Sec.~\ref{sec:LO}, the integral over
longitudinal loop momenta $l^-$ and $l^+$ factorizes and can be directly
associated with the $qr^*r^* \to qg$ and $qr^*r^* \to q$
subprocesses. Generalizing the analysis carried out in
Sec.~\ref{sec:LO} we therefore consider the process $q(p_a) + r^*(l) +
r^*(k-l) \to q (p) + g (q)$ with the following Sudakov decomposition
of external momenta
\begin{align}
  \label{eq:Suda_central}
 p_a & = p_a^+ \frac{n^-}{2}, \qquad     \qquad  \qquad 
   k  = k^- \frac{n^+}{2} + {\bm k}, &
  l & =  l^- \frac{n^+}{2} + {\bm l}, \notag  \\ p & = (1-z)p_a^+ \frac{n^-}{2} + \frac{{\bm p}^2}{(1-z) p_a^+} \frac{n^+}{2} +  {\bm p}, &  
 q & = z p_a^+  \frac{n^-}{2} + \frac{{\bm q}^2}{z p_a^+} \frac{n^+}{2} + {\bm q} \,.
\end{align}
 The
necessary set of Feynman diagrams is depicted in
Fig.~\ref{fig:realNLOdiag}.
At amplitude level we obtain
\begin{align}
  \label{eq:ampliutde}
  {i}\phi_{qqg}  & = \int \frac{d l^-}{8 \pi}  i\mathcal{M}^{c b_1 b_2}_{q 2 r^* qg} P^{b_1 b_2}  = {g^3}  t^c  
\sum_{i=1}^5 a_i .
% check the details of this equation and its agreement with subsequent expressions 
\end{align}

For the evaluation of the integral over $l^-$, we combined diagrams
with adjacent reggeized gluons (without emission of a real gluon in between), similar to the LO case
Eq.~(\ref{eq:impa_def}). In this way, the convergence of all
integrals is verified in a straightforward manner and the integral can be evaluated by taking residues. As a result we  obtain the following five amplitudes,
\begin{align}
  \label{eq:A1to5}
a_1 & = -\frac{C_f}{2\sqrt{N_c^2 -1}}
\cdot
  \frac{z(1-z)}{{\bm \Delta}^2}
 \cdot 
 { \bar{u} (p)  {\fdag{\epsilon} (\fdag{p}_a + \fdag{k}) \fdag{n}^+}u(p_a) },\notag\\
a_2 &= \frac{C_f}{2\sqrt{N_c^2 -1}} \cdot
\frac{z}{{\bm q}^2} \cdot
{  \bar{u} (p){\fdag{n}^+ (\fdag{p}_a - \fdag{q})  \fdag{\epsilon}}   u(p_a)},\notag\\
a_3 &= \frac{C_a}{\sqrt{N_c^2 -1}} \cdot \frac{1-z}{{\bm p}^2} \cdot {   \bar{u} (p)  \gamma^\rho (z p_a^+ \epsilon_{\rho} + (n^+)^\rho k\cdot \epsilon)   u(p_a)},\notag \\
a_4 &=  \frac{C_a}{4 \sqrt{N_c^2 -1}} \cdot   \frac{1  }{  p_a^+ {\bm \Sigma}_1^2} \cdot  \bar{u} (p) { \fdag{n}^+ (\fdag{l}_i + \fdag{p}_a - \fdag{q}) (z p_a^+ \epsilon_\rho + n^+_\rho l_i\cdot\epsilon) \gamma^\rho} u(p_a), \notag \\
a_5 &= \frac{C_a}{4\sqrt{N_c^2 -1}}\cdot \frac{1}{ p_a^+ {\bm \Upsilon}_1^2 } \cdot 
 \bar{u} (p){\fdag{n}^+ (\fdag{k} - \fdag{l}_i + \fdag{p}_a - \fdag{q})(z p_a^+ \epsilon_\rho + n^+_\rho ( k\cdot \epsilon - l_i\cdot\epsilon)) \gamma^\rho}   u(p_a),
\end{align}
where  $l_i$, $i=1,2$ is the loop momenta of the reggeized gluon
loop with $i=1$ assigned to the amplitude and $i=2$ to its complex
conjugate.  We also defined the transverse momenta
\begin{align}
  \label{eq:sigmausp}
 {\bm \Delta} & = {\bm q} - z {\bm k}, &
  {\bm \Sigma}_i & = {\bm q} - {\bm l}_i,  & {\bm \Upsilon}_i & = {\bm q} - {\bm k} + {\bm l}_i & i &= 1,2.
\end{align}
With the  2-particle phase space
\begin{align}
  \label{eq:phi2}
  d\Phi^{(2)} &= \frac{1}{(4 \pi)^{1 + \epsilon}} \int \frac{d z}{2 p_a^+ z(1-z)} \int \frac{d^{2 + 2\epsilon} {\bm q}}{  \pi^{1 + \epsilon}} \delta\left(k^- - \frac{{\bm \Delta}^2 + z(1-z) {\bm k}^2}{(1-z)z p_a^+} \right)
\end{align}
and  the invariant mass of the final state quark-gluon system
\begin{align}
  \label{eq:MXconstrint}
 \hat{M}_X^2 & = (p_a + k)^2 = \frac{{\bm \Delta}^2}{ z(1-z)} < x M_{X, \text{max}}^2 - (1-x){\bm k}^2 \equiv  \hat{M}_{X, \text{max}}^2,
\end{align}
we obtain
\begin{align}
  \label{eq:resultss}
& h^{(1)}_{r, qg} d [{\bm q}] d z = \frac{1}{2} \sum_{\text{spin}} \frac{1}{N_c} \sum_{\text{color}} 
\int d k^- \frac{ \Theta\left(M_{X,\text{max}}^2 - M_X^2 \right)}{2^{-\epsilon} p_a^+ (2\pi) (4\pi)^{2 + 2\epsilon}}
\left(
\sum_i^5 a_i \right)\left(
\sum_i^5 a_i^\dagger \right) d \Phi^{(2)}
\notag \\
&=
h^{(0)}
\frac{\alpha_{s, \epsilon}}{ 2\pi }
 \frac{ P_{gq}(z, \epsilon)}{\Gamma(1-\epsilon) \mu^{2\epsilon}}
%\notag \\
%&
 \left[   C_f   
 \left(
\frac{{\bm \Delta}}{{\bm \Delta}^2} - \frac{{\bm q}}{{\bm q}^2}
\right)
-
{C_a}\left( \frac{{\bm p}}{{\bm p}^2} + \frac{1}{2} \frac{{\bm \Sigma}_1}{{\bm \Sigma}_1^2} + \frac{1}{2} \frac{{\bm \Upsilon}_1}{{\bm \Upsilon}^2_1} \right)
  \right]
 \cdot
 \bigg[  C_f    
\left(
\frac{{\bm \Delta}}{{\bm \Delta}^2} - \frac{{\bm q}}{{\bm q}^2}
\right)
\notag \\
& 
-
{C_a}{}\left( \frac{{\bm p}}{{\bm p}^2} + \frac{1}{2} \frac{{\bm \Sigma}_2}{{\bm \Sigma}_2^2} + \frac{1}{2} \frac{{\bm \Upsilon}_2}{{\bm \Upsilon}^2_2} \right)
  \bigg] \Theta\left( x {M_{X, \text{max}}^2 - (1-x) {\bm k}^2} - \frac{{\bm \Delta}^2}{z(1-z)} \right) d [{\bm q}] d z,
\end{align}
where 
\begin{align}
  \label{eq:splitting}
  P_{gq}(z,\epsilon) & = C_f \frac{1 + (1-z)^2 + \epsilon z^2}{z}
\end{align}
is the real part of the $q \to g$ splitting function and  we used the shorthand  expression $d [{\bm k}] \equiv d^{2 + 2\epsilon} {\bm k}/\pi^{1 + \epsilon}$. Organizing the terms according to their color coefficient we arrive at 
\begin{align}
  \label{eq:resulx2}
  h^{(1)}_{r,qg} =   &
h^{(0)}
\frac{\alpha_{s, \epsilon}}{ 2\pi }
 \frac{ P_{gq}(z, \epsilon)}{\Gamma(1-\epsilon) \mu^{2\epsilon}}
%\notag \\
%&
\Theta\left(\hat{M}_{X,{\rm max}}^2- \frac{{\bm \Delta}^2}{z(1-z)} \right)
\notag \\
& \qquad \qquad 
 \bigg\{
C_f^2  \frac{z^2 {\bm k}^2}{{\bm \Delta}^2 {\bm q}^2}  
+
 {C_a}{ C_f} \bigg( J_{1} ({\bm q}, {\bm k}, {\bm l}_1, z)
+ 
J_{1} ({\bm q}, {\bm k}, {\bm l}_2, z) 
   \bigg)
 + 
C_a^2      J_{2} ({\bm q}, {\bm k}, {\bm l}_1, {\bm l}_2) 
\bigg\},
\end{align}
where
\begin{align}
  \label{eq:Jcf2}
 J_{1} ({\bm q}, {\bm k}, {\bm l}_i, z)   & = \frac{1}{4}
\bigg[
 2 \frac{{\bm k}^2}{{\bm p}^2}
\bigg(\frac{(1-z)^2}{{\bm \Delta}^2} - \frac{1}{{\bm q}^2} \bigg)
-
\frac{1}{{\bm \Sigma}_i^2}
\bigg(
\frac{({\bm l}_i - z {\bm k})^2}{{\bm \Delta}^2} - 
\frac{{\bm l}_i^2}{{\bm q}^2}
\bigg)
\notag \\
& \qquad \qquad \qquad \qquad 
-
\frac{1}{{\bm \Upsilon}_i^2}
\bigg( 
  \frac{({\bm l}_i - (1-z) {\bm k})^2}{{\bm \Delta}^2}  
-
\frac{({\bm l}_i - {\bm k})^2}{{\bm q}^2}
\bigg)
\bigg]; 
\notag \\
 J_{2} ({\bm q}, {\bm k}, {\bm l}_1, {\bm l}_2)  &=
\frac{1}{4} \bigg[
\frac{{\bm l}_1^2}{ {\bm p}^2 {\bm \Upsilon}^2_1} 
+ 
\frac{( {\bm k} - {\bm l}_1)^2}{ {\bm p}^2 {\bm \Sigma}^2_1}
 +
\frac{{\bm l}_2^2}{ {\bm p}^2 {\bm \Upsilon}^2_2} 
+ 
\frac{( {\bm k} - {\bm l}_2)^2}{ {\bm p}^2 {\bm \Sigma}^2_2}
\notag \\
& 
- \frac{1}{2}
\bigg(
\frac{({\bm l}_1 - {\bm l}_2)^2}{{\bm \Sigma}_1^2 {\bm \Sigma}_2^2}
+
\frac{({\bm k} - {\bm l}_1 - {\bm l}_2)^2}{ {\bm \Upsilon}_1^2 {\bm \Sigma}_2^2   }
+
\frac{({\bm k} - {\bm l}_1 - {\bm l}_2)^2}{  {\bm \Sigma}_1^2 {\bm \Upsilon}_2^2  }
+
\frac{({\bm l}_1 - {\bm l}_2)^2}{{\bm \Upsilon}_1^2 {\bm \Upsilon}_2^2}
\bigg)
 \bigg].
\end{align}

With
\begin{align}
  \label{eq:MXdef}
  \hat{M}_X^2 &  =   \left( \frac{{\bm q}^2}{z} + \frac{{\bm p}^2}{1-z} - {\bm k}^2 \right)\;,
\end{align}
large partonic diffractive mass corresponds to the limits $z \to 0$ and $z \to
1$ at fixed transverse gluon and quark momentum respectively. For $z
\to 1$ we find that Eq.~(\ref{eq:resultss}) is finite and no high
energy singularity is present. This is to be expected as this case
corresponds to highly negative rapidities of the real quark, which are
power suppressed in the high energy limit.  For $z \to 0$ we find, on
the other hand, that the term proportional to the color factor $C_a^2$ contains a
high energy singularity $1/z$. Meanwhile, the terms proportional to $C_f^2$ and
$C_f C_a$ vanish in the limit $z \to 0$ and hence
cancel the singularity present in $P_{gq}(z, \epsilon)$.  It is then
straightforward to check that the singular term agrees
precisely with the high energy factorized cross-section of 
Eq.~\eqref{eq:Hfact} derived in the Appendix \ref{sec:central_prod},
thus validating the correctness of our result in this limit.

\section{The jet  vertex  for quark induced jets with rapidity gap}
\label{sec:jet}

To obtain from the partonic real NLO corrections in 
Eq.~(\ref{eq:resulx2}) for the jet vertex, we need to combine this result
with the corresponding virtual corrections, add a jet definition and
absorb initial state singularities into parton distribution
functions. We follow here closely the corresponding treatment in the
case of Mueller-Navelet jets discussed in \cite{Bartels:2001ge,
  Caporale:2011cc}.

\subsection{Virtual corrections and renormalization}
\label{sec:virtual}
The virtual corrections have been calculated in \cite{Fadin:1999df}.
Unlike the present calculation, the authors of \cite{Fadin:1999df} make no  use of Lipatov's effective  action, but calculate the corresponding corrections
directly from QCD Feynman diagrams with the help of dispersion
relations, employing analyticity and unitarity of QCD scattering
amplitudes.  The virtual corrections are then given as the sum of
quark-intermediate state impact factor and quark-gluon-intermediate
quark impact factor, where the terminology appears natural from the
calculational method of \cite{Fadin:1999df}. The result for the quark-gluon-intermediate state is given in Eq.~(6.19) of  \cite{Fadin:1999df}. Projected on the color singlet it reads at cross-section level
\begin{align}
  \label{eq:H1vqg}
  h^{(1)}_{v, a}  ({{\bm k}, {\bm l}_1}, {\bm l}_2) & = C_f^2 h^{(0)} {\frac{(4 \pi)^{1 + \epsilon}\alpha_{s,\epsilon}}{\mu^{2\epsilon}\Gamma(1-\epsilon)}}  \bigg\{
 - C_f  
I_B^{(+)} ( {\bm l}_1, {\bm k} )  
 -  \frac{C_a}{2} \big[ 
\tilde{I}_A^{(+)} ( {\bm l}_1, {\bm k} )
\notag \\
& \qquad \qquad \qquad  \qquad \qquad   
  - 
I_B^{(+)} ( {\bm l}_1, {\bm k} ) 
+ 
\tilde{I}_C^{(+)} ( {\bm l}_1, {\bm k} )  
\big] + ({\bm l}_1) \leftrightarrow ({\bm l}_2)\bigg\}.
\end{align}
The functions $\tilde{I}_A^{(+)}$, ${I}_B^{(+)}$ and
$\tilde{I}_C^{(+)}$ are given in Eqs.~(6.11), (6.18) and (6.15) of
\cite{Fadin:1999df}\footnote{A factor $\delta_{\lambda_{A'}\lambda_A}$
  which denotes helicity conservation at amplitude level, present in
  the definition of  $\tilde{I}_A^{(+)}$, ${I}_B^{(+)}$ and
$\tilde{I}_C^{(+)}$  in \cite{Fadin:1999df}, has been already extracted  from
  our functions and summed/averaged  over.}.  The quark intermediate state reads at cross-section
level, after projection on the color singlet
\begin{align}
  \label{eq:quark_imediate}
  h^{(1)}_{v,b} ({\bm k}, {\bm l}_1, {\bm l}_2) & = C_f^2 h^{(0)}  \frac{\alpha_{s, \epsilon}\Gamma^2(1 + \epsilon) }{ 4 \pi \Gamma(1 +2\epsilon) (-\epsilon)} \bigg\{
\left[
\left( \frac{{\bm l}_1^2}{ \mu^2}\right)^\epsilon + \left( \frac{ ({\bm k} - {\bm l}_1)^2}{ \mu^2}\right)^\epsilon 
 \right]
\bigg[ 
\frac{-n_f (1 + \epsilon)}{ (1 + 2 \epsilon) (3 + 2\epsilon) } 
\notag \\
&
+ (2 C_f - C_a) \left( \frac{1}{\epsilon (1 + 2 \epsilon)} + \frac{1}{2} \right) 
+
 C_a\bigg(
 \psi(1-\epsilon) - 2 \psi(\epsilon) + \psi(1)   
\notag \\
&
+ \frac{1}{4(1 + 2 \epsilon) (3 + 2\epsilon)}- \frac{1}{\epsilon (1 + 2\epsilon)}
 - \frac{7}{4(1 + 2\epsilon)}
\bigg) \bigg]
\notag \\
&  \qquad   +  
C_a \bigg[ \ln \frac{s_0}{{\bm l}_1^2}
\left( \frac{{\bm l}_1^2}{ \mu^2}\right)^\epsilon +  \ln \frac{s_0}{ ({\bm k} - {\bm l}_1)^2} \left( \frac{ ({\bm k} - {\bm l}_1)^2}{ \mu^2}\right)^\epsilon 
\bigg]  +  ({\bm l}_1) \leftrightarrow ({\bm l}_2)
\bigg\}.
\end{align}
$s_0$ denotes here the reggeization scale, which sets the scale of the
energy logarithms, resummed by the  non-forward
BFKL Green's function; $\mu^2$ is the scale of dimensional
regularization and $\beta_0 = \frac{11}{3}N_c  - \frac{2}{3} n_f$. Expanding in
$\epsilon$ we find  for the virtual  corrections
\begin{align}
  \label{eq:virtual}
  h^{(1)}_v & =  h_{v, a}^{(1)} ({\bm k}, {\bm l}_1, {\bm l}_2)   + h_{v, b}^{(1)} ({\bm k}, {\bm l}_1,  {\bm l}_2),
\end{align}
the following terms
\begin{align}
  \label{eq:H1div}
 & h^{(1)}_v 
=
{h^{(0)}} C_f^2  \frac{\alpha_{s,\epsilon}}{4 \pi} \bigg\{
-2 \frac{\beta_0}{ \epsilon} 
 {-\frac{\beta_0}{2}\left[\left\{\ln\left(\frac{\bm{l}_1^2}{\mu^2}\right)+\ln\left(\frac{(\bm{l}_1-\bm{k})^2}{\mu^2}\right)+\{1\leftrightarrow 2\}\right\}-\frac{20}{3}\right]} \notag \\ &+
      2 {C_f} \left[- \frac{2}{\epsilon^2} 
+ \frac{1}{\epsilon} \left(3-2\ln\left(\frac{{\bm k}^2}{\mu^2}  \right)\right){-\ln^2\left(\frac{\bm{k}^2}{\mu^2}\right)+3\ln\left(\frac{\bm{k}^2}{\mu^2}\right)+\frac{\pi^2}{3}-8}\right]
\notag \\ &
                {+C_a\bigg[\bigg\{\frac{3}{2\bm{k}^2}\left\{\bm{l}_1^2\ln\left(\frac{(\bm{l}_1-\bm{k})^2}{\bm{l}_1^2}\right)+(\bm{l}_1-\bm{k})^2\ln\left(\frac{\bm{l}_1^2}{(\bm{l}_1-\bm{k})^2}\right)-4|\bm{l}_1||\bm{l}_1-\bm{k}|\phi_1\sin\phi_1\right\}}
\notag \\
&\quad\quad {-\frac{3}{2}\left[\ln\left(\frac{\bm{l}_1^2}{\bm{k}^2}\right)+\ln\left(\frac{(\bm{l}_1-\bm{k})^2}{\bm{k}^2}\right)\right]-\ln\left(\frac{\bm{l}_1^2}{\bm{k}^2}\right)\ln\left(\frac{(\bm{l}_1-\bm{k})^2}{s_0}\right)}\notag \\
&\quad\quad{-\ln\left(\frac{(\bm{l}_1-\bm{k})^2}{\bm{k}^2}\right)\ln\left(\frac{\bm{l}_1^2}{s_0}\right)-2\phi_1^2+\{1\leftrightarrow 2\}\bigg\}+2\pi^2+\frac{14}{3}\bigg]}+{\cal{O}(\epsilon)\bigg\}}.
\end{align}
Here
\begin{align}
  \label{eq:defohi}
  \phi_i & = \arccos  \frac{{\bm k}^2 - {\bm l}_i^2 - ({\bm k - {\bm l}_i)^2}}{{2}|{\bm l}_i| |{\bm l}_i-{\bm k}|},
& i &= 1,2,
\end{align}
denotes the angle between the reggeized gluon momenta with $|\phi_i| \leq \pi$, $i = 1,2$.
The first divergent term is of ultraviolet origin and comes multiplied by the first term of the QCD $\beta$ function. Employing renormalization of the QCD Lagrangian within the
$\overline{\text{MS}}$ scheme 
\begin{align}
  \label{eq:MSbar}
  \alpha_s(\mu) & = \alpha_{s, \epsilon} \left[1- \frac{\alpha_{s, \epsilon} \beta_0}{4 \pi \epsilon}  \right],
\end{align}
this term will be canceled. The remaining divergences are of soft or collinear origin. They will be partly canceled by corresponding singularities in the real corrections, with the remainder to be absorbed by collinear factorization. 

\subsection{The  jet vertex at partonic and hadronic level at leading order}
\label{sec:LOvertex}

To extract the jet vertex at partonic level, we need to combine the results obtained so far 
with a jet function, following
Eq.~\eqref{eq:djetsigma}. Due to high energy factorization of the
cross-section, it is possible to carry out this analysis separately
for each impact factor. To be more precise, we write  the differential partonic jet cross-section in its most general form as
\begin{align}
  \label{eq:jetpartonic}
  \frac{d \hat{\sigma}}{d J_1 \, d J_2 \, \, d^2 {\bm k}} & =   
 \int \frac{d^{2} {\bm l}_1}{\pi} 
\int \frac{d^{2 } {\bm l}'_1}{\pi} 
 \int \frac{d^{2} {\bm l}_2}{\pi}
\int \frac{d^{2 } {\bm l}'_2}{\pi}
\frac{d \hat{V}({\bm l}_1, {\bm l}_2, {\bm k}, {\bm p}_{J,1}, { y}_1, s_0)}{d J_1}
\notag \\
& \qquad \qquad \qquad 
G\left( {\bm l}_1, {\bm l}_1', {\bm k},  \frac{\hat{s}}{s_0}\right)
G\left( {\bm l}_2, {\bm l}_2', {\bm k},  \frac{\hat{s}}{s_0} \right)
 \frac{d \hat{V}({\bm l}'_1, {\bm l}'_2, {\bm k}, {\bm p}_{J,2}, { y}_2, s_0)}{d J_2},
\end{align}
where $G$ denotes the non-forward BFKL Green's function which is
either taken in the asymptotic limit $\ln \hat{s}/s_0 \to \infty$ or
implies a suitable infrared regulator.  If the final state is given
by a single quark, the jet definition is trivial and given by
Eq.~\eqref{eq:S0}.  We find in that case
\begin{align}
  \label{eq:dVq}
  \frac{d \hat{V}_q}{d J} &=  v^{(0)}_q S^{(2)}_J({\bm {\bm k}}, x),  &&\text{with}& 
v_q^{(0)}&= \frac{\alpha_s^2 C_f^2}{N_c^2 -1}.
\end{align}

An identical expression holds for the virtual corrections in 
Eq.~\eqref{eq:H1div}, but with $h^{(0)}_q$ replaced by $h^{(1)}_v$.
In the following we assume that the reggeization scale $s_0$ is chosen
such that the BFKL Green's functions do not explicitly depend on the
proton momentum fractions $x_1$ and $x_2$ of the initial quarks.
Examples of such choices for $s_0$ are $\log s/s_0 = \Delta \eta$
where $\eta$ denotes multiples of either the separation of the jets in
rapidity $\Delta y$ or the size of the gap $\Delta
y_{\text{gap}}$. For such scenarios we can define
\begin{align}
  \label{eq:dVqH}
  \frac{d V_q^{(0)}}{d J} & = \int_{x_0}^1 d x \, \, f_{q/H}(x, \mu_F^2) \,  h_q^{(0)} \, S^{(2)}_J ({\bm p}, x)
\notag \\
&
 =   v_q^{(0)}  \cdot  x_J f_{q/H}(x_J, \mu_F^2)\, \delta^{(2)}({\bm p} - {
\bm k}_J), \qquad \qquad \qquad  x_0 <  x_J = \frac{|{\bm k}_J|{e^{y_J}}}{\sqrt{s}}<1 ,  
\end{align}
and the  corresponding hadronic cross-section is  given by Eq.~\eqref{eq:jetpartonic} with all `hats' removed. 

\subsection{Next-to-leading order vertex: jet function}
\label{sec:jetfunction}

As soon as the final state is no longer given by a single quark, the
jet function is no longer trivial and some dependence on the chosen jet
algorithm enters.  Since the additional final state gluon may be soft
or collinear to either initial or final state quark, the jet function
needs to fulfill the following set of requirements \cite{jetdef}, to
guarantee infrared finiteness of the cross-section.  For a general
partonic process with momenta $p_a + p_b \to p_1 + \ldots p_n$ the jet
function for $n$ final state particles $S_J^{n}(p_1 ..., p_n, ;p_a,
p_b)$ reduces to the jet function of $n-1$ final state particles in
the following way.  If the particle $j$ is soft,
\begin{align}
  \label{eq:softJet}
  \lim_{p_j \to 0}  S_J^{n}( p_1 \ldots, p_j, \ldots, p_n ;p_a, p_b)  =  S_J^{n-1}(p_1,  \ldots, \hat{p}_j, \ldots,  p_n ;p_a, p_b),    
\end{align}
where  $\hat{p}_j$ indicates omission of the $j$-th particle. If two final state partons with index  $i$ and  $j$  are collinear, $p_i = a\cdot  p$ and $p_j = b \cdot p$,  
\begin{align}
  \label{eq:finalcoll}
   S_J^{n}( \ldots , a\cdot p, \ldots,  b\cdot p,  \ldots , ;p_a, p_b)  =  S_J^{n-1}(p_1 \ldots , (a +b)\cdot p,  \ldots p_n ; p_a, p_b)\; ,
\end{align}
and  if a final state parton with index $i$ is collinear to an initial state parton, $p_i = a \cdot p_a$
\begin{align}
  \label{eq:finalint}
   S_J^{n}(p_1, \ldots, a\cdot p_a, \ldots, p_n ;p_a, p_b)  =  S_J^{n-1}(p_1,  \ldots, \hat{p}_i, \ldots  p_n; (1-a)\cdot p_a, p_b).
\end{align}
In the present case, with the phase space of the  final quark-gluon system parametrized  both  by longitudinal momentum fraction,  carried forward from the initial quark with momentum fraction $x$ by gluon  ($z$)  and quark (${\bar z} = 1-z$), and gluon (${\bm p}$)  and quark (${\bm q}$)  transverse momentum, these conditions can be expressed as follows 
\begin{align}
  \label{eq:limitsS2}
  S_J^{(3)} ({\bm p}, {\bm q}, z x, x) & \to S_J^{(2)}({\bm p}, x) &&
  {\bm q} \to 0, z \to 0  \notag\\
  S_J^{(3)} ({\bm p}, {\bm q}, z x, x) & \to S_J^{(2)}({\bm k}, x) &&
 \frac{ {\bm q}}{z} \to  \frac{ {\bm p}}{1-z}  \notag\\
  S_J^{(3)} ({\bm p}, {\bm q}, z x, x) & \to S_J^{(2)}({\bm k},(1-z)
  x) &&
  {\bm q}\to  0  \notag\\
  S_J^{(3)} ({\bm p}, {\bm q}, z x, x) & \to S_J^{(2)}({\bm k}, z x)
  &&
  {\bm p}\to  0,  \notag\\
\end{align}
together with  symmetry of $S^{(3)}$ under  the simultaneous swapping of ${\bm p} \leftrightarrow {\bm q}$ and $z \leftrightarrow 1-z$.  While finiteness of the jet impact factor is generally expected due to these particular constraints imposed onto the jet definition, we note that the verification of the latter is non-trivial  in the present case due to high energy factorization  of the partonic cross-section into jet impact factors and two reggeized gluon exchange.

\subsection{Next-to-leading order jet vertex: different contributions}
\label{sec:jet_ctJet}

The virtual part of the one-loop corrections to the jet vertex follows exactly the tree-level result. % and reads
% \begin{align}
%   \label{eq:dVqHv}
%   \frac{d V_v^{(1)}}{d J} & = \int_0^1 d x \, \,  f_{q/H}(x, \mu_F^2) \,  
% \frac{d \hat{V}_v^{(1)}}{d J}
% &
% \frac{d \hat{V}_v^{(1)}}{d J} &=
% h_v^{(1)} \,  S^{(2)}_J ({\bm p}, x).
% \end{align}
After renormalization within the $\overline{\text{MS}}$
scheme, following Eq.~\eqref{eq:MSbar}, we split the virtual corrections into a finite term and a term which gathers the entire set of so-far uncanceled soft and collinear singularities,
\begin{align}
  \label{eq:splitvirutal}
  \frac{d \hat{V}_v^{(1)}}{d J} & =   \frac{d \hat{V}_{v, \text{sc}}^{(1)}}{d J} +    \frac{d \hat{V}_{v, \text{finite}}^{(1)}}{d J} ,
\end{align}
with
\begin{align}
  \label{eq:sc}
    \frac{d \hat{V}_{v, \text{sc}}^{(1)}}{d J}  & = S_J^{(2)}({\bm k}, x) \cdot {h^{(0)}} C_f^3  \frac{\alpha_s}{2 \pi} 
      \bigg(- \frac{2}{\epsilon^2} 
+ \frac{3}{\epsilon} 
                - \frac{2}{\epsilon} \ln \frac{{\bm k}^2}{\mu^2}  
          \bigg)
\end{align}
and
\begin{align}
  \label{eq:virt_finite}
     \frac{d \hat{V}_{v, \text{finite}}^{(1)}}{d J} 
&=  S_J^{(2)}({\bm k}, x) \cdot  v^{(0)}_q \cdot \frac{\alpha_{s}}{4 \pi} \bigg\{ 
 {-\frac{\beta_0}{2}\left[\left\{\ln\left(\frac{\bm{l}_1^2}{\mu^2}\right)+\ln\left(\frac{(\bm{l}_1-\bm{k})^2}{\mu^2}\right)+\{1\leftrightarrow 2\}\right\}-\frac{20}{3}\right]} \notag \\ &+
      2 {C_f} \left[{-\ln^2\left(\frac{\bm{k}^2}{\mu^2}\right)+3\ln\left(\frac{\bm{k}^2}{\mu^2}\right)+\frac{\pi^2}{3}-8}\right]
\notag \\ &+{C_a\bigg[\bigg\{\frac{3}{2\bm{k}^2}\left\{\bm{l}_1^2\ln\left(\frac{(\bm{l}_1-\bm{k})^2}{\bm{l}_1^2}\right)+(\bm{l}_1-\bm{k})^2\ln\left(\frac{\bm{l}_1^2}{(\bm{l}_1-\bm{k})^2}\right)-4|\bm{l}_1||\bm{l}_1-\bm{k}|\phi_1\sin\phi_1\right\}}
\notag \\
&\quad\quad { -\frac{3}{2}\left[\ln\left(\frac{\bm{l}_1^2}{\bm{k}^2}\right)+\ln\left(\frac{(\bm{l}_1-\bm{k})^2}{\bm{k}^2}\right)\right]-\ln\left(\frac{\bm{l}_1^2}{\bm{k}^2}\right)\ln\left(\frac{(\bm{l}_1-\bm{k})^2}{s_0}\right)}\notag \\
&\quad\quad{ -\ln\left(\frac{(\bm{l}_1-\bm{k})^2}{\bm{k}^2}\right)\ln\left(\frac{\bm{l}_1^2}{s_0}\right)-2\phi_1^2+\{1\leftrightarrow 2\}\bigg\}+2\pi^2+\frac{14}{3}\bigg]\bigg\}}.
\end{align}

To obtain from the NLO partonic cross-section a finite NLO collinear
coefficient, we further need to absorb initial state collinear
singularities into parton distribution functions. This can be achieved
 by adding suitable counterterms to the partonic NLO
cross-section. At the level of the jet vertex in Eq.~\eqref{eq:dVqH} the
counterterms read (in the $\overline{\text{MS}}$-scheme)
\begin{align}
  \label{eq:rewritect}
 \frac{d V^{(1)}_{ct}}{d J} &=  \int_{x_0}^1  d x \, \,  f_q\left( x,  \mu_F^2   \right)  \frac{d \hat{V}^{(1)}_{ct}}{d J}, \qquad 
\frac{d \hat{V}^{(1)}_{ct}}{d J} = \frac{d \hat{V}^{(1)}_{ct,q}}{d J} + \frac{d \hat{V}^{(1)}_{ct,g}}{d J};
 \notag \\
\frac{d \hat{V}^{(1)}_{ct,q}}{d J}
&=
- \frac{\alpha_{s, \epsilon}}{2\pi} \left( \frac{1}{\epsilon} +  \ln \frac{\mu_F^2}{\mu^2}\right)   \int_{z_0}^1 {dz} \,\, S_J^{(2)} ({\bm k}, zx) 
\cdot 
  h_q^{(0)} P_{qq}^{(0)}(z), 
\notag \\
\frac{d \hat{V}^{(1)}_{ct,g}}{d J}
&=
- \frac{\alpha_{s, \epsilon}}{2\pi} \left( \frac{1}{\epsilon} +  \ln \frac{\mu_F^2}{\mu^2}\right)   \int_{z_0}^1 {dz} \,\, S_J^{(2)} ({\bm k}, zx) 
\cdot    h^{(0)}_g { P_{gq}^{(0)}}(z)  \, ,
\end{align}
with the LO splitting functions
\begin{align}
  \label{eq:splittings}
  P_{qq}^{(0)}(z) & = C_f \left(\frac{1+z^2}{1-z} \right)_+, 
& P_{gq}^{(0)}(z) &= C_f \frac{1 + (1-z)^2}{z}\, ,
\end{align}
and the  plus distribution
\begin{align}
  \label{eq:plus_distr}
  \int_\alpha^1 dx \, \,  f(x) [g(x)]_+ & \equiv  \int_\alpha^1 dx \, \,  \big( f(x) - f(1) \big) g(x)
-
f(1) \int_0^\alpha dx \, \, g(x)\,.
\end{align}
For the  lower bound $z_0$  we notice that we can use the combination of splitting function and  LO partonic cross-section  $\hat{M}_X^2 = \frac{{\bm k}^2(1-z)}{z}$ and write 
\begin{align}
  \label{eq:z0}
  z_0 & = \frac{{\bm k}^2/x}{M_{X, \text{max}}^2 + {\bm k}^2} \,\, .
\end{align} 

The real corrections are finally given by
\begin{align}
  \label{eq:resuJJJ}
  \frac{d V^{(1)}_r}{d J}  & = \int_{x_0}^1 dx \, \, f_{q/H}(x, \mu_F^2) \, 
 \frac{d \hat{V}^{(1)}_r}{d J} ,
\notag \\
\frac{d \hat{V}^{(1)}_r}{d J}   &=
 \int_0^1 d z \int \frac{d^{2 + 2 \epsilon} {\bm q}}{\pi^{1 + \epsilon}}  \, \, 
h^{(1)}_{r, qg} 
S_J^{(3)} ({\bm p}, {\bm q}, z x, x).
\end{align}

To extract the soft and collinear singularities from the latter, we first decompose $h^{(1)}_r$ according to its color structure, following Eqs.~\eqref{eq:resulx2}, \eqref{eq:Jcf2}. We start with the terms proportional to  $C_f^2$. 
 Substituting  $z \to 1-z$ and  rescaling   the gluon transverse momentum ${\bm q} \to (1-z) {\bm q}$, where $z$ indicates  now the momentum fraction carried by the final state quark,  we have 
\begin{align}
  \label{eq:Cf2J}
 \left(\frac{d \hat{V}^{(1)}_r}{d J} \right)_{C_f^2}
&=
 \int_0^1 d z \int \frac{d^{2 + 2 \epsilon} {\bm q}}{\pi^{1 + \epsilon}} 
h^{(0)}
\frac{\alpha_{s, \epsilon}}{ 2\pi }
 \frac{ C_f^3 }{\Gamma(1-\epsilon) \mu^{2\epsilon}}
\frac{1 + z^2 + \epsilon (1-z)^2}{(1-z)^{1-2\epsilon}}
\frac{{\bm k}^2}{{\bm q}^2  ({\bm q} - {\bm k})^2}
\notag \\
\times S_J^{(3)} ({\bm k} &- (1-z){\bm q}, (1-z){\bm q}, (1-z) x, x) 
\Theta \left(\frac{\hat{M}_{X,{\rm max}}^2}{(1-z)} - \frac{( {\bm q } - {\bm k})^2}{z} \right).
\end{align}
The next step is to decompose the denominator in the first line
\begin{align}
  \label{eq:rewrite}
 C_f \frac{1 + z^2 + \epsilon (1-z)^2}{(1-z)^{1-2\epsilon}} & = C_f  \left(\frac{1}{\epsilon} - \frac{3}{2} \right)\delta(1-z) 
+
P_{qq}(z) +
\notag \\
& \qquad 
 \epsilon \cdot C_f \cdot \left[ (1-z)^{1 + 2 \epsilon} + 2 (1 + z^2) \left(\frac{\ln (1-z)}{1-z} \right)_+ \right] + \mathcal{O}(\epsilon^2),
\end{align}
 using
the identity
\begin{align}
  \label{eq:id_papa}
  (1-z)^{2\epsilon -1} & = \frac{1}{2\epsilon} \delta(1-z) + \frac{1}{(1-z)_+} + 2\epsilon \left( \frac{\ln(1-z)}{1-z} \right)_+ + \mathcal{O}(\epsilon^2), 
\end{align}
and split Eq.~\eqref{eq:resuJJJ} into the three corresponding terms
\begin{align}
  \label{eq:cV123CF2}
   \left(\frac{d \hat{V}^{(1)}_r}{d J} \right)_{C_f^2} &= 
 \left(\frac{d \hat{V}^{(1)}_{r}}{d J} \right)_{C_f^2, a} +
 \left(\frac{d \hat{V}^{(1)}_r}{d J} \right)_{C_f^2, b} +
 \left(\frac{d \hat{V}^{(1)}_r}{d J} \right)_{C_f^2, c} + \mathcal{O}(\epsilon) \, .
\end{align}
For the first term the jet function turns out to be trivial and we obtain (up to ${\cal O}(\epsilon)$)
\begin{align}
  \label{eq:soft}
 \left(\frac{d \hat{V}^{(1)}_{r}}{d J} \right)_{C_f^2, a}
  &=
 h^{(0)} 
\frac{\alpha_{s, \epsilon}  C^3_f }{ 2 \pi } \left(\frac{2}{\epsilon^2} - \frac{3}{\epsilon} + \frac{2}{\epsilon} \ln \frac{{\bm k}^2}{\mu^2} - \frac{\pi^2}{3}   - 3 \ln \frac{{\bm k}^2}{\mu^2}  +   \ln^2 \frac{{\bm k}^2}{\mu^2} \right)
S_J^{(2)} ({\bm k},  x).
\end{align}
The emerging poles in $1/\epsilon$ of this term cancel precisely against  the corresponding singularities in the virtual corrections in Eq.~(\ref{eq:sc}). For the second term we find
\begin{align}
  \label{eq:coll}
 & \left(\frac{d \hat{V}^{(1)}_{r}}{d J} \right)_{C_f^2, b } =
 \int_0^1 d z  \int \frac{d^{2 + 2 \epsilon} {\bm q}}{\pi^{1 + \epsilon}} 
h^{(0)}
\frac{\alpha_{s,\epsilon}}{ 2\pi }
 \frac{ C_f^2 }{\Gamma(1-\epsilon) \mu^{2\epsilon}}
  P_{qq}(z) \cdot  \frac{{\bm k}^2}{{\bm q}^2  ({\bm q} - {\bm k})^2  } 
\notag \\
&   \quad%  \qquad \qquad 
 \cdot \Theta \left(\frac{\hat{M}_{X,{\rm max}}^2}{(1-z)} - \frac{ ({\bm q} - {\bm k})^2}{z} \right) \cdot \,   S_J^{(3)} ({\bm k} -  (1-z){\bm q}, (1-z){\bm q}, (1-z) x, x) 
.
\end{align}

To isolate singular configurations with a final state gluon (${\bm q}^2
= 0$) and a final state quark ($({\bm q} - {\bm k})^2 = 0$) collinear to
the initial quark, we introduce a phase space slicing parameter
$\lambda$.  Since
\begin{align}
  \label{eq:S3toS2}
 \lim_{{\bm q}^2 \to 0}  S_J^{(3)} ({\bm k} -  (1-z){\bm q}, (1-z){\bm q}, (1-z) x, x) &= S_J^{(2)}({\bm k}, zx), 
\end{align}
 we find for ${\bm q}^2 < \lambda^2$ with ${\bm k}^2 \gg \lambda^2 \to 0$ 
\begin{align}
  \label{eq:collq2Cf}
    \left(\frac{d \hat{V}^{(1)}_{r}}{d J} \right)_{C_f^2, b, \lambda }
&=
 \int \frac{d^{2 + 2 \epsilon} {\bm q}}{\pi^{1 + \epsilon}} \frac{\Theta({ \lambda^2} - {\bm q}^2)}{{\bm q}^2} 
h^{(0)}  C_f^2
\frac{\alpha_{s,\epsilon}}{ 2\pi }
\notag \\
&
 \qquad\times   \int_0^1 d z  \frac{P_{qq}(z)  S_J^{(2)}({\bm k}, z x) }{\Gamma(1-\epsilon) \mu^{2\epsilon}}
 %
%\notag \\
%& \hspace{6cm}
 \Theta\left(\frac{\hat{M}_{X,{\rm max}}^2}{1-z} - \frac{ {\bm k}^2}{z} \right)
\notag \\
& = 
\frac{\alpha_{s,\epsilon}}{ 2\pi }
\left( \frac{1}{\epsilon} + \ln \frac{\lambda^2}{\mu^2} \right)
\int_{z_0}^1 d z  \,\, h_q^{(0)} 
 S_J^{(2)}({\bm k}, z x)%
P_{qq}(z)
  + \mathcal{O}(\epsilon).
\end{align}
Adding the first collinear counterterm,
\begin{align}
  \label{eq:canceledcounter}
   \left(\frac{d \hat{V}^{(1)}_{r}}{d J} \right)_{C_f^2, b, \lambda } +  \left(\frac{d \hat{V}^{(1)}_{r}}{d J} \right)_{ct,q}
&=
v_q^{(0)}\cdot \frac{\alpha_{s} }{ 2\pi}
 \ln \frac{\lambda^2}{\mu^2_F} 
\int_{z_0}^1 d z  \,\, 
 S_J^{(2)}({\bm k}, z x)%
P_{qq}(z) + \mathcal{O}(\epsilon),
\end{align}
this contribution turns out to be finite.
 Since
\begin{align}
  \label{eq:S3toS2}
 \lim_{ ({\bm q} - {\bm k})^2 \to 0}  S_J^{(3)} ({\bm k} -  (1-z){\bm q}, (1-z){\bm q}, (1-z) x, x) &= S_J^{(2)}({\bm k}, x),
\end{align}
and
\begin{align}
  \label{eq:and}
    \int_0^1 dz \,\, P_{qq}(z)=0,
\end{align}
the coefficient of the second collinear pole is absent; the finite  remainder of the second term reads 
\begin{align}
  \label{eq:coll_finte}
& v_q^{(0)} \cdot  \frac{\alpha_{s}}{ 2\pi }  \int_0^1 d z \int \frac{d^{2} {\bm q}}{\pi} 
%
% \bigg[
  P_{qq}(z) \cdot  \Theta \left(\hat{M}_{X,{\rm max}}^2 - \frac{ ({\bm p} -z {\bm k})^2}{z(1-z)} \right)
\notag \\
&   \qquad \qquad  \qquad   \Theta\left(\frac{|{\bm q}|}{1-z} - \lambda \right)  \cdot \,   S_J^{(3)} ({\bm p}, {\bm q}, (1-z) x, x) 
%r
%\bigg]
\frac{{\bm k}^2}{{\bm q}^2  ({\bm p} - z {\bm k})^2  },
\end{align}
 where we inverted the initial rescaling through ${\bm q} \to {\bm q}/(1-z)$ and used ${\bm p} = {\bm k} - {\bm q}$. The third term is only non-zero if the transverse integral is divergent. We find
 \begin{align}
   \label{eq:Cf2c}
     \left(\frac{d \hat{V}^{(1)}_{r}}{d J} \right)_{C_f^2, c} 
=
v_q^{(0)} \cdot  \frac{\alpha_{s}}{ 2\pi }
 \bigg\{&
\int_{z_0}^1 d z  \,\, 
 S_J^{(2)}({\bm k}, z x)%
 \cdot \left[ (1-z) + 2 (1 + z^2) \left(\frac{\ln (1-z)}{1-z} \right)_+ \right] 
\notag \\ &
  \qquad \qquad  \qquad \qquad   \qquad \qquad +
4 S_J^{(2)}({\bm k}, x)    \bigg\} + \mathcal{O}(\epsilon),
 \end{align}
 where the first and second line arise due to the initial and final state
 collinear singularity respectively.  The terms with color factor $C_f
 C_a$ read
\begin{align}
  \label{eq:CFCA}
 \left( \frac{d \hat{V}^{(1)}_r}{d J}  \right)_{C_fC_a}  &=
 \int_0^1 d z \int \frac{d^{2 + 2 \epsilon} {\bm q}}{\pi^{1 + \epsilon}}  \, \, 
h^{(0)} C_a C_f \frac{\alpha_{s, \epsilon}}{2 \pi}  \Theta\left(\hat{M}_{X,{\rm max}}^2- \frac{{\bm \Delta}^2}{z(1-z)} \right)
\notag \\
&
\qquad \qquad   \frac{P_{gq}(z, \epsilon)}{\Gamma(1-\epsilon) \mu^{2\epsilon}} 
 \big[  J_1({\bm q}, {\bm k}, {\bm l}_1,z)  +   J_1({\bm q}, {\bm k}, {\bm l}_2,z)  \big]\,\,
S_J^{(3)} ({\bm p}, {\bm q}, z x, x)\, ,
\end{align}
with the function $J_1$ given in
Eq.~(\ref{eq:Jcf2}).  Unlike the $C_f^2$ term, all divergent
transverse integrals cancel in this expression and the result is
finite. This is also true for the limit $z \to 0$ where the function
vanishes identically. While an analytic treatment of finite terms is
not possible due to the presence of the jet function, we point out that
the inclusive analysis (with $S_J \to 1$) carried out in Appendix
\ref{sec:incl} confirms the finiteness of this term, revealing at the
same time the presence of single and double logarithms in the
$t$-channel gluon momenta ${\bm l}_i^2$ and $({\bm k} - {\bm l}_i)^2$,
$i = 1,2$.  The final result for the jet case hence reads
\begin{align}
  \label{eq:CFCA_final}
 \left( \frac{d \hat{V}^{(1)}_r}{d J}  \right)_{C_fC_a}  &=
 \int_0^1 d z \int \frac{d^{2} {\bm q}}{\pi}  \, \, 
 \frac{C_a v_q^{(0)}}{C_f} \frac{\alpha_{s}  P_{gq}(z)}{2 \pi}
\big[ J_1({\bm q}, {\bm k}, {\bm l}_1,z) +  J_1({\bm q}, {\bm k}, {\bm l}_2,z) \big]\,\,
\notag \\
& \hspace{2cm}
\Theta\left(\hat{M}_{X,{\rm max}}^2 - \frac{{\bm \Delta}^2}{z(1-z)} \right) S_J^{(3)} ({\bm p}, {\bm q}, z x, x) \, .
\end{align}

The terms with color factor $C_a^2$ read 
\begin{align}
  \label{eq:CaCA}
 \left( \frac{d \hat{V}^{(1)}_r}{d J}  \right)_{C^2_a}  &=
 \int_0^1 d z \int \frac{d^{2 + 2 \epsilon} {\bm q}}{\pi^{1 + \epsilon}}  \, \, 
h^{(0)} C_a^2 \frac{\alpha_{s, \epsilon}}{2 \pi} \Theta\left(\hat{M}_{X,{\rm max}}^2- \frac{{\bm \Delta}^2}{z(1-z)} \right)
\notag \\
&
\qquad \qquad   \frac{P_{gq}(z, \epsilon)}{\Gamma(1-\epsilon) \mu^{2\epsilon}} 
  J_2({\bm q}, {\bm k}, {\bm l}_1, {\bm l}_2) \, \cdot \,
S_J^{(3)} ({\bm p}, {\bm q}, z x, x)\, ,
\end{align}
with the function $J_2$ given in Eq.~(\ref{eq:Jcf2}). As for $J_{1}$
the transverse integral is finite for ${\bm q}^2 \to 0$, the
singularity at $z \to 0$, present in the overall splitting function,
is regulated by the constraint on the diffractive mass. Among all of
the transverse denominators in $J_2$, only the limit ${\bm p}^2 \to 0$
leads to an actual divergence, while all other singularities are
canceled against each other. Introducing a phase space slicing
parameter $\lambda$ to isolate this singularity, and using
\begin{align}
  \label{eq:S_simple}
   \lim_{{\bm p}^2 \to 0} S_J^{(3)} ({\bm p}, {\bm q}, z x, x) & = S^{(2)}_J({\bm k}, zx)
\end{align}
together with
\begin{align}
  \label{eq:MXp20}
  \lim_{{\bm p}^2 \to 0}  \frac{{\bm \Delta}^2}{z(1-z)} = \frac{1-z}{z} {\bm k}^2,
\end{align}
we  find 
\begin{align}
  \label{eq:CaCaJcoll}
 \left( \frac{d \hat{V}^{(1)}_r}{d J}  \right)_{C^2_a, \lambda}   &=   \int_{0}^1 d z \int \frac{d^{2 + 2 \epsilon} {\bm q}}{\pi^{1 + \epsilon}} 
h^{(0)}C_a^2
\frac{\alpha_{s,\epsilon}}{ 2\pi }
 \frac{ P_{gq}(z, \epsilon)}{\Gamma(1-\epsilon) \mu^{2\epsilon}} 
S_J^{(2)} ({\bm k}, z x)
\notag \\
& \hspace{2.5cm }
\cdot 
 \frac{\Theta (\lambda^2 - {\bm p}^2)}{{\bm p}^2}  \Theta\left(\hat{M}_{X,{\rm max}}^2- \frac{(1-z)  {\bm k}^2}{z} \right)
\notag \\
&=
  \frac{\alpha_{s, \epsilon}}{2 \pi} \left(\frac{1}{\epsilon} + { \ln}\frac{\lambda^2}{\mu^2} \right) \, \int_{z_0}^1 d z  \, \, 
h^{(0)}_g
 S_J^{(2)} ({\bm k}, z x ) \cdot P_{gq}(z)
\notag \\
& \qquad \qquad \qquad   \qquad 
+
\frac{\alpha_s^3 C_a^2 C_f}{ \pi (N_c^2 -1)}  \, \int_{z_0}^1 d z  \, \, 
 S_J^{(2)} ({\bm k}, z x ) \cdot  \frac{z-1}{z}   + \mathcal{O}(\epsilon)\, .
\end{align}
Adding the second collinear counterterm we obtain
\begin{align}
  \label{eq:canceledcounter2}
   \left(\frac{d \hat{V}^{(1)}_{r}}{d J} \right)_{C_a^2, b, \lambda } +  \left(\frac{d \hat{V}^{(1)}_{r}}{d J} \right)_{ct,g}
&=
{ \frac{\alpha_s}{2\pi}v_q^{(0)}\frac{C_a^2}{C_f^2}\bigg[\ln\frac{\lambda^2}{\mu_F^2}\int_{z_0}^1dz\,S_J^{(2)}(\bm{k},zx)\,P_{gq}(z)}  
\notag \\
& \qquad 
{ +2\int_{z_0}^1 dz\,\frac{z-1}{z}S_J^{(2)}(\bm{k},zx)\bigg]}   + \mathcal{O}(\epsilon)\, .
\end{align}
To obtain the full result for the terms proportional to $C_a^2$, this contribution must  be added to the remainder, {\it i.e.},
\begin{align}
  \label{eq:CaCA_finite}
 \left( \frac{d \hat{V}^{(1)}_r}{d J}  \right)_{C^2_a, \text{finite}}  &= v_q^{(0)}  \frac{\alpha_{s} }{2 \pi} \frac{C_a^2}{C_f^2}
 \int_0^1 d z \int \frac{d^{2} {\bm q}}{\pi}  \, \,   P_{gq}(z)
 J_2({\bm q}, {\bm k}, {\bm l}_1, {\bm l}_2)  \cdot  S_J^{(3)} ({\bm p}, {\bm q}, z x, x)\,\,
\notag \\
& \qquad 
  \cdot
\Theta\left(\hat{M}_{X,{\rm max}}^2- \frac{{\bm \Delta}^2}{z(1-z)} \right)
\cdot \Theta\left({\bm p}^2 - \lambda^2 \right)
 \, .
\end{align}

\subsection{Final result for the jet impact factor}
\label{sec:final}

Having verified the cancellation of all singular terms, the final result for the jet vertex reads
\begin{align}
  \label{eq:combine}
 \frac{d\hat{V}^{(1)}( {\bm k}, {\bm l}_1, {\bm l}_2; x_J, {\bm k}_J; M_{X,\text{max}}, s_0, \mu_F, \mu)}{d J} \notag \\
 & 
\hspace{-5cm}
= \int_0^1 d x\, \, f_{q/H}(x, \mu_F^2) \cdot \frac{d\hat{V}^{(1)}(x, {\bm k}, {\bm l}_1, {\bm l}_2; x_J, {\bm k}_J; M_{X,\text{max}}, s_0, \mu_F, \mu)}{d J},
\end{align}
with
\begin{align}
  \label{eq:alles_finite}
 &   \frac{d\hat{V}^{(1)}(x, {\bm k}, {\bm l}_1, {\bm l}_2; x_J, {\bm k}_J; M_{X,\text{max}}, s_0)}{d J}  = \notag \\%  \frac{d \hat{V}^{(1)}}{d J} 
&= v_q^{(0)}\frac{\alpha_s}{2 \pi}  \Bigg [  S_J^{(2)}({\bm k}, x) \cdot    { \Bigg[-\frac{\beta_0}{4}\left[\left\{\ln\left(\frac{\bm{l}_1^2}{\mu^2}\right)+\ln\left(\frac{(\bm{l}_1-\bm{k})^2}{\mu^2}\right)+\{1\leftrightarrow 2\}\right\}-\frac{20}{3}\right]-8C_f} \notag \\ &+{ \frac{C_a}{2}\bigg[\bigg\{\frac{3}{2\bm{k}^2}\left\{\bm{l}_1^2\ln\left(\frac{(\bm{l}_1-\bm{k})^2}{\bm{l}_1^2}\right)+(\bm{l}_1-\bm{k})^2\ln\left(\frac{\bm{l}_1^2}{(\bm{l}_1-\bm{k})^2}\right)-4|\bm{l}_1||\bm{l}_1-\bm{k}|\phi_1\sin\phi_1\right\}}
\notag \\
&\quad\quad { -\frac{3}{2}\left[\ln\left(\frac{\bm{l}_1^2}{\bm{k}^2}\right)+\ln\left(\frac{(\bm{l}_1-\bm{k})^2}{\bm{k}^2}\right)\right]-\ln\left(\frac{\bm{l}_1^2}{\bm{k}^2}\right)\ln\left(\frac{(\bm{l}_1-\bm{k})^2}{s_0}\right)}\notag \\
&\quad\quad{ -\ln\left(\frac{(\bm{l}_1-\bm{k})^2}{\bm{k}^2}\right)\ln\left(\frac{\bm{l}_1^2}{s_0}\right)-2\phi_1^2+\{1\leftrightarrow 2\}\bigg\}+2\pi^2+\frac{14}{3}\bigg]}\Bigg]\notag \\&
+
 { \ln \frac{\lambda^2}{\mu^2_F} 
\int_{z_0}^1 d z  \,\, 
 S_J^{(2)}({\bm k}, z x)%
\bigg[
P_{qq}(z)+\frac{C_a^2}{C_f^2}P_{gq}(z)\bigg]}
\notag \\ 
&
{ +\int_0^1 dz\int\frac{d^2\bm{q}}{\pi}\bigg[P_{qq}(z)
\Theta\left(\hat{M}_{X,{\rm max}}^2 -\frac{(\bm{p}-z\bm{k})^2}{z(1-z)}\right)\Theta\left(\frac{|\bm{q}|}{1-z}-\lambda\right)}\notag \\
&
\qquad\qquad\qquad\qquad\quad
{ \times \frac{\bm{k}^2}{\bm{q}^2(\bm{p}-z\bm{k})^2}S_J^{(3)}(\bm{p},\bm{q},(1-z)x,x)}
\notag\\
&
\qquad\qquad\qquad\quad{ +\Theta\left(\hat{M}_{X,{\rm max}}^2 -\frac{\bm{\Delta}^2}{z(1-z)}\right)S_J^{(3)}(\bm{p},\bm{q},zx,x)P_{gq}(z)}
\notag \\
&\qquad\qquad\qquad\qquad\quad{\times \left\{\frac{C_a}{C_f}[J_1(\bm{q},\bm{k},\bm{l}_1,z)+J_1(\bm{q},\bm{k},\bm{l}_2,z)]+\frac{C_a^2}{C_f^2}J_2(\bm{q},\bm{k},\bm{l}_1,\bm{l}_2)\Theta(\bm{p}^2-\lambda^2)\right\}\bigg]}
\notag\\
&
{+4\int_{z_0}^1 dz
\left\{\left[\frac{1-z}{4}\left[1-\frac{2}{z}\frac{C_a^2}{C_f^2}\right]
+\frac{(1+z^2)}{2}\left(\frac{\ln(1-z)}{1-z}\right)_+\right]S_J^{(2)}(\bm{k},zx)+S_J^{(2)}(\bm{k},x)\right\}\bigg],}
\end{align}
and
\begin{align}
  \label{eq:J1=CaCf}
  J_{1} ({\bm q}, {\bm k}, {\bm l}, z) &= \frac{1}{2} \frac{{\bm k}^2}{({\bm q}- {\bm k})^2} \left( \frac{(1-z)^2}{({\bm q} - z  {\bm k})^2} - \frac{1}{{\bm q}^2} \right)
 - \frac{1}{4}
  \frac{1}{ ({\bm q} - {\bm l})^2} \left( \frac{({\bm l} - z\cdot {\bm k})^2}{({\bm q} - z {\bm k})^2} - \frac{{\bm l}^2}{{\bm q}^2} \right)
\notag \\
&
- \frac{1}{4} \frac{1}{({\bm q} - {\bm k} + {\bm l})^2}
\left(
\frac{({\bm l} - (1-z) {\bm k})^2}{({\bm q}- z  {\bm k} )^2}
-
\frac{({\bm l} - {\bm k})^2}{{\bm q}^2}
 \right);
\notag \\
J_{2}({\bm q}, {\bm k}, {\bm l}_1, {\bm l}_2) &= 
\frac{1}{4} \bigg[
\frac{{\bm l}_1^2}{({\bm q} - {\bm k})^2 ({\bm q} - {\bm k} + {\bm l}_1)^2} + 
\frac{({\bm k} -{\bm l}_1)^2}{({\bm q} - {\bm k})^2 ({\bm q} -  {\bm l}_1)^2} 
\notag \\
&
\qquad +
\frac{{\bm l}_2^2}{({\bm q} - {\bm k})^2 ({\bm q} - {\bm k} + {\bm l}_2)^2} 
+ 
\frac{({\bm k} -{\bm l}_2)^2}{({\bm q} - {\bm k})^2 ({\bm q} -  {\bm l}_2)^2} 
-\frac{1}{2} \bigg(
\frac{({\bm l}_1 - {\bm l}_2)^2}{({\bm q} -  {\bm l}_1)^2({\bm q} -  {\bm l}_2)^2}  
\notag \\
&
\qquad 
+
\frac{({\bm k} - {\bm l}_1 - {\bm l}_2)^2}{({\bm q} -  {\bm k}+  {\bm l}_1)^2({\bm q} -  {\bm l}_2)^2}
+
\frac{({\bm k} - {\bm l}_1 - {\bm l}_2)^2}{({\bm q} -  {\bm k}+  {\bm l}_2)^2({\bm q} -  {\bm l}_1)^2} 
\notag \\
& \qquad +
\frac{({\bm l}_1 - {\bm l}_2)^2}{({\bm q} -  {\bm k} +  {\bm l}_1)^2({\bm q} -  {\bm k} +  {\bm l}_2)^2} 
\bigg)
\bigg].
\end{align}
The collinear splitting functions are given in Eqs.~\eqref{eq:splittings}.
%%%%%%%%%%%%%%%%%%%%%%%

\section{Summary \& Outlook}
\label{sec:concl}

We have presented the details of our calculation of the one-loop corrections to the quark induced Mueller-Tang jet vertex within high energy factorization \cite{letter}, making use
of Lipatov's high energy effective action and previous results for the virtual
corrections present in the literature \cite{Fadin:1999df}. Our NLO jet vertex 
can be used for phenomenological studies of non-forward BFKL evolution
in jet-gap-jet events at next-to-leading order accuracy. We find that
the one-loop corrections to the quark induced impact factors are well
defined within collinear factorization, given that a suitable treatment of infrared divergences of Coulomb/Glauber gluon exchange 
in the $t$-channel is provided. In a forthcoming work \cite{toappear} we will present the corresponding  calculation of the next-to-leading order corrections to the gluon initiated jet vertex which are needed for a complete NLO phenomenology of jets events with associated rapidity gaps.

\subsubsection*{Acknowledgments}\label{5}
We thank J. Bartels, V. Fadin and L. Lipatov for constant support for
many years. We are further grateful to D.~Ivanov for a comment at the
meeting ``Scattering Amplitudes \& the Multi-Regge Limit 2014''
concerning contributions of the proton remanent to the diffractive
system.  We acknowledge partial support by the Research Executive
Agency (REA) of the European Union under the Grant Agreement number
PITN-GA-2010-264564 (LHCPhenoNet), the Comunidad de Madrid through
Proyecto HEPHACOS ESP-1473, by MICINN (FPA2010-17747), by the Spanish
Government and EU ERDF funds (grants FPA2007-60323, FPA2011-23778 and
CSD2007- 00042 Consolider Project CPAN) and by GV
(PROMETEUII/2013/007).  M.H.  acknowledges support from the
U.S. Department of Energy under contract number DE-AC02-98CH10886 and
a ``BNL Laboratory Directed Research and Development'' grant (LDRD
12-034).  The research of J.D.M. is supported by the European Research
Council under the Advanced Investigator Grant ERC-AD-267258.

\appendix

\section{The central production vertex}
\label{sec:central_prod}

\begin{figure}[htb]
  \centering
   \parbox{2cm}{\includegraphics[width=2cm]{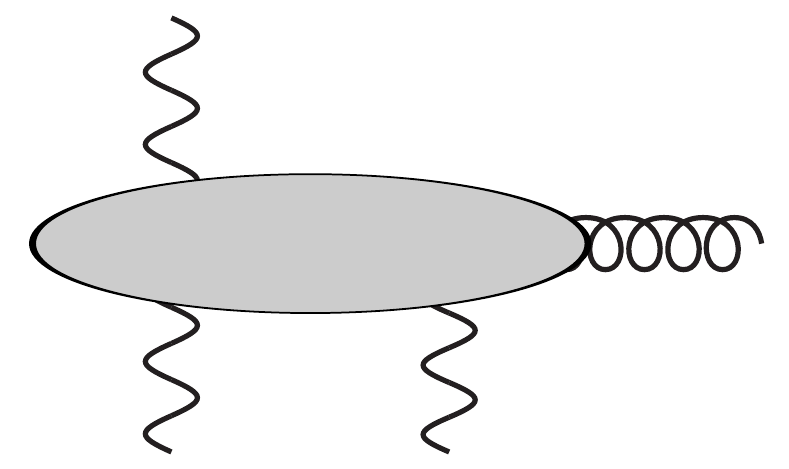}} = 
\parbox{2cm}{\includegraphics[width=2cm]{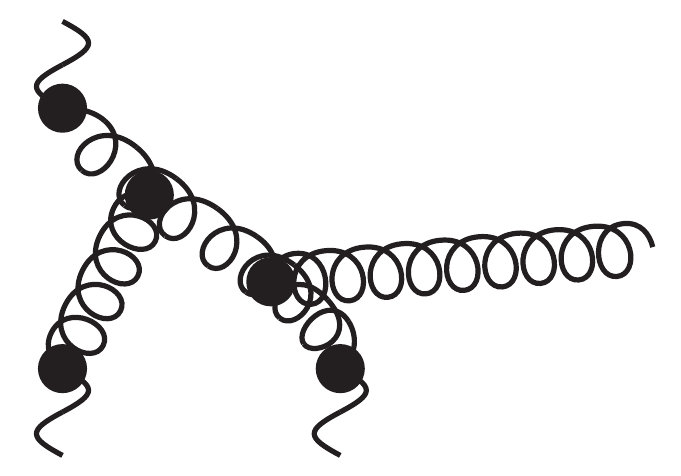}} +
%\parbox{2cm}{\includegraphics[width=2cm]{BV1b.pdf}} +
\parbox{2cm}{\includegraphics[width=2cm]{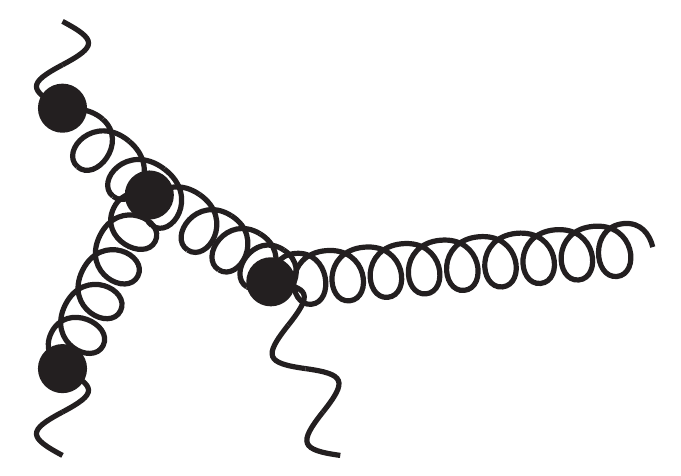}} +
\parbox{2cm}{\includegraphics[width=2cm]{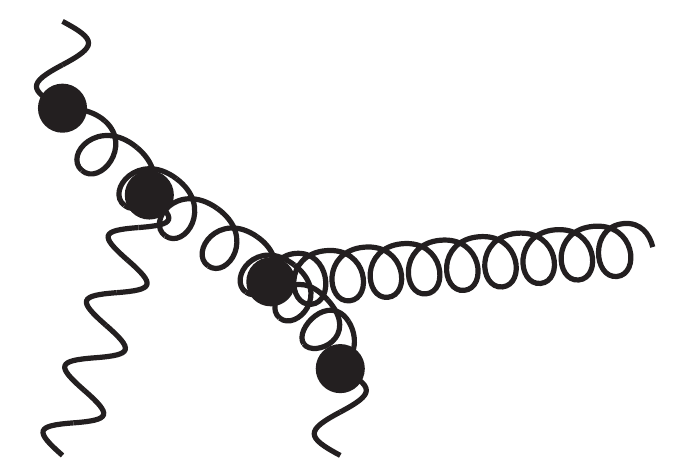}} +
\parbox{2cm}{\includegraphics[width=2cm]{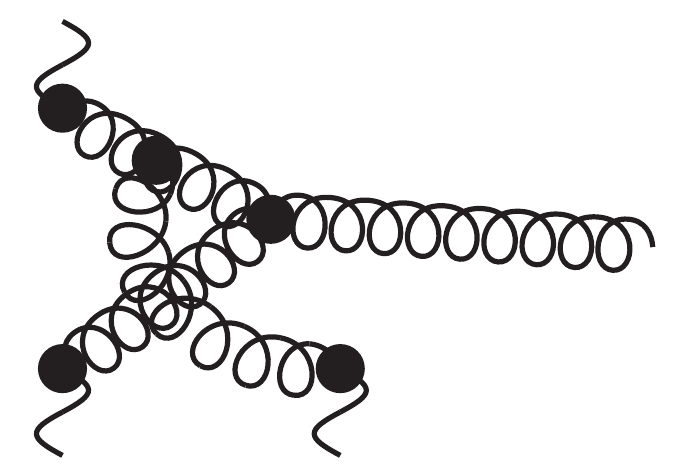}} +
\parbox{2cm}{\includegraphics[width=2cm]{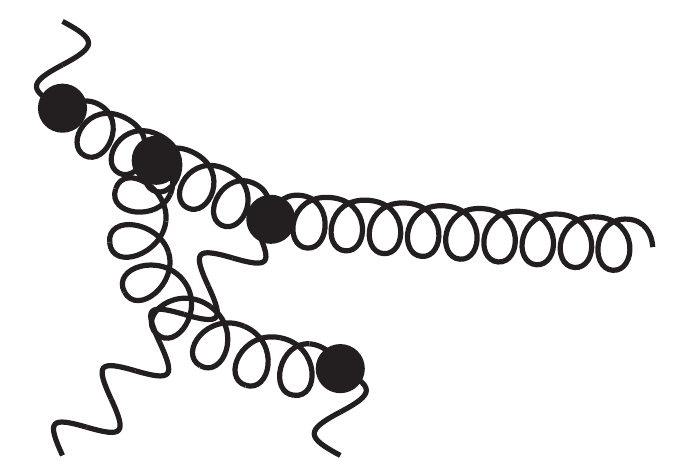}} +
\parbox{2cm}{\includegraphics[width=2cm]{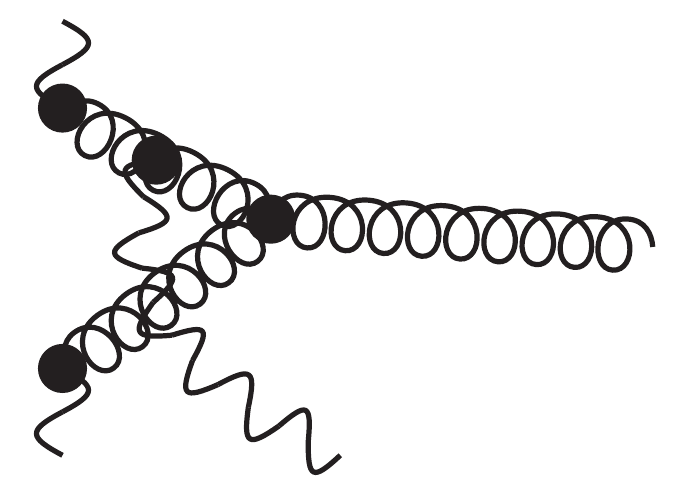}} +
\parbox{2cm}{\includegraphics[width=2cm]{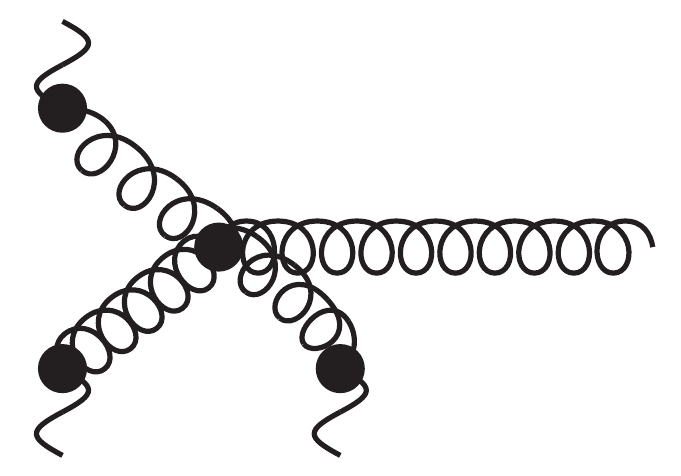}} +
\parbox{2cm}{\includegraphics[width=2cm]{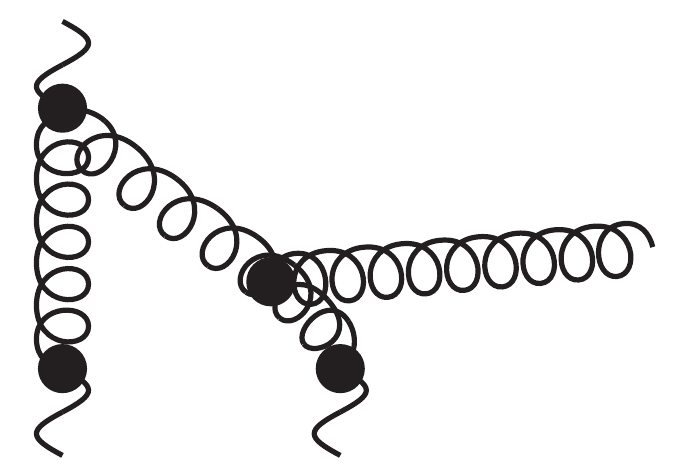}} +
\parbox{2cm}{\includegraphics[width=2cm]{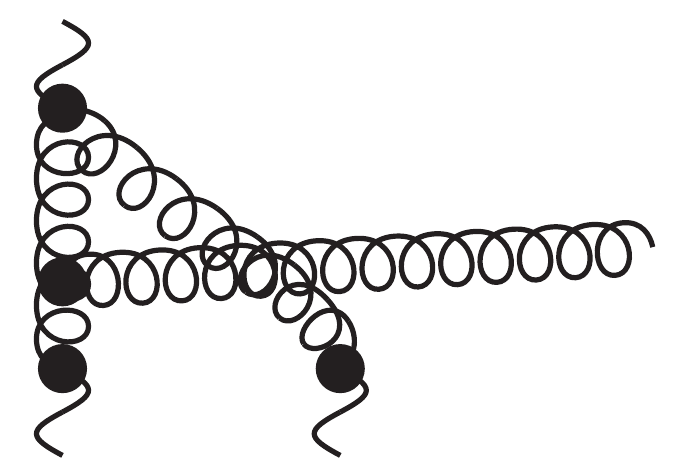}} +
\parbox{2cm}{\includegraphics[width=2cm]{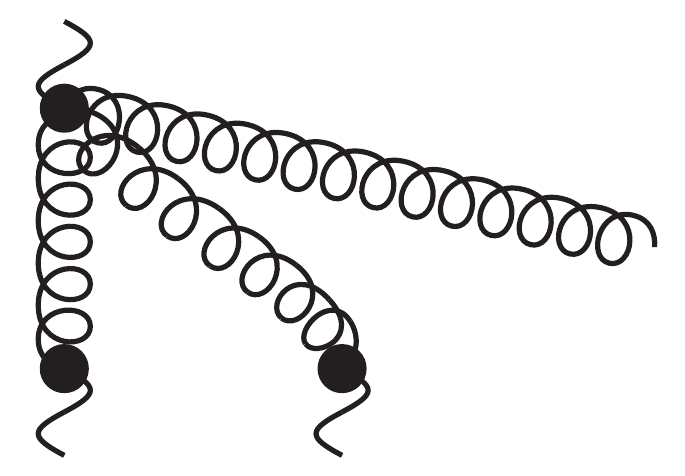}}

  \caption{real NLO diagrams: the reggeized gluon + 2 reggeized gluon $\to$ gluon vertex}
  \label{fig:RG2R}
\end{figure}

 Feynman diagrams for the determination of the 
$r_-^*(p) + r_+^*(l) + r_+^*(k-l) \to g(q)$ amplitude are given in
Fig.~\ref{fig:RG2R}.
From Eq.~(\ref{eq:kinematic}), the momenta have the following Sudakov decomposition
\begin{align}
  \label{eq:Suda_central}
  p & = p^+ \frac{n^-}{2} + {\bm p} & 
  l & =  l^- \frac{n^+}{2} + {\bm l} \notag  \\
  k & = k^- \frac{n^+}{2} + {\bm k} & q & = q^+ \frac{n^-}{2} + q^- \frac{n^+}{2} + {\bm q} \,.
\end{align}
With
\begin{align}
  \label{eq:def_ampRG2R}
  i\mathcal{M}_{r^*g2r^*}^{ab_1b_2c} & =  \parbox{3cm}{\includegraphics[width=2.5cm]{BV.pdf}} % prepare a diagram with label
\end{align}
we obtain
\begin{align}
  \label{eq:ampRg2Rintproje}
a_{r^*2r^*  g}^{ac} =  \int \frac{d l^-}{8 \pi}  i\mathcal{M}_{r^*2r^* g}^{ab_1b_2c}  P^{b_1b_2} & =  -
\frac{g^2 C_a \delta^{ac} {\bm p}^2}{(N_c^2 - 1)^{1/2}} \left[
2 \frac{{\bm p} \cdot {\bm \epsilon}}{{\bm p}^2}
 - 
\frac{ ({\bm p} - {\bm l}_1 )\cdot {\bm \epsilon} }{({\bm p} - {\bm l}_1 )^2}
 + 
\frac{({\bm q} - {\bm l}_1 ) \cdot {\bm \epsilon} }{({\bm q} - {\bm l}_1 )^2}
\right],
\end{align}
where we used the polarization vectors of Eqs.~\eqref{eq:rhg1},
\eqref{eq:rhgpara} for the real gluon with momentum $q$.  Absorbing
also half of the propagators of the internal reggeized gluon line into
the RP2R vertex, we obtain at cross-section level for the
`$RG2R$'-vertex
\begin{align}
  \label{eq:Vcross-section}
  V_{r^*2r^*g}  ({\bm k}, {\bm q})  &  d \ln q^+ = \int
 \frac{d k^- d{p}^+}{(2 \pi)^2 (4 \pi)^{2 + 2\epsilon}}
\left(\frac{1/4}{{\bm p}^2}\right)
P^{aa'}
\left| a_{r^*2r^*  g}^{ac} \left( a_{r^*2r^*  g}^{a'c}\right)^* \right| 
d\Phi^{(1)} \notag \\
= & 
\frac{\alpha_s^2 C_a^2 {\bm p}^2}{2 \pi \mu^{4\epsilon} \Gamma^2(1-\epsilon)  (N_c^2 -1)^{3/2} } \left[
 \frac{{\bm p}}{{\bm p}^2}
 - 
\frac{1}{2}\frac{ ({\bm p} - {\bm l}_1 ) }{({\bm p} - {\bm l}_1 )^2}
 + 
\frac{1}{2}
\frac{({\bm q} - {\bm l}_1 )  }{({\bm q} - {\bm l}_1 )^2}
\right]
\notag \\
&
\hspace{5cm}
\cdot
\left[
 \frac{{\bm p} }{{\bm p}^2}
 - 
\frac{1}{2}\frac{ ({\bm p} - {\bm l}_2 )}{({\bm p} - {\bm l}_2 )^2}
 + 
\frac{1}{2}
\frac{({\bm q} - {\bm l}_2 )  }{({\bm q} - {\bm l}_2 )^2}
\right] d \ln q^+.
\end{align}
Here the  1-particle phase space has been taken to be
\begin{align}
  \label{eq:1ptspace_central}
  d\Phi^{(1)} & = 2 \pi \delta(p^+ k^- - {\bm q}^2).
\end{align}
 Momentum conservation ${\bm k} = {\bm q} + {\bm p}$ is also  implied.
For the coupling of a single reggeized gluon to a quark, we obtain at cross-section level
\begin{align}
  \label{eq:qark_singRatXsec}
  \tilde{h}^{(0)}({\bm p}^2) & = \frac{1}{2} \sum_{\text{spin}} \frac{1}{N_c} \sum_{\text{color}} \int \frac{d p^-}{ 4 p_a^+ (2 \pi)^{2 + \epsilon} {\bm p}^2}  P^{aa'}\left| \mathcal{M}^a_{qr^*  q} \left(\mathcal{M}^{a'}_{qr^*  q} \right)\right|^2  d \Phi^{(1)} \notag \\
&=   \frac{\alpha_s C_f 2^{1 + \epsilon}}{\Gamma(1-\epsilon) \mu^{2\epsilon}}\frac{C_f}{(N_c^2 -1)^{1/2} {\bm p}^2},
\end{align}
where the 1-particle phase space is understood in this case as
\begin{align}
  \label{eq:1ptspace_central}
  d\Phi^{(1)} & = 2 \pi \delta(p_a^+ p^- - {\bm p}^2).
\end{align}
 The complete high energy factorized cross-section for the process $q + 2r^* \to q + g$ is given by
 \begin{align}
   \label{eq:Hfact}
   h^{(1), \text{fact.}}_{qg} d^{2 + 2\epsilon} {\bm q}\, d \ln q^+  & = \frac{1}{\pi^{1 + \epsilon}}  \tilde{h}^{(0)} ({\bm p}^2) \cdot V_{r^*2r^*g}({\bm k}, {\bm q})  
d^{2 + 2\epsilon} {\bm q}\, d\ln q^+ \, .
 \end{align}
 This provides the starting point for a resummation of logarithms in the
 partonic diffractive mass $\hat{M}_X^2 = p_a^+ k^- - {\bm k}^2$ build
 from the quark-gluon final state, see \cite{tripleReggeStuff} for a
 related study.

\section{The inclusive Pomeron quark impact factor}
\label{sec:incl}

In the following we determine the inclusive analog to the Mueller-Tang
jet impact factor. To ease the calculation we take the cut-off on the
diffractive mass in the limit $M_{X. \text{max}}^2\to \infty$.  This
is sufficient to have a simple analytic check on the exclusive
Mueller-Tang impact factor determined in Sec.~\ref{sec:jet}.
The collinear counterterm reads in this  case
\begin{align}
  \label{eq:ct_coll}
&   \lim_{M_{X, \text{max}}^2 \to \infty} \bigg[ - \frac{\alpha_s}{2\pi} \left( \frac{1}{\epsilon} +  \ln \frac{\mu_F^2}{\mu^2}\right)  h^{(0)}_g   C_f 
\int_{x_0}^1  d x
%\notag \\ & \qquad \qquad \qquad \qquad \qquad 
 %
  f_q\left( x,  \mu_F^2   \right)
\left(2 \ln \frac{x \cdot M_{X, \text{max}}^2}{ {\bm k}^2} - \frac{3}{2} \right) \bigg]\; ,
\end{align}
with $x_0 = {\bm k}^2/M^2_{X, \text{max}} $
The inclusive real corrections are  at partonic level given by 
\begin{align}
  \label{eq:resultss}
 h^{(1)}_{r, qg}& =  \lim_{M^2_{X, \text{max}} \to \infty}
\int_0^1 d z \int \frac{d^{2 + 2 \epsilon} {\bm q}}{\pi^{1 + \epsilon}}  \theta\left(xM_{X, \text{max}}^2 -  \hat{M}_X^2  - (1-x){\bm k}^2 \right)
\notag \\
&
h^{(0)}
\frac{\alpha_{s,\epsilon}}{ 2\pi }
 \frac{ P_{gq}(z, \epsilon)}{\Gamma(1-\epsilon) \mu^{2\epsilon}}
 \bigg[   C_f   
 \left(
\frac{{\bm \Delta}}{{\bm \Delta}^2} - \frac{{\bm q}}{{\bm q}^2}
\right)
%\notag \\
%& \qquad 
-
{C_a}\left( \frac{{\bm p}}{{\bm p}^2} + \frac{1}{2} \frac{{\bm \Sigma}_1}{{\bm \Sigma}_1^2} + \frac{1}{2} \frac{{\bm \Upsilon}_1}{{\bm \Upsilon}^2_1} \right)
  \bigg] 
\notag \\
& \hspace{4cm}
 \cdot
 \left[  C_f    
\left(
\frac{{\bm \Delta}}{{\bm \Delta}^2} - \frac{{\bm q}}{{\bm q}^2}
\right)
-
{C_a}{}\left( \frac{{\bm p}}{{\bm p}^2} + \frac{1}{2} \frac{{\bm \Sigma}_2}{{\bm \Sigma}_2^2} + \frac{1}{2} \frac{{\bm \Upsilon}_2}{{\bm \Upsilon}^2_2} \right)
  \right],
\end{align}
with $\hat{M}_X^2 = (p_a + k)^2 =(p + q)^2$. The evaluation of the integrals over the terms proportional to $C_f^2$ is straightforward and yields
\begin{align}
  \label{eq:JCfintz}
 C_f I_{C_f^2}  =  \int_0^1 dz P_{gq}(z, \epsilon)  \int \frac{d^{2 + 2\epsilon} {\bm q}}{ \pi^{1 + \epsilon}} \frac{z^2 {\bm k}^2}{{\bm \Delta}^2 {\bm q}^2} & =  2 C_f \frac{c_\Gamma(\epsilon)}{\epsilon}\bigg[\frac{1}{\epsilon} - \frac{3 - 2\epsilon}{2 + 4 \epsilon} \bigg] ({\bm k}^2)^\epsilon,
\end{align}
with
\begin{align}
  \label{eq:cG}
  c_\Gamma& = \frac{\Gamma(1-\epsilon) \Gamma^2(1 + \epsilon)}{\Gamma(1 + 2 \epsilon)}.
\end{align}
To evaluate the $C_a C_f$ contributions we make use of the integral $K_2'$ defined  and calculated up to order $\epsilon$ in \cite{Fadin:1999df}, Eqs.~(6.17)  and (6.18), 
\begin{align}
  \label{eq:k2'}
  K_2' ({\bm q}^2, {\bm {\bm q}_1}, {\bm q}_2, \epsilon) & =   \int_0^1 \frac{dz}{z} [2(1-z) + (1 + \epsilon) z^2]
\bigg[
 \bigg( \left\{\left[ z {\bm q}_1 + (1-z) {\bm q}_2 \right]^2\right\}^\epsilon - \left[   (1-z)^2 {\bm q}_2^2 \right]^\epsilon \bigg) \notag \\
& \hspace{8cm} + ({\bm q}_2 \to {\bm q}_1) \bigg]
\notag \\
&
\hspace{-2cm}= \epsilon \left [ 1 + \frac{1}{2} \ln^2 \left(\frac{{\bm q}_1^2}{{\bm q}_2^2} \right) 
 - \frac{3}{2} \frac{({\bm q}_1^2 - {\bm q}_2^2)}{{\bm q}^2} \ln \left(\frac{{\bm q}_1^2}{{\bm q}_2^2} \right) - 6 \frac{|{\bm q}_1| |{\bm q}_2|}{{\bm q}^2} \theta 
 \sin \theta + 8 \psi'(1) - 2 \theta^2
\right],
\end{align}
with ${\bm q} = {\bm q}_1 - {\bm q}_2$ and $\theta$ the angle between ${\bm q}_1$ and ${\bm q}_2$ such that $|\theta| \leq \pi$. We further introduce a second integral
\begin{align}
  \label{eq:K2''}
  K_2^{''}({\bm q}_1^2, \epsilon) & =  \int_0^1 \frac{dz}{z} [2(1-z) + (1 + \epsilon) z^2] \bigg( \left[ (1-z)^2 {\bm q}_1^2 \right]^\epsilon  - \left[ {\bm q}_1^2\right]^\epsilon \bigg)
\notag \\
&=
\left[2 \psi(1) - 2 \psi (3 + 2\epsilon) + \frac{ 6 + 13 \epsilon + 3 \epsilon^2 - 2 \epsilon^3}{(1 + 2\epsilon) (2 + 2\epsilon)}     \right]\, .
\end{align}
We  obtain
\begin{align}
  \label{eq:JCaCf2}
 C_f I_{C_f C_a} (i) = \int_0^1 dz P_{gq}(z,\epsilon )  \int \frac{d^{2 + 2\epsilon} {\bm q}}{ \pi^{1 + \epsilon}} &  J_1({\bm q}, {\bm k}, {\bm l}_i, z)
= 
C_f \frac{c_\Gamma(\epsilon)}{2\epsilon}   \bigg[
2 K_2^{''} ({\bm k}^2, \epsilon)
 - K_2^{''} ({\bm l}_i^2, \epsilon) 
\notag \\
&
- K_2^{''} (({\bm k} - {\bm l}_i)^2,\epsilon)
 -
K_2' ({\bm k}^2, {\bm l_i}^2, ({\bm l_i} - {\bm k})^2, \epsilon)
\bigg],
\end{align}
where the angle $\theta$ in $K_2'$ must be defined for vectors ${\bm q}_1 = {\bm l}_i$ and  ${\bm q}_2 = {\bm k} - {\bm l}_i$ such that ${\bm q}_1 - {\bm q}_2 = {\bm k}$ holds.
For the terms proportional to $C_a^2$ we have for  non-zero ${\bm q}^2$
\begin{align}
  \label{eq:cutoff_z}
  z > \bar{z}_0 =  \frac{{\bm q}^2}{x M_{X, \text{max}}^2}.
\end{align}
The region $z \to 1$ is on the other hand already finite and the limit$M_{X, \text{max}}^2 \to \infty$ can be taken immediately in this case.
We  obtain
\begin{align}
  \label{eq:int_PqgX}
  \lim_{ M_{X, \text{max}}^2\to \infty}\int \limits_{\bar{z}_0}^1 P_{gq}(z, \epsilon) & =   2 \ln  \frac{x M_{X, \text{max}}^2}{{\bm q}^2} - \frac{3}{2} + \frac{\epsilon}{2}.
\end{align}
 The appearance of a  $\ln1/{\bm q}^2$ in  Eq.~(\ref{eq:int_PqgX}) requires a
new integral which, up to order $\epsilon^0$, has been evaluated in
\cite{Fadin:1999de}, Eqs.(A1)-(A13). It reads
\begin{align}
  \label{eq:IA}
  \tilde{I}_1 \left(  {\bm l}^2, {\bm k}^2_1, {\bm k}_2^2 \right)
&= \int \frac{d^{2 + 2\epsilon} {\bm q}}{ \pi^{1 + \epsilon}} \ln \left( \frac{1}{{\bm q}^2}\right)  \frac{{\bm l}^2}{ ({\bm q} - {\bm k}_1)^2 ({\bm q} - {\bm k}_2)^2}
\notag  \\
&=
\frac{\Gamma(1-\epsilon) \Gamma^2(1 + \epsilon)}{\Gamma(1 + 2\epsilon)} 
 \left(  {\bm l}^2 \right)^\epsilon
\bigg\{
\frac{1}{\epsilon^2} \left[ 2
 -
 \left(\frac{{\bm k}_1^2}{{\bm l}^2} \right)^\epsilon 
 - 
\left(\frac{ {\bm k}_2^2}{{\bm l}^2} \right)^\epsilon 
 \right]
\notag \\
& \hspace{6cm }  + 4 \psi''(1)\epsilon
+ \ln \frac{{\bm k}_1^2}{{\bm l}^2} \ln \frac{ {\bm k}_2^2}{{\bm l}^2}
+
\frac{2}{\epsilon} \ln \frac{1}{ {\bm l}^2}
\bigg\},
\end{align}
with ${\bm l}^2 = ({\bm k}_1 - {\bm k}_2)^2$. Altogether we obtain
  \begin{align}
    \label{eq:KA}
 &   K_A \left( ({\bm k}_1 - {\bm k}_2)^2, {\bm k}_1^2, {\bm k}_2^2, x \cdot M_{X, \text{max}}^2, \epsilon \right) 
=    
\bigg[ 
 \int_{\bar{z}_0}^1 d z  P_{gq}(z,\epsilon) 
 \bigg]
\frac{({\bm k}_1 - {\bm k}_2)^2}{({\bm q} - {\bm k}_2)^2({\bm q} - {\bm k}_1)^2  } =c_\Gamma(\epsilon)
\notag \\
& \quad\times \left[ ({\bm k}_1 - {\bm k}_2)^2 \right]^\epsilon  \bigg \{ 
\frac{2}{\epsilon} \left [  2 \ln \frac{x M_{X, \text{max}}^2}{ ({\bm k}_1 - {\bm k}_2)^2 }
 - \frac{3}{2} + \frac{\epsilon}{2}  \right]
+ 
2 \ln \frac{ {\bm k}_1^2}{  ({\bm k}_1 - {\bm k}_2)^2  } \ln \frac{ {\bm k}_2^2}{  ({\bm k}_1 - {\bm k}_2)^2  }
\notag \\
& \qquad \qquad \qquad \qquad + \frac{2}{\epsilon^2}
 \bigg[ 2 - \left( \frac{ {\bm k}_1^2}{  ({\bm k}_1 - {\bm k}_2)^2   } \right)^\epsilon - \left( \frac{ {\bm k}_2^2}{  ({\bm k}_1 - {\bm k}_2)^2   } \right) \bigg] 
 + 8 \psi^{''} (1) \epsilon 
\bigg\}.
  \end{align}
This integral then allows to express the $C_a^2$ contribution in the following way,
\begin{align}
  \label{eq:Ca^2result}
C_f  I_{C_a^2} & = \int_{\bar{z}_0}^1 d z   P_{gq}(z,\epsilon)   
 J_2({\bm q}, {\bm k}, {\bm l}_1, {\bm l}_2) \notag \\
= & 
\frac{C_f}{4} \bigg\{ 
K_A \left( {\bm l}_1^2, ({\bm l}_1 - {\bm k})^2, {\bm k}^2 , x M_{X, \text{max}}^2, \epsilon \right)  
+
K_A \left( ({\bm l}_2 - {\bm k})^2, {\bm k}^2,  {\bm l}_2^2   , x M_{X, \text{max}}^2, \epsilon \right) 
\notag \\
& +
 K_A \left(  ({\bm l}_1 - {\bm k})^2, {\bm l}_1^2,  {\bm k}^2   , x M_{X, \text{max}}^2, \epsilon \right) 
+
 K_A \left(   {\bm l}_2^2, {\bm k}^2, ({\bm l}_2 - {\bm k})^2,   x M_{X, \text{max}}^2, \epsilon \right) 
\notag \\ & -
\frac{1}{2}
\bigg[
 K_A \left( ({\bm l}_1 - {\bm l}_2)^2, {\bm l}_1^2,  {\bm l}_2^2  ,x M_{X, \text{max}}^2, \epsilon \right)  
\notag \\
& 
 \hspace{3cm}
+
  K_A \left( ({\bm l}_1 + {\bm l}_2 - {\bm k})^2, ( {\bm l}_1 - {\bm k})^2,  {\bm l}_2^2   ,x M_{X, \text{max}}^2, \epsilon \right) 
\notag \\
& 
+
  K_A \left(  ({\bm l}_1 + {\bm l}_2 - {\bm k})^2, {\bm l}_1^2, x M_{X, \text{max}}^2, \epsilon \right)  
\notag \\
& 
 \hspace{3cm}+
  K_A \left( ({\bm l}_1 - {\bm l}_2)^2,  ( {\bm l}_2 - {\bm k})^2, ( {\bm l}_1 - {\bm k})^2  ,x M_{X, \text{max}}^2, \epsilon \right) 
\bigg]   
\bigg\}.
\end{align}
Our final result then reads
\begin{align}
  \label{eq:result_final}
 h_{r}^{(1)} &= h^{(0)} \frac{\alpha_s}{2\pi} \frac{\mu^{-2\epsilon}}{\Gamma(1-\epsilon)} \bigg[
C_f^3 I_{C_f^2} +C^2_f C_a I_{C_fC_a}(1)+  C_f^2C_a I_{C_fC_a}(2) + C_a^2 C_f  I_{C_a^2}
\bigg] .
\end{align}
Expanding in $\epsilon$ we find the following divergent terms
\begin{align}
  \label{eq:H1divx}
  \frac{h_{r}^{(1)}}{h^{(0)}} & = \alpha_s \frac{C^3_f}{\pi} \left( \frac{1}{\epsilon^2} - \frac{3}{2\epsilon} + \frac{\ln {\bm k}^2}{\epsilon} \right) + 
  \frac{\alpha_s C_a^2 C_f}{ \pi \epsilon}
 \left( \ln \frac{x M_{X, \text{max}}^2}{{\bm k}^2} - \frac{3}{4} 
 \right) + \mathcal{O}({\epsilon^0}).
\end{align}
Combining these terms with the virtual corrections in 
Eq.~\eqref{eq:H1div}, including ultraviolet renormalization, and the
collinear counterterm of Eq.~(\ref{eq:ct_coll}), we find that all poles
in $\epsilon$ cancel and the result is finite in the limit
$\epsilon \to 0$.

\end{document}